\documentclass[aps,superscriptaddress,showpacs,nofootinbib,tightenlines]{revtex4-1}
\usepackage{mathrsfs}
\usepackage{latexsym,bm}
\usepackage{graphicx}
\usepackage{indentfirst}
\usepackage{slashed}
\usepackage{amssymb}
\usepackage{amsmath}
\usepackage{bbm}
\usepackage{color}
\usepackage[dvipsnames]{xcolor}
\usepackage{epsfig}
\usepackage[titletoc]{appendix}
\usepackage{multirow}%
\usepackage{rotating}
\usepackage{epstopdf}
\usepackage{extarrows}%add text under or on the arrows
\usepackage{tabularx}
\usepackage{soul} % use this (many fancier options)
\usepackage{xcolor}
\usepackage{lipsum}
\usepackage[colorlinks=true,allcolors=BlueViolet,hyperfootnotes=false]{hyperref}

\usepackage[utf8]{inputenc}

\newcommand{\be}{\begin{equation}} \newcommand{\ee}{\end{equation}}
\newcommand{\ba}{\begin{array}{c}} \newcommand{\ea}{\end{array}}
\newcommand{\bea}{\begin{eqnarray}} \newcommand{\eea}{\end{eqnarray}}

\newcommand\tstrut{\rule{0pt}{2.0ex}}       % "top" strut
\begin{document}
\title{\Large $\bar B_c \to \eta_c$, $\bar B_c \to J/\psi$ and $\bar B \to D^{(*)}$ semileptonic decays including new physics.}

\author{Neus Penalva}
\affiliation{Departamento de F\'\i sica Fundamental,\\ Universidad de Salamanca, E-37008 Salamanca, Spain}
\affiliation{Instituto de F\'{\i}sica Corpuscular (centro mixto CSIC-UV), Institutos de Investigaci\'on de Paterna,
Apartado 22085, 46071, Val\`encia, Spain}

\author{Eliecer Hern\'andez}
\affiliation{Departamento de F\'\i sica Fundamental 
  e IUFFyM,\\ Universidad de Salamanca, E-37008 Salamanca, Spain}

\author{Juan~Nieves}
\affiliation{Instituto de F\'{\i}sica Corpuscular (centro mixto CSIC-UV), Institutos de Investigaci\'on de Paterna,
Apartado 22085, 46071, Val\`encia, Spain}

\date{\today}
% \date{December 4, 2018}
\begin{abstract}
We apply the general formalism derived in  N. Penalva {\it et al.} 
[Phys. Rev. D 101, 113004 (2020)]
to the semileptonic decay of pseudoscalar mesons containing a $b$ quark. 
While present $\bar B\to D^{(*)}$ data give the strongest evidence in favor of
 lepton flavor  universality
violation, the observables that are normally considered are not able to distinguish
between different new physics (NP) scenarios. In the above reference we discussed
the relevant role that the various contributions to the double differential decay widths $d^2\Gamma/(d\omega d\cos\theta_\ell)$ 
and $d^2\Gamma/(d\omega dE_\ell)$ could play to this end. Here $\omega$ is the product of
the two hadron four-velocities, $\theta_\ell$ is the angle made by the final lepton
and final hadron three-momenta in the center of mass of the final 
two-lepton system, and $E_\ell$ is the final charged lepton
energy in the laboratory system. The formalism was applied in  N. Penalva {\it et al.}  to  the analysis of the
 $\Lambda_b\to\Lambda_c$ 
semileptonic decay, showing  the new observables were able to 
 tell apart different NP scenarios. Here we analyze  the $\bar B_c\to \eta_c \tau\bar\nu_\tau$, 
$\bar B_c\to J/\psi\tau\bar\nu_\tau$, 
$\bar B\to D\tau\bar\nu_\tau$ and $\bar B\to D^*\tau\bar\nu_\tau$ semileptonic decays. We find that, as a general rule,  the $\bar B_c \to J/\psi$ observables, even including $\tau$ polarization, 
are less optimal for distinguishing between NP scenarios than those obtained from $\bar B_c \to \eta_c$ decays, or those presented in N. Penalva {\it et al.} for the related $\Lambda_b \to \Lambda_c$ semileptonic decay. Finally, we show that  $\bar B\to D$ and  $\bar B_c\to \eta_c$, and $\bar B\to D^*$ and  $\bar B_c\to J/\psi$ decay observables exhibit similar behaviors.
\end{abstract}
%\keywords{...}

\maketitle
\section{Introduction}

\vspace{1cm}

The  present values of the
${\cal R}_{D^{(*)}}$ ratios ($\ell=e,\mu$)
\bea
{\cal R}_{D}
=\frac{\Gamma(\bar B\to D\tau\bar\nu_\tau)}
{\tstrut\Gamma(\bar B\to D \ell\bar\nu_\ell)}=0.340\pm0.027\pm0.013\ \ ,\ \
{\cal R}_{D^*}
=\frac{\Gamma(\bar B\to D^*\tau\bar\nu_\tau)}
{\Gamma(\bar B\to D^*\ell\bar\nu_\ell)}=0.295\pm0.011\pm0.008.
\eea
are the strongest experimental evidence 
for the possibility of lepton flavor universality violation (LFUV).
These values have been obtained by the Heavy Flavour Averaging Group 
 (HFLAV)~\cite{Amhis:2019ckw} 
 (see also Ref.~\cite{Amhis:2016xyh} for earlier results), from a combined
  analysis of different
experimental data  by the 
BaBar~\cite{Lees:2012xj,Lees:2013uzd}, 
Belle~\cite{Huschle:2015rga,Sato:2016svk,Hirose:2016wfn,Belle:2019rba} 
and LHCb~\cite{Aaij:2015yra,Aaij:2017uff} collaborations together with standard model (SM) 
predictions~\cite{Aoki:2016frl, Bigi:2016mdz, Bigi:2017jbd,
Jaiswal:2017rve, Bernlochner:2017jka},
and they show a tension  with the SM at the level of  $3.1\,\sigma$. 
However, taking only the latest Belle experiment from Ref.~\cite{Belle:2019rba}
the  tension with  SM predictions
reduces to $0.8\,\sigma$ so that new experimental analyses seem to be
 necessary to confirm or rule out
LFUV in $\bar B$ meson decays.
Another source of tension with the SM predictions is in the ratio
\bea
{\cal R}_{J/\psi}=\frac{\Gamma(\bar B_c\to J/\psi\tau\bar\nu_\tau)}
{\Gamma(\bar B_c\to J/\psi\mu\bar\nu_\mu)}=0.71\pm0.17\pm0.18 \label{eq:RpsiLHCB}
\eea
recently measured by the LHCb Collaboration~\cite{Aaij:2017tyk}. This shows 
a  $1.8\,\sigma$ disagreement with SM results that are in the range
$R^{\rm SM}_{J/\psi} \sim 0.25-0.28$~\cite{Anisimov:1998uk,Ivanov:2006ni,
Hernandez:2006gt,Huang:2007kb,Wang:2008xt,Wen-Fei:2013uea, Watanabe:2017mip, Issadykov:2018myx,Tran:2018kuv,
Hu:2019qcn,Leljak:2019eyw,Azizi:2019aaf,Wang:2018duy}.

If the anomalies seen in the data persist, they will be a clear indication 
of LFUV and new physics (NP) beyond the SM will be necessary to explain it. Since the data
for the two first generations of quarks and leptons is in agreement with
SM expectations, NP is assumed to affect just the last quark and lepton 
generation. Its
effects can be studied in a  phenomenological  way  by following an
effective field theory model-independent analysis that includes  
different $b\to c \tau \bar\nu_\tau$  effective operators:
scalar, pseudo-scalar and tensor NP terms, 
 as well as corrections to the SM vector and axial contributions~\cite{Fajfer:2012vx}. Considering only left-handed neutrinos, in the notation
 of Ref.~\cite{Murgui:2019czp} one writes
\bea
H_{\rm eff}&=&\frac{4G_F|V_{cb}|^2}{\sqrt2}[(1+C_{V_L}){\cal O}_{V_L}+
C_{V_R}{\cal O}_{V_R}+C_{S_L}{\cal O}_{S_L}+C_{S_R}{\cal O}_{S_R}
+C_{T}{\cal O}_{T}]+h.c.,
\label{eq:hnp}
\eea
with fermionic operators given by ($\psi_{L,R}=\frac{1 \mp \gamma_5}{2}\psi$)
\be
{\cal O}_{V_{L,R}} = (\bar c \gamma^\mu b_{L,R}) 
(\bar \ell_L \gamma_\mu \nu_{\ell L}), \quad {\cal O}_{S_{L,R}} = 
(\bar c\,  b_{L,R}) (\bar \ell_R  \nu_{\ell L}), \quad {\cal O}_{T} = 
(\bar c\, \sigma^{\mu\nu} b_{L}) (\bar \ell_R \sigma_{\mu\nu} \nu_{\ell L}).
\label{eq:hnp2}
\ee 
The corrections to the SM  are assumed to be generated 
by NP that enter at a much higher energy scale, and 
which strengths at the SM scale
  are governed by  unknown, complex in general, Wilson coefficients ($C_{V_L},C_{V_R},C_{S_L},C_{S_R}$ and $C_{T}$  in Eq.~(\ref{eq:hnp}) ) that 
  should be  fitted to data. For the numerical part of the present work, we take the values for the Wilson coefficients from the analysis carried out in Ref.~\cite{Murgui:2019czp}.

The  findings of these phenomenological studies  show that in fact NP can solve some of the
present discrepancies. However, it is also found that different combinations of NP terms
 could give very similar  results for the ${\cal R}_{D^{(*)}}, {\cal R}_{J/\psi}$ ratios.
Thus, even though those ratios are our present best experimental evidence
 for  the possible existence
of NP beyond the SM, they are not good observables for distinguishing between
 different NP scenarios. 
 
 The  relevant role   that  the various contributions to the two  differential decay widths $d^2\Gamma/(d\omega d\cos\theta_\ell)$ and $d^2\Gamma/(d\omega dE_\ell)$ could play to this end was  analyzed
 in detail in 
Refs.~\cite{Penalva:2019rgt,Penalva:2020xup}. Here, $\omega$ is the product of
the two hadron four-velocities, $\theta_\ell$ is the angle made by the final lepton
and final hadron three-momenta in the center of mass of the final 
two-lepton pair (CM), and $E_\ell$ is the final charged lepton
energy in the laboratory frame (LAB). 

Even in the presence of NP, it is shown that for any charged current semileptonic 
decay with an unpolarized final charged lepton
one can write~\cite{Penalva:2020xup}
\bea
 \frac{d^2\Gamma}{d\omega d\cos\theta_\ell}& =& \frac{G^2_F|V_{cb}|^2
    M'^3M^2}{16\pi^3}
  \sqrt{\omega^2-1}\left(1-\frac{m_\ell^2}{q^2}\right)^2
   A(\omega,\theta_\ell),\nonumber\\  &&A(\omega,\theta_\ell)=
\frac{2\overline{\sum}|{\cal M}|^2}{M^2(1-\frac{m_\ell^2}{q^2})}\Bigg|_{\rm unpolarized}=
a_0(\omega)+a_1(\omega)\cos\theta_\ell
+a_2(\omega)(\cos\theta_\ell)^2,\label{eq:aux1}\\
   \frac{d^2\Gamma}{d\omega dE_\ell}& =& \frac{G^2_F|V_{cb}|^2
    M'^2M^2}{8\pi^3} C(\omega,E_\ell),\nonumber\\ 
&&C(\omega,E_\ell)= \frac{2\overline{\sum}|{\cal M}|^2}{M^2}
\Bigg|_{\rm unpolarized}=c_0(\omega)+c_1(\omega)\frac{E_\ell}M
+c_2(\omega)\frac{E^2_\ell}{M^2},
\eea
where $M,M'$ and $m_\ell$ are the masses of the initial and final hadrons and the final charged
lepton respectively, 
$q^2$ is the four momentum transferred squared (related to $\omega$ via
$q^2=M^2+M^{\prime 2}-2MM'\omega$) and 
${\cal M}$ is the invariant amplitude for the decay. Note that at zero recoil $\theta_\ell$ is not longer defined and thus 
 $a_1(\omega=1)$ and $a_2(\omega=1)$ vanish  accordingly. 
The $a_{0,1,2}$ 
  CM  and $c_{0,1,2}$ LAB  expansion coefficients are 
 scalar functions that depend on $\omega$ and the masses of the particles involved in
 the decay. In the general tensor formalism 
 developed in Refs.~\cite{Penalva:2019rgt,Penalva:2020xup}, it is shown how
 they are determined  in terms  of  the
 16  Lorentz scalar $\widetilde W's$ structure functions  (SFs) that
 parameterize all the hadronic input. These $\widetilde W's$ 
 SFs depend on the  Wilson coefficients 
 ($C's$) and the  genuine hadronic responses ($W's$), the latter
 being scalar functions of the actual  form factors that  parameterize the hadronic
 transition matrix elements for a given decay. 
 The general  expressions for he $a_{0,1,2}$ 
  CM  and $c_{0,1,2}$ LAB  expansion coefficients
  in terms of the $\widetilde W's$ SFs can be found in
   Ref.~\cite{Penalva:2020xup}, where the  hadron tensors associated with
    the different
 SM and NP   contributions (including all possible interferences) are also
  explicitly given\footnote{In fact, full general expressions for both LAB and CM decay distributions, decomposed in  
  helicity contributions of the outgoing charged lepton, can also be found in  \cite{Penalva:2020xup}. }. 

  The fully developed formalism 
 was applied in Ref.~\cite{Penalva:2020xup} to the 
 analysis of the $\Lambda_b\to\Lambda_c\tau\bar\nu_\tau$ decay. The shape
 of the $d\Gamma(\Lambda_b\to\Lambda_c
 \mu\bar\nu_\mu)/d\omega$ differential decay width has already been
  measured by
 the LHCb Collaboration~\cite{Aaij:2017svr} and there are expectations 
 that  the ${\cal
R}_{\Lambda_c}=\frac{\Gamma(\Lambda_b\to\Lambda_c\tau\bar\nu_\tau)}{
\Gamma(\Lambda_b\to\Lambda_c\mu\bar\nu_\mu)}$ ratio may
reach the precision  obtained for ${\cal R}_D$ and 
${\cal R}_{D^*}$~\cite{Cerri:2018ypt}. With the use of Wilson coefficients
 from Ref.~\cite{Murgui:2019czp}, fitted to experimental data in the
  $\bar B$-meson sector,  it is shown in   
Ref.~\cite{Penalva:2020xup} that, with the exception of $a_0$, all the
 other $a_{1,2}$   CM  and $c_{0,1,2}$ LAB  expansion coefficients are able 
 to disentangle 
   between   different NP scenarios, i.e. different fits to the available data
  that   otherwise give   very similar  values for  the ${\cal R}_{\Lambda_c}$, 
    and ${\cal R}_{D^{(*)}},{\cal R}_{J/\psi}$ ratios, or  the  corresponding
     $d\Gamma/d\omega$ distributions.
     
\vspace{0.15cm}
In this work we apply  the general formalism of Ref.~\cite{Penalva:2020xup} to the study of
the  semileptonic $P_b\to P_c $ and $P_b\to P_c^*$ decays, with  
$P_b$ and $P_c$ pseudoscalar mesons ($\bar B_c$ or $\bar B$ and $\eta_c$ or $D$, respectively) and 
$P_c^*$ a vector meson ($J/\psi$ or $D^*$). 
%\vspace{0.05cm}

For the case of $\bar B\to D^{(*)}$ decays, the hadronic matrix elements are relatively well known.  
In fact, there exist some experimental $q^2-$shape information~\cite{Lees:2013uzd, Huschle:2015rga}, which 
can be used to constrain the transition form factors. They are then computed using a heavy quark effective theory parameterization that includes  corrections of order $\alpha_s$, $\Lambda_{QCD}/m_{b,c}$ and partly $(\Lambda_{QCD}/m_c)^2$~\cite{Bernlochner:2017jka}. Moreover,  some inputs from lattice quantum Chromodynamics (LQCD)~\cite{Bailey:2014tva, Lattice:2015rga,Na:2015kha, Harrison:2017fmw}, light-cone \cite{Faller:2008tr} and QCD sum rules ~\cite{Neubert:1992wq,Neubert:1992pn, Ligeti:1993hw} are also available.  In addition, a considerable number of phenomenological studies~\cite{Datta:2012qk, Duraisamy:2013pia,Duraisamy:2014sna, Ligeti:2016npd,Becirevic:2016hea, Bhattacharya:2018kig,Blanke:2018yud,Murgui:2019czp,Blanke:2019qrx,Alok:2019uqc,Jaiswal:2020wer,Iguro:2020cpg,Kumbhakar:2020jdz, Bhattacharya:2020lfm} have already discussed some specific details of the CM  $d^2\Gamma/(d\omega d\cos\theta_\ell)$ distribution, as for instance the $\tau-$forward-backward and polarization asymmetries\footnote{Indeed, Eq.~\eqref{eq:aux1} for the  CM angular distribution of the semileptonic decays of a pseudoscalar meson to a daughter pseudoscalar or vector meson is well known. The coefficient functions are commonly given
in terms of the mediator helicity amplitudes, and  several studies propose a series of quantities  to check for the presence of NP,  see for instance Ref.~\cite{Becirevic:2016hea}.}. Other observables present in the full four-body
$\bar B\to D^*(D\pi)\tau\bar\nu_\tau$ angular distribution, and their power to distinguish between different NP scenarios, have also been  addressed in Refs.~\cite{Duraisamy:2013pia,Duraisamy:2014sna} and ~\cite{Becirevic:2016hea}, with the emphasis in the first two works focused on CP violating quantities, while in the latter one the possible  pollution of $\bar B\to D^*(D\pi)_{S}\ell\bar\nu_\ell$ by the $\bar B\to D^*_0(2400)\ell\bar\nu_\ell$, with $D^*_0(2400)$ a broad isoscalar $S-$wave meson, is also analyzed.  In Ref.~\cite{Ligeti:2016npd}, the $\bar B\to D^*(D Y)\tau(X \nu_\tau)\bar\nu_\tau$, with $Y = \pi$ or $\gamma$ and $X=\ell \bar\nu_\ell$ or $\pi$, reactions are studied, paying attention to interference effects in the full phase space of the visible $\tau$ and $D^*$ decay products in the presence of NP.  Such effects are missed in analyses that treat the $\tau$ or $D^*$ or both as stable, and in addition, it is argued in \cite{Ligeti:2016npd}  that analyses including more differential kinematic information can provide greater  discriminating power for NP, than single kinematic variables alone.
The full five-body $\bar B\to D^*(D\pi)\tau(\pi\nu_\tau)\bar\nu_\tau$ angular distribution has also been  analyzed
in Ref.~\cite{Bhattacharya:2020lfm},  where  it is claimed that  magnitudes and  relative phases of all the NP Wilson coefficients can be extracted from a fit to this full five-body angular distribution.

In this work, with respect to the $\bar B \to D^{(*)}$ transitions,  we have used the set of form factors and Wilson coefficients found in \cite{Murgui:2019czp} and, in addition to the CM distribution, we present in 
Sec.~\ref{sec:B2D} for the first time details of the  LAB $d^2\Gamma/(d\omega E_\ell)$ differential decay width and its usefulness to distinguish between different NP scenarios.

The analysis of the $\bar B_c\to \eta_c,J/\psi$ transitions is more novel, with a less abundant previous 
literature~\cite{Dutta:2017xmj,Tran:2018kuv,Leljak:2019eyw}. These works analyze  NP effects on the CM angular distribution observables,  with right-handed neutrino terms  also considered in \cite{Dutta:2017xmj}.   Here, in Sec.~\ref{sec:Bc2cc}, we discuss the relevance  of NP in the 
$d^2\Gamma/(d\omega d\cos\theta_\ell)$ and $d^2\Gamma/(d\omega dE_\ell)$ distributions for both decays, 
highlighting  the  observables  that are able to tell apart different NP fits among those preferred in \cite{Murgui:2019czp}. We also show  
results with a polarized final $\tau$ lepton (Subsec.~\ref{app:ascspol}).

As for the $\bar B_c\to \eta_c$ and $\bar B_c\to J/\psi$ hadronic matrix elements, different theoretical schemes were examined in  
Ref.~\cite{Tran:2018kuv}. Form-factors obtained within the non-relativistic (NRQM), the covariant light-front and the covariant 
confined quark models of Refs.~\cite{Hernandez:2006gt}, \cite{Wang:2008xt} and \cite{Tran:2018kuv} respectively, together with those 
derived in perturbative QCD (pQCD)~\cite{Wen-Fei:2013uea} and the QCD and non-relativistic QCD sum rule approaches of Refs.~\cite{Kiselev:1999sc,Kiselev:2002vz} were compared in ~\cite{Tran:2018kuv}.  On the other hand, a  model independent global study of only the form factors involved in the SM matrix elements was conducted in Ref.~\cite{Cohen:2019zev}. It exploited preliminary lattice-QCD data from
Ref.~\cite{Colquhoun:2016osw}, dispersion relations and heavy-quark symmetry. Importantly, such analysis provided realistic uncertainty bands for the relevant form factors. Finally, very recently the HPQCD collaboration has reported a LQCD determination of the SM vector and axial form factors for the  $\bar B_c\to J/\psi$ semileptonic decay~\cite{Harrison:2020gvo}. These LQCD results have been used  in Ref.~\cite{Harrison:2020nrv} to evaluate  ${\cal R}_{J/\psi}$ and angular distributions observables both, within the SM and including $C_{V_L}$ and $C_{V_R}$ NP terms. We note, however,  that in order to calculate the effect of all NP terms in Eq.~(\ref{eq:hnp}), some additional form factors, not determined in Refs.~\cite{Cohen:2019zev,Harrison:2020gvo}, are also needed.

In summary, there are different theoretical  determinations of the form factors but, to our knowledge, there exist
 neither  shape-measurements nor  systematic LQCD calculations, except for the very recent work of the HPQCD collaboration, and only for the SM vector and axial form factors of the $\bar B_c\to J/\psi$ decay.  For our numerical calculations, we will not use the incomplete LQCD input, and we  
 shall employ the form factors obtained within the NRQM scheme of Ref.~\cite{Hernandez:2006gt}. One of the advantages of such choice is consistency, since all the form factors needed to compute the full NP effects  encoded  in Eq.~(\ref{eq:hnp}) will be obtained within the same scheme, and without having to rely on quark field level equations of motion. Furthermore, the effects of NP on $\bar B_c\to \eta_c$ and $\bar B_c\to J/\psi$ decays will be consistently compared in this way, since there is no   LQCD information for the $\bar B_c\to \eta_c$ reaction either. The
form factors computed in \cite{Hernandez:2006gt} follow  a pattern  consistent with heavy quark spin symmetry (HQSS) and its expected breaking corrections.  Moreover, five different reasonable inter-quark potentials were considered in \cite{Hernandez:2006gt}, 
and the range of results obtained from them  allow us to provide  theoretical uncertainties to our predictions.
Additionally, we shall also consider  the form factors from the pQCD factorization approach of Ref.~\cite{Wen-Fei:2013uea} that  have recently been used in Refs.~\cite{Murgui:2019czp, Watanabe:2017mip} to predict the ${\cal R}_{J/\psi}$ ratio within different NP scenarios. However, as we shall see below, these latter form factors do not respect a kinematical constraint at $q^2=0$ and they  display large violations of HQSS. 

Since SM LQCD vector and axial form factors are now available  for the $\bar B_c\to J/\psi$ decay, we have  
systematically compared, both in the CM and LAB frames,  SM observables  computed 
  with the LQCD input~\cite{Harrison:2020gvo} and using the phenomenological NRQM. In general, though there appear overall normalization inconsistencies, we find quite good agreements for $\omega$ (or $q^2$) shapes, which become much better for observables constructed out of ratios of distributions, 
like the $\tau-$forward-backward [${\cal A}_{FB}$] and $\tau-$polarization [${\cal A}_{\lambda_\tau}$] asymmetries, as well as the  ratios between predictions obtained in $\tau$ and ($e/\mu$) modes like
${\cal R}_{J/\psi}$ or   ${\cal R}({\cal A}_{FB})$. 

This work is organized as follows: in Sec.~\ref{sec:Bc2cc} we present the results for the  $\bar B_c\to \eta_c,J/\psi$ semileptonic decays both for unpolarized and polarized (well defined helicity in the CM or LAB frames) final $\tau$'s. The corresponding results for the
$\bar B\to D^{(*)}$ reactions are given in Sec.~\ref{sec:B2D} for the unpolarized cases, and in Appendix~\ref{app:bddspol} for  the decays with polarized outgoing leptons. We find that the qualitative characteristics of the observables and the main extracted conclusions are similar to those discussed  for the  $\bar B_c\to\eta_c, J/\psi$ transitions. The most relevant findings of this work are summarized in Sec.~\ref{sec:conclusions}. Besides, the definition of the form factors appropriate  for these processes are given in Appendix~\ref{app:ff}, while the expressions for the 
16 $\widetilde W$ SFs in terms of the form factors and Wilson coefficients 
are compiled in Appendices~\ref{app:ff1} and \ref{app:ff2} for decays
into pseudoscalar and vector mesons, respectively. Finally, in Appendix~\ref{app:NRQM} we collect the expressions for the 
$\bar B_c\to \eta_c$ and $\bar B_c\to  J/\psi$ semileptonic decay form factors  obtained within the NRQM of Ref.~\cite{Hernandez:2006gt}.
\section{$\bar B_c\to\eta_c$ and $\bar B_c\to J/\psi$ semileptonic decay results}
\label{sec:Bc2cc}
In this section  we present the results for the $\bar B_c\to\eta_c\tau\bar\nu_\tau$ and
$\bar B_c\to J/\psi\tau\bar\nu_\tau$ semileptonic decays. For the NP terms we 
 use  the Wilson coefficients
corresponding to Fits 6 and 7 in Ref.~\cite{Murgui:2019czp}. Among the different scenarios studied on that reference, only Fits 4, 5, 6 and 7 include all the NP terms  in Eq.~\eqref{eq:hnp}. However,  Fits 4 and 5 lead to an unlikely physical situation in which the SM coefficient is almost  canceled and its effect is replaced by  NP contributions. The numerical values of the Wilson coefficients (fitted parameters) are compiled in Table 6 of  Ref.~\cite{Murgui:2019czp}.  The data used for the fits include  the ${\cal R}_D$ and ${\cal R}_{D^*}$ ratios,
the normalized experimental distributions of $d\Gamma(\bar B\to D \tau \bar\nu_\tau)/dq^2$ and $d\Gamma(\bar B\to D^* \tau \bar\nu_\tau)/dq^2$ measured by Belle and BaBar as well as the  longitudinal polarization fraction 
$F_L^{D^*}=\Gamma_{\lambda_{D^*}=0}(\bar B\to D^{*}\tau \bar \nu_\tau)/\Gamma(\bar B\to D^{*}\tau \bar \nu_\tau)$ provided by Belle. 
The $\chi^2$ merit function is defined in  Eq.~(3.1)  of Ref.~\cite{Murgui:2019czp}, and it is constructed with  
the above data inputs and some prior knowledge of the $\bar B \to D $  and $\bar B \to D^*$ semileptonic form-factors. Some upper bounds   on the leptonic decay rate 
$\bar B_c\to \tau \bar\nu_\tau$ are also imposed.  The corresponding $\chi^2_{\rm min}/{\rm d.o.f.}$
are 37.6/53 and 38.9/53 for Fits 6 and 7 respectively.

As already mentioned, for the form factors, defined in Appendix~\ref{app:ff}, we shall use two different sets obtained within two independent theoretical approaches. 

The first one is determined from the NRQM calculations of Ref.~\cite{Hernandez:2006gt}. There, five different inter-quark
potentials  are used: AL1, AL2, AP1 and AP2 taken from 
Refs.~\cite{Semay:1994ht,SilvestreBrac:1996bg}, and the BHAD potential from 
Ref.~\cite{Bhaduri:1981pn}. All the form factors  are obtained 
 without the need
to rely  on quark field level equations of motion and their expressions in terms of the quark wave-functions can be found in 
Appendix~\ref{app:NRQM}. As in Ref.~\cite{Hernandez:2006gt},
we will take as central values of the computed quantities the results corresponding to  the 
AL1 potential. The deviations from this result obtained with the other four
potentials are used to estimate the theoretical error associated to the form-factors determination
  in this type of models. These errors will be shown in the corresponding figures below as  uncertainty bands.
  
\begin{figure}[tbh]
\includegraphics[height=5cm]{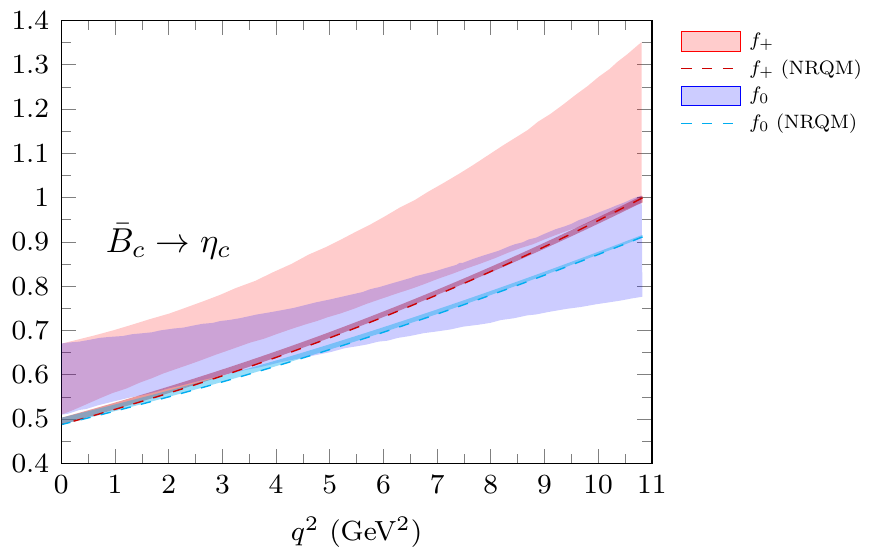} \hspace{.15cm}
\includegraphics[height=5cm]{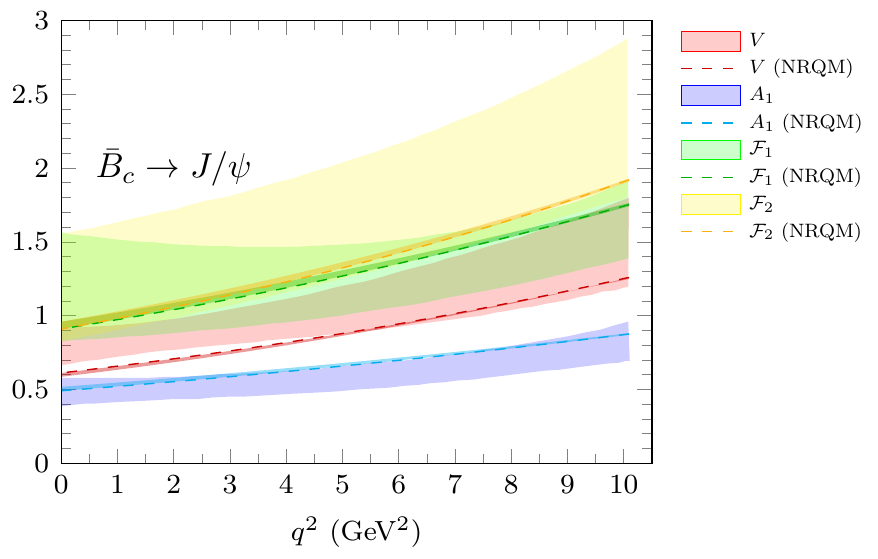}\\
\caption{ Comparison of the SM form factors for the $\bar B_c\to \eta_c$ (left) and $\bar B_c\to J/\psi$ (right) transitions, obtained with the NRQM calculation of Ref.~\cite{Hernandez:2006gt} (dashed lines plus narrow uncertainty bands around them) and in the model independent analysis carried out in Ref.~\cite{Cohen:2019zev} (shown as bands). For a definition of the  depicted form  factors  see  Eqs.~(2)--(3) and (7)--(9) of Ref.~\cite{Cohen:2019zev}. As in Figs.6 and 7 of this latter reference, 
the dimension-full $f_0$ and ${\cal F}_1$ form-factors are given in units of $(M^2-M^{\prime\,2})$  and $\frac{1}2(M^2-M^{\prime\,2})$, respectively.
}  
\label{fig:ffcohen}
\end{figure}

  A comparison of the  SM form factors thus obtained with the ones from 
  Ref.~\cite{Cohen:2019zev} is presented in Fig.~\ref{fig:ffcohen}.  
  We see that NRQM results comfortably lie, best for the $J/\psi$ decay and in the large $q^2$ region (close to zero recoil), within the colored
bands which show  the one-standard-deviation $(1\sigma)$ best-fit regions obtained
from the global dispersive analysis carried out in \cite{Cohen:2019zev}.

  To evaluate the theoretical error associated to the  Wilson coefficients  for each of Fits 6 and 7, we use different sets of  coefficients  obtained through 
successive small steps in the multiparameter space, with each step leading to a moderate $\chi^2$ enhancement.
We use  1$\sigma$ sets, i.e. values  of the Wilson coefficients for which $\Delta\chi^2\le 1$ with respect to its minimum value, to generate the distribution of each observable, taking into account in this way statistical correlations. From this derived distributions, we determine the maximum 
deviation above and below its central value, the latter
obtained with the values of the Wilson coefficients corresponding to the minimum of $\chi^2$ and the AL1 form factors. These deviations define the, asymmetric in general, uncertainty associated with the NP Wilson coefficients.
  The two type of errors are then added in quadrature and they are shown in the figures 
  as an extra,   larger in size, uncertainty band.
  
The second set of form factors we shall use  are the ones evaluated in
Ref.~\cite{Wen-Fei:2013uea} within a
perturbative QCD (pQCD) factorization  approach. In this latter case only
vector and axial-vector form factors have been obtained\footnote{Note that in Ref.~\cite{Wen-Fei:2013uea} they work with  different  form factor decompositions than those used here. The relations between our form factors and theirs can be obtained straightforwardly.}. They have been evaluated 
in the low $q^2$ region and   extrapolated to higher $q^2$ values 
using a model dependent 
parameterization. These form factors have been used in the two recent calculations of
Refs.~\cite{Murgui:2019czp,Watanabe:2017mip} where  the rest of form factors needed 
(scalar, pseudoscalar or tensor ones) were determined
using  the quark level equations of
motion of Ref.~\cite{Sakaki:2014sea}. In Ref.~\cite{Wen-Fei:2013uea}, the authors
give the theoretical uncertainties for the vector and axial form factors
at $q^2=0$. However neither correlations, nor errors for the parameters used in the $q^2-$extrapolation are provided. Besides, it is not clear what errors are introduced in the calculation through the use of the quark level equations of
motion. Thus, in this case
we will only show the error band stemming from the Wilson 
coefficients, even though  larger uncertainties are to be expected.

As mentioned, for the $\bar B_c\to  J/\psi$ decay, we shall compare our SM results 
with the ones reported in Ref.~\cite{Harrison:2020nrv}, and obtained with the LQCD axial and vector form factors determined in Ref.~\cite{Harrison:2020gvo}. In this  case the $1\sigma$ uncertainty bands are obtained with the use of the correlation matrix provided in this latter reference.

\subsection{Results with an unpolarized final $\tau$ lepton}
\begin{figure}[b!]
\includegraphics[height=5cm]{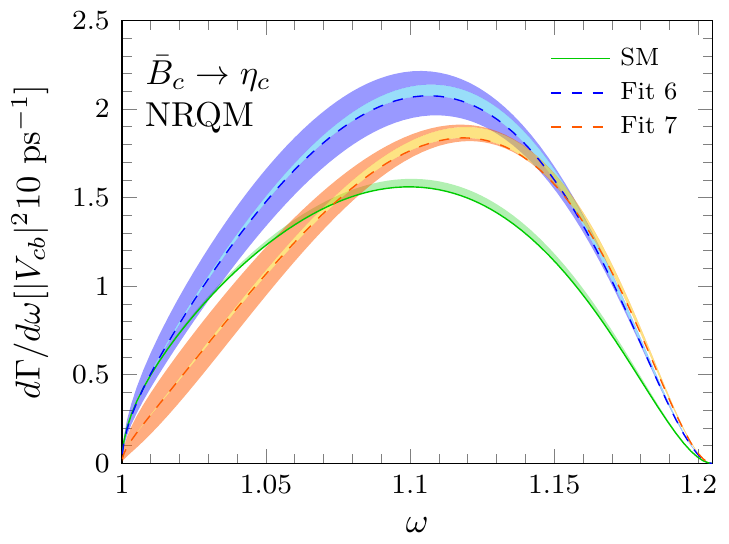} \hspace{.15cm}
\includegraphics[height=5cm]{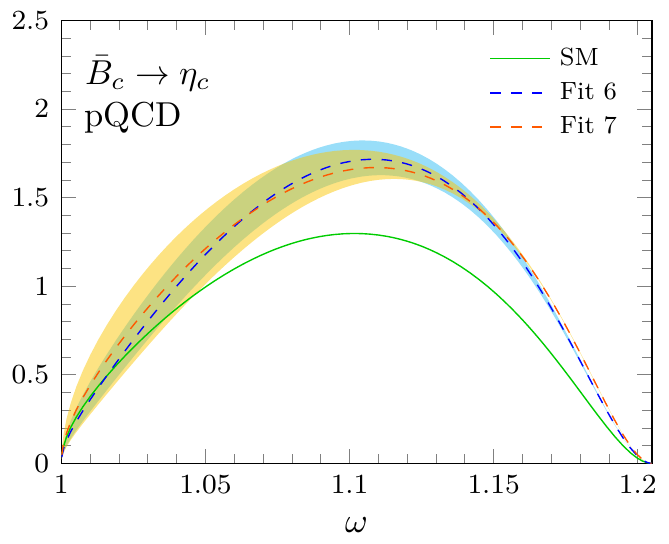}\\
\includegraphics[height=5cm]{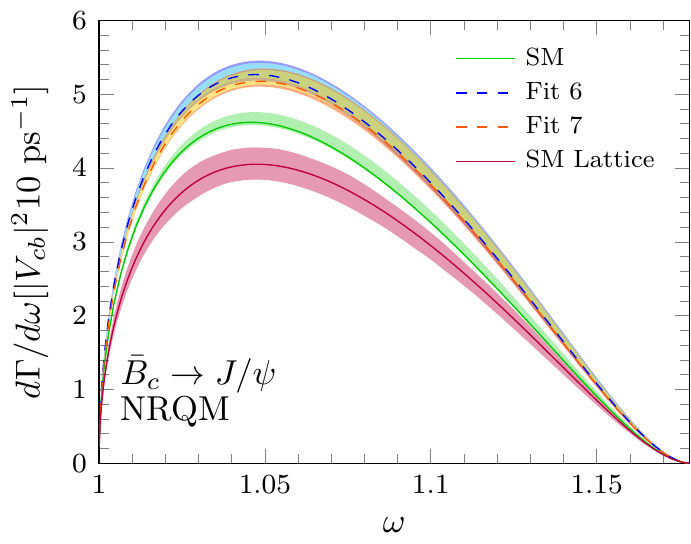} \hspace{.15cm}
\includegraphics[height=5cm]{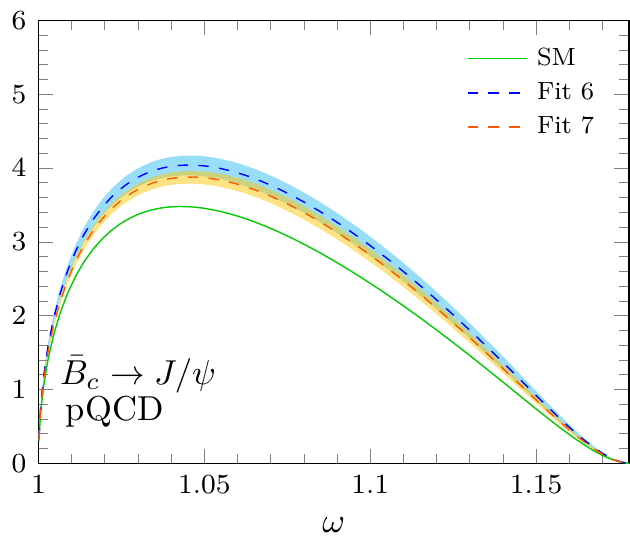}%
\caption{ Top: $d\Gamma(\bar B_c\to\eta_c \tau\bar \nu_\tau)/d\omega$ 
differential decay width, as a function of $\omega$ 
and in units of $10 |V_{cb}|^2 {\rm ps}^{-1}$. 
We show SM predictions and full NP results obtained using the Wilson coefficients from 
Fits 6  and  7  of Ref.~\cite{Murgui:2019czp} and form factors from the 
NRQM  (left panel) and the pQCD (right panel) approaches of  Refs.~\cite{Hernandez:2006gt} and \cite{Wen-Fei:2013uea}, respectively. Bottom: Same as before but for the
$\bar B_c\to J/\psi \tau\bar \nu_\tau$ decay. In the left-bottom plot, we also show the
SM lattice result from Ref.~\cite{Harrison:2020nrv}. Uncertainty bands  obtained 
as detailed in the main text. 
}  
\label{fig:dgdw}
\end{figure}

We begin with the results corresponding to an unpolarized final
$\tau$ lepton. In Fig.~\ref{fig:dgdw}, we show
the $d\Gamma/d\omega$ differential distribution for  $\bar B_c\to\eta_c \tau\bar \nu_\tau$
and $\bar B_c\to J/\psi \tau\bar \nu_\tau$ reactions. As can be seen in the plots, the $\omega$ values accessible in the transitions are around  $\omega\sim 1.2$ at most,  while for the similar  $\bar B\to D^{(*)} \tau\bar \nu_\tau$ reactions the available phase-space is larger, and $\omega$ varies from 1 to 1.35--1.40.   
 
 In both $\bar B_c$ decays, the  NRQM form factors from Ref.~\cite{Hernandez:2006gt} lead to larger
 total widths. Looking at the SM results for the $\bar B_c\to J/\psi$ decay, one sees that the LQCD prediction from Ref.~\cite{Harrison:2020nrv} is in between the  NRQM and pQCD distribution, somewhat closer 
 to the former one, but still showing a tension of around $2\sigma$ in most of the phase space. Since relativistic effects increase\footnote{The kinematical treatment is  fully relativistic, but close to $q^2=0$, the transition matrix elements are sensitive to  large momentum components of the non-relativistic meson wave-functions.} as one departs from zero-recoil to  $q^2=0$, they might be responsible for some of the NRQM-LQCD discrepancies exhibited in the figure far from the vicinity of $\omega=1$.  
 
For the decay into $\eta_c$ computed  with the NRQM form factors we note  that already $d\Gamma/d\omega$ discriminate between NP Fits 6 and 7. For the rest of cases shown in the figure, though NP effects are clearly visible, we see that this observable  would not be able to distinguish between the two NP scenarios examined in this work.   
\begin{table}[t!]
\begin{center}
\begin{tabular}{c|ccc|cc|cc}%\cline{2-7}
 \multicolumn{1}{c}{}&\multicolumn{3}{|c|}{SM}&
 \multicolumn{2}{c|}{NP Fit 6} & 
 \multicolumn{2}{c}{NP Fit 7}\\%\cline{2-7}
\multicolumn{1}{c|}{} & [NRQM] &
 [pQCD]& [HPQCD]&
  [NRQM] & [pQCD]
 & [NRQM] &  [pQCD]\\
  \hline
  &&&&&&\\
${\cal R}_{\eta_c}=\frac{\Gamma(\bar B_c\to \eta_c \tau\bar \nu_\tau)}
{\tstrut\Gamma(\bar B_c\to \eta_c \mu\bar \nu_\mu)}$ & $0.349^{+0.000}_{-0.007}$ & 0.309& & $0.452^{+0.034}_{-0.030}$ & 
$0.40^{+0.03}_{-0.03}$&$0.384^{+0.024}_{-0.018}$&$0.40^{+0.04}_{-0.03}$
\\&&&&&&\\\hline&&&&&&\\
 ${\cal R}_{J/\psi}=\frac{\Gamma(\bar B_c\to J/\psi \tau\bar \nu_\tau)}
{\tstrut\Gamma(\bar B_c\to J/\psi \mu\bar \nu_\mu)}$ &$0.266^{+0.000}_{-0.004}$ & 0.289&$0.2601\pm0.0036$&$0.306^{+0.007}_{-0.007}$
 & $0.342^{+0.013}_{-0.015}$&$0.301^{+0.005}_{-0.007}$&
 $0.326^{+0.008}_{-0.009}$ \\&&&&&& \\\hline
\end{tabular}
\end{center}
\caption{${\cal R}_{\eta_c}$ and ${\cal R}_{J/\psi}$ ratios obtained in the SM and with NP effects from Ref.~\cite{Murgui:2019czp}. We give   
results using the [NRQM]~\cite{Hernandez:2006gt} and [pQCD]~\cite{Wen-Fei:2013uea} sets of form-factors, and additionally  ${\cal R}_{J/\psi}^{\rm SM}$ from the LQCD analysis carried out by the HPQCD collaboration~\cite{Harrison:2020nrv}.  }
\label{tab:ratios}
\end{table}

Evaluating the SM   predictions for a final massless charged lepton ($\mu$ or $e$), we  obtain the ${\cal R}_{\eta_c}$ and ${\cal R}_{J/\psi}$ 
ratios collected in Table~\ref{tab:ratios}.  The systematic uncertainties due to the inter-quark potential in the NRQM scheme are largely canceled out in the ratios, as can be inferred from the SM predictions. 
For the SM we find a nice agreement of the NRQM determination of ${\cal R}_{J/\psi}$
and the lattice evaluation of Ref.~\cite{Harrison:2020nrv}, pointing out also to a compensation in the ratio of the overall-normalization discrepancies noted in Fig.~\ref{fig:dgdw}.   
On the other hand, predictions with the NRQM and pQCD form factors differ by approximately 10\%, except for NP Fit 7 $ {\cal R}_{\eta_c} $, where the change is only of 4\%. In fact, the form-factor systematic uncertainties are reduced compared to those observed  in some regions of the differential distributions in Fig.~\ref{fig:dgdw}. We also note that the NRQM  ${\cal R}_{\eta_c}$ and ${\cal R}_{J/\psi}$ ratios are systematically bigger and smaller, respectively, than those obtained with pQCD form factors. For the latter ratio, we mentioned above that $\Gamma^{\rm NRQM}(\bar B_c\to J/\psi \tau\bar\nu_\tau) \ge \Gamma^{\rm pQCD}(\bar B_c\to J/\psi \tau\bar \nu_\tau)$ and thus, the massless lepton modes of the $\bar B_c\to J/\psi $ semileptonic decay calculated with NRQM form factors must also be larger than when pQCD form factors are used. Moreover, the difference has to be greater than for the $\tau$ mode to explain $R^{\rm NRQM}_{J/\psi} \le R^{\rm pQCD}_{J/\psi}$. 
  
The ratios  including NP are greater than pure SM expectations, around  30\% and 15\% for ${\cal R}_{\eta_c}$ and  ${\cal R}_{J\psi}$, respectively, except for the NRQM  ${\cal R}_{\eta_c}$ case evaluated with the NP Fit 7 where  an increase of only 10\% is found. In fact, at the level of ratios, only the NRQM ${\cal R}_{\eta_c}$  discriminates between NP Fits 6 and 7.  One can 
compare the values for ${\cal R}_{J/\psi}$ with the only  available experimental measurement quoted above. In this
case we see all predictions fall short of the present central experimental value by almost $2\sigma$, adding in quadratures the errors given   in Eq.~\eqref{eq:RpsiLHCB}. The agreement is slightly better  when using the pQCD form-factor set. The improvement is not significant, however, within the present accuracy in the data  and, as we explain below, there are some inconsistencies in the corresponding pQCD form factors for this decay.

\begin{figure}[tbh]
\includegraphics[height=4.cm]{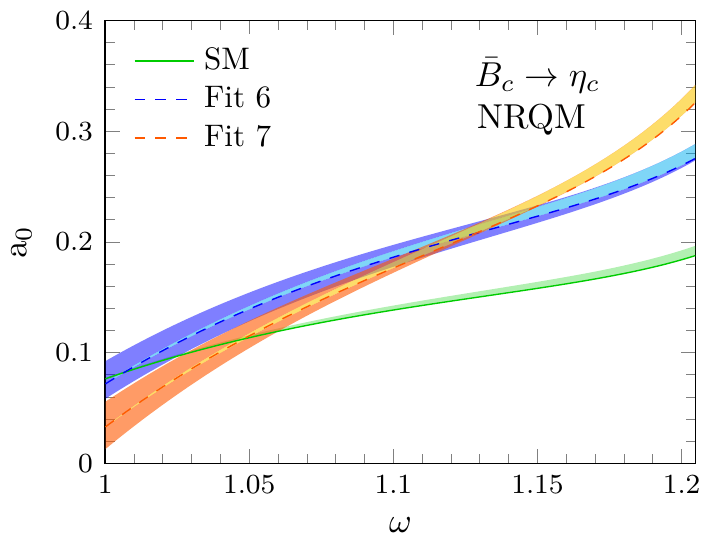} \hspace{.15cm}
\includegraphics[height=4.cm]{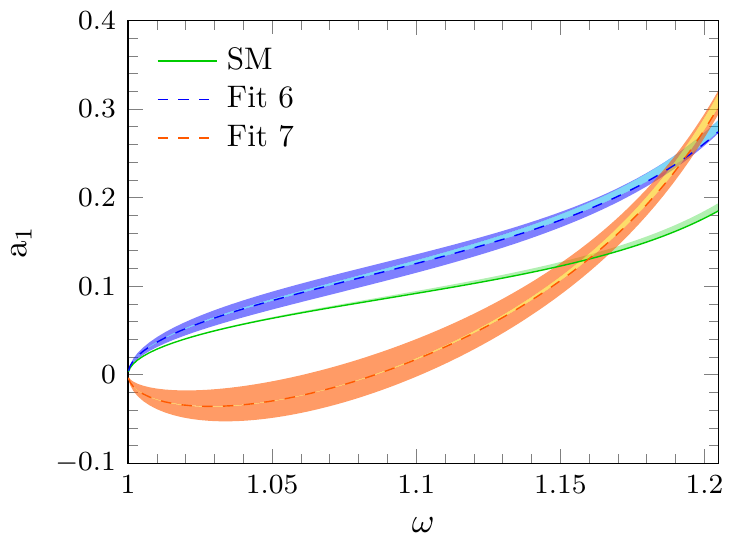}\hspace{.15cm}
\includegraphics[height=4.cm]{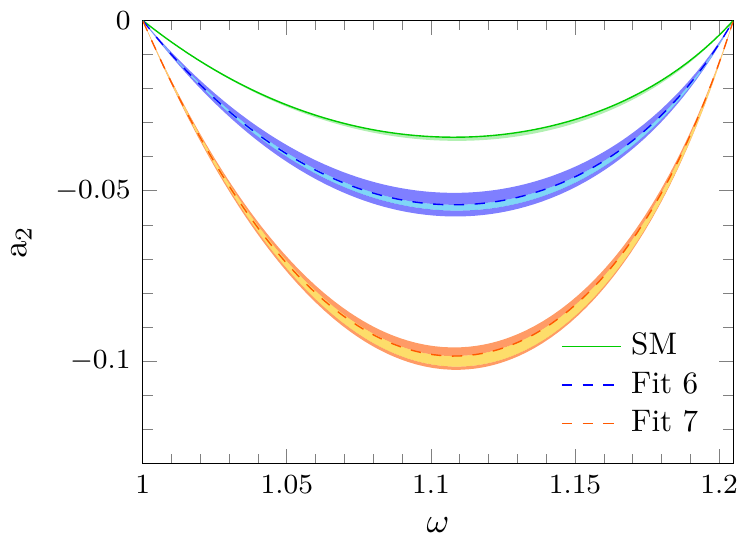}\hspace{.15cm}
\\
\includegraphics[height=4.cm]{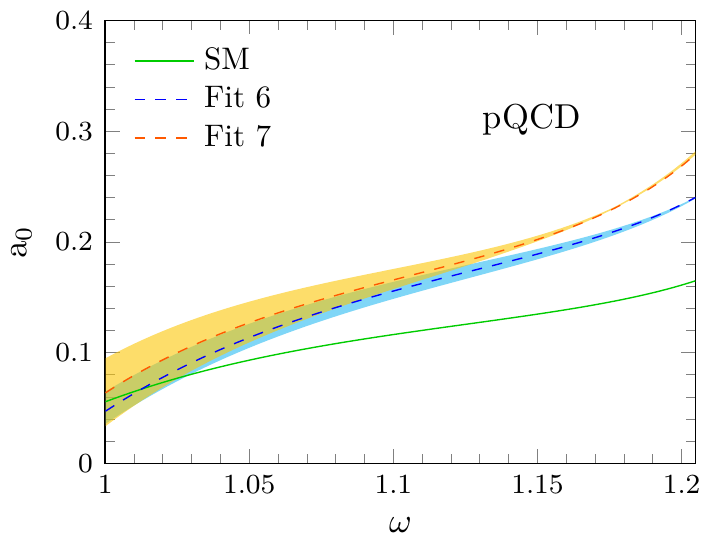} \hspace{.15cm}
\includegraphics[height=4.cm]{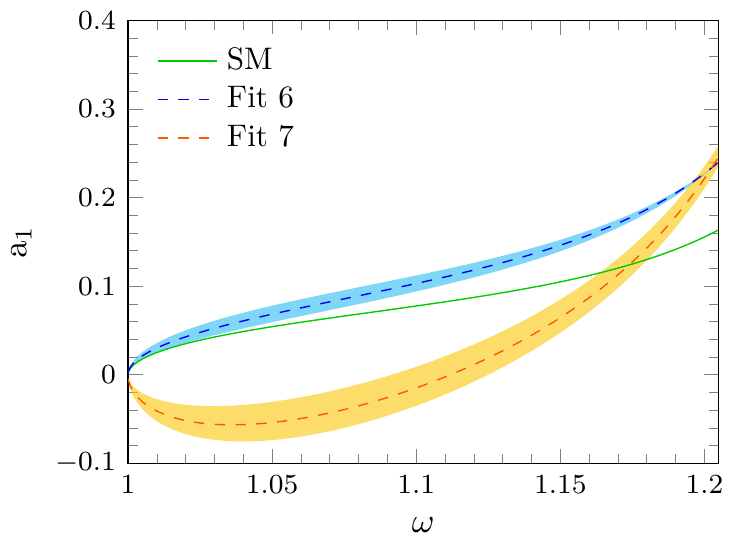}\hspace{.15cm}
\includegraphics[height=4.cm]{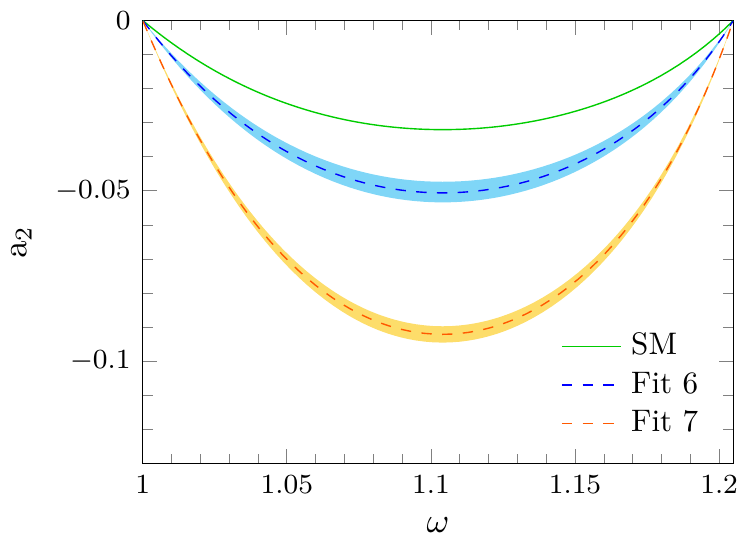}\hspace{.15cm}
\caption{ $\bar B_c\to\eta_c\tau\bar\nu_\tau$ decay: CM $a_{0,1,2}$  angular expansion coefficients  
as a function of $\omega$. We show results obtained with both,  NRQM (upper panels) and pQCD (lower panels) form factors from Refs.~\cite{Hernandez:2006gt} and \cite{Wen-Fei:2013uea}, respectively. The beyond of the SM scenarios Fits 6 and 7 are taken from Ref.~\cite{Murgui:2019czp}. Uncertainty bands as in Fig.~\ref{fig:dgdw}. 
}  
\label{fig:asetac}
\end{figure}
\begin{figure}[tbh]
\includegraphics[height=4.cm]{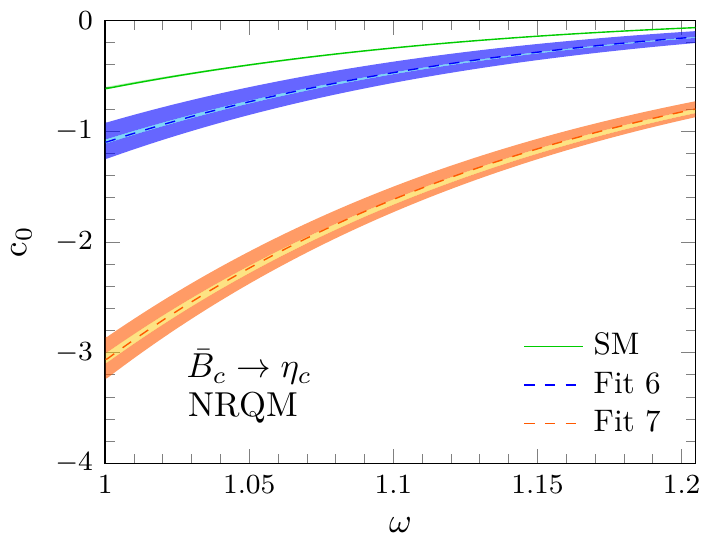} \hspace{.15cm}
\includegraphics[height=4.cm]{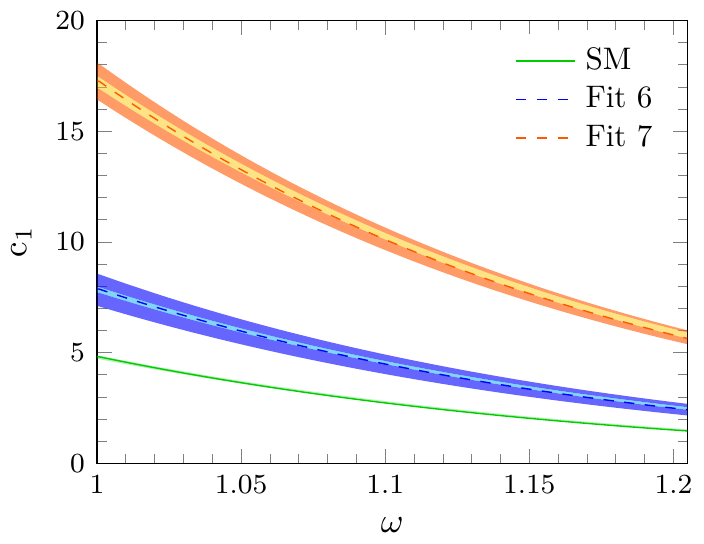}\hspace{.15cm}
\includegraphics[height=4.cm]{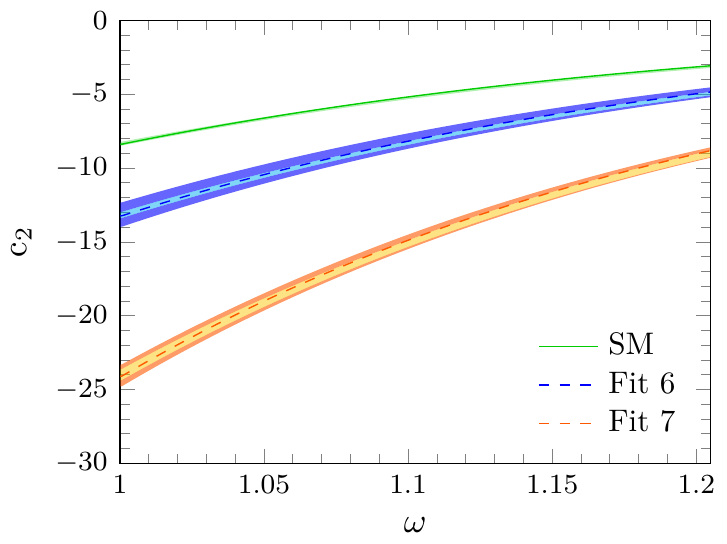}\hspace{.15cm}
\\
\includegraphics[height=4.cm]{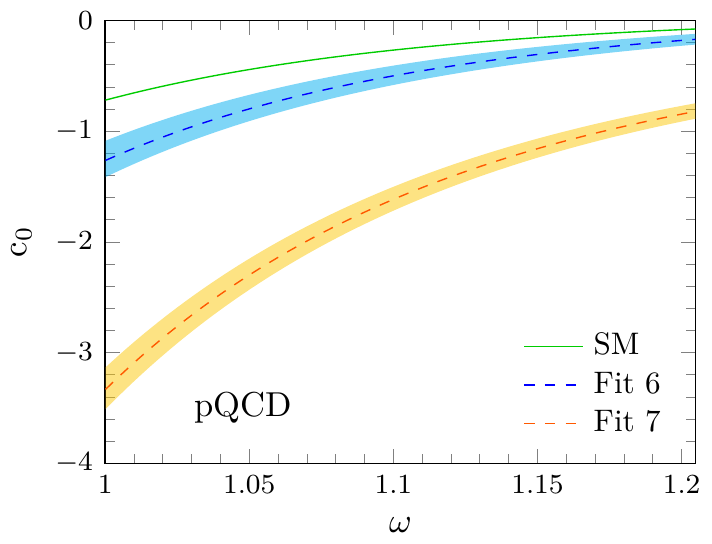} \hspace{.15cm}
\includegraphics[height=4.cm]{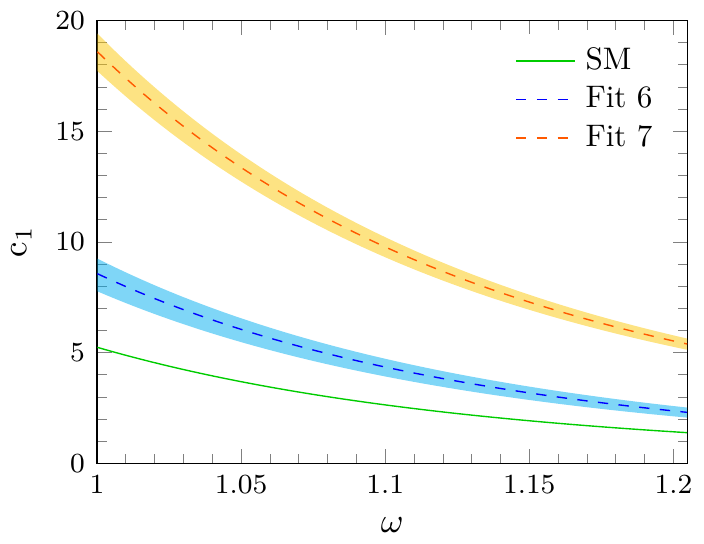}\hspace{.15cm}
\includegraphics[height=4.cm]{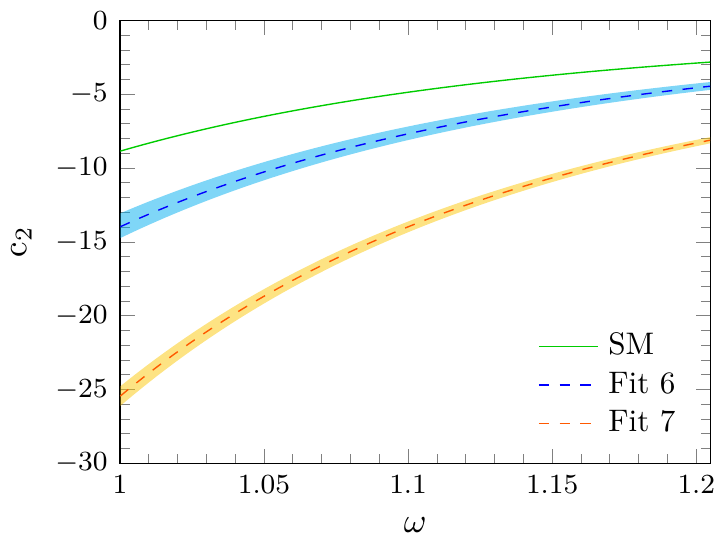}\hspace{.15cm}
\caption{ $\bar B_c\to\eta_c\tau\bar\nu_\tau$ decay: Same as in Fig.~\ref{fig:asetac} but
for the LAB $c_{0,1,2}$  energy expansion coefficients. 
}  
\label{fig:csetac}
\end{figure}

\begin{figure}[tbh]
\includegraphics[height=4.cm]{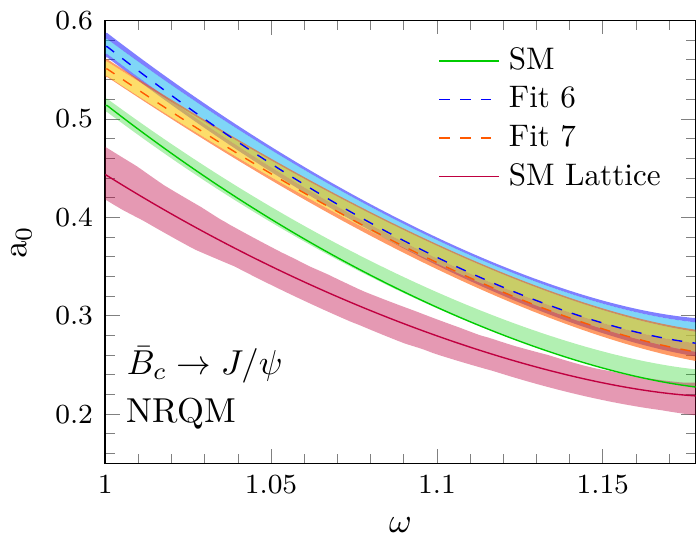} \hspace{.15cm}
\includegraphics[height=4.cm]{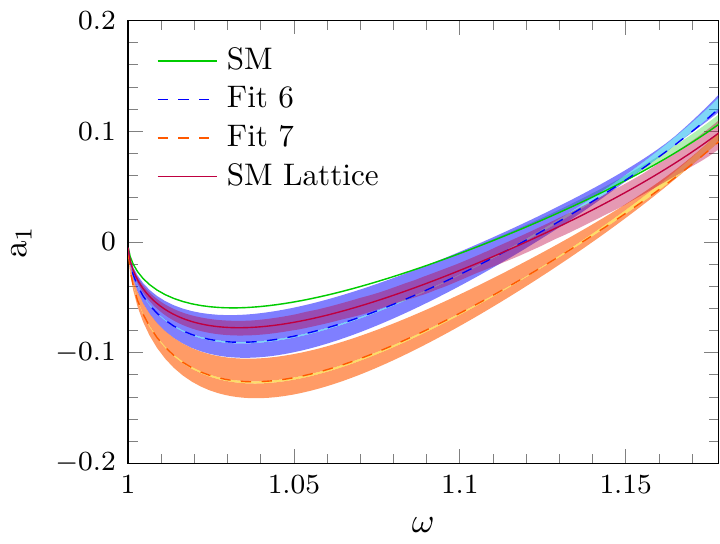}\hspace{.15cm}
\includegraphics[height=4.cm]{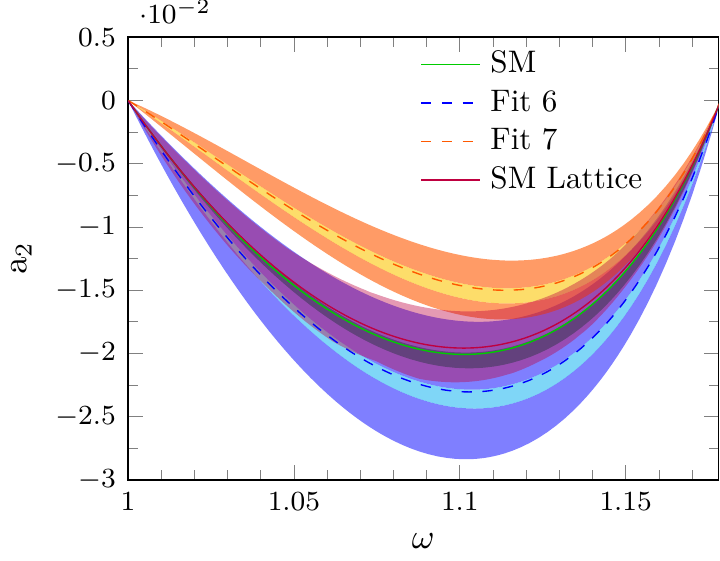}\hspace{.15cm}
\\
\includegraphics[height=4.cm]{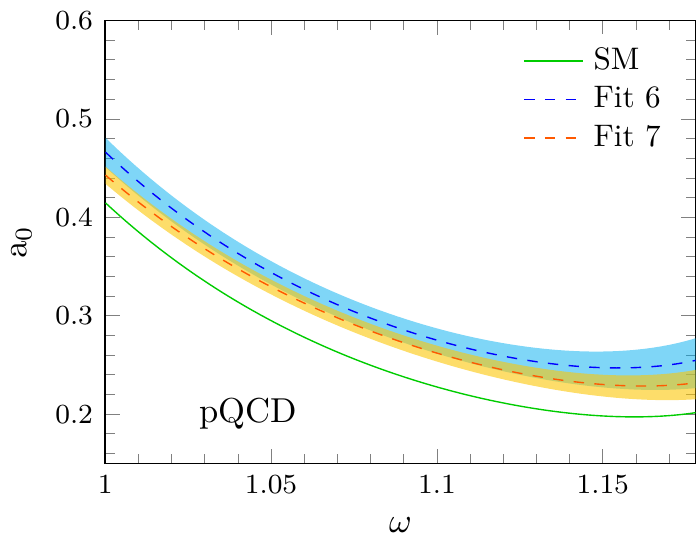} \hspace{.15cm}
\includegraphics[height=4.cm]{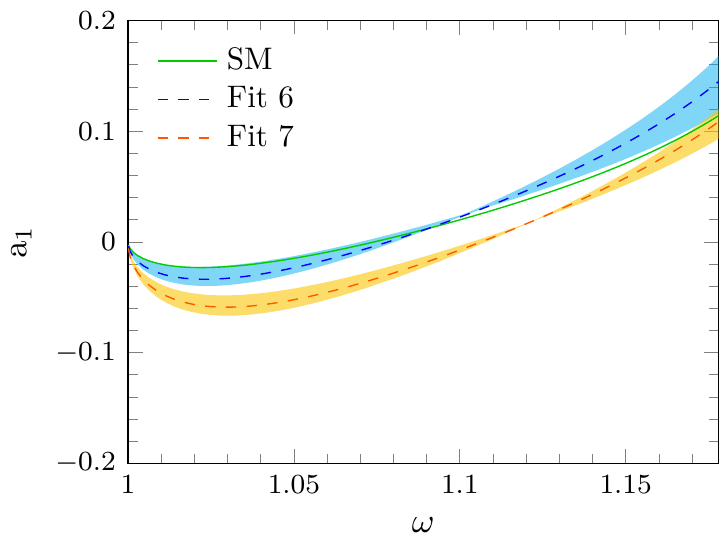}\hspace{.15cm}
\includegraphics[height=4.cm]{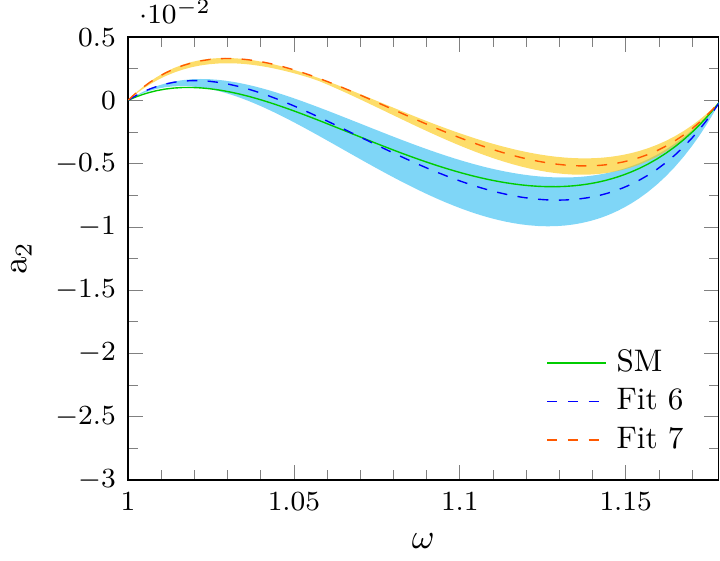}\hspace{.15cm}
\caption{$\bar B_c\to J/\psi\tau\bar\nu_\tau$ decay: 
CM $a_{0,1,2}$  angular expansion coefficients  
as a function of $\omega$. We show results obtained with both,  NRQM (upper panels) and pQCD (lower panels) form factors from Refs.~\cite{Hernandez:2006gt} and \cite{Wen-Fei:2013uea}, respectively. In the upper panels, SM  results for these observables 
obtained with the LQCD form-factors reported in Ref.~\cite{Harrison:2020gvo} are also displayed. The beyond of the SM scenarios Fits 6 and 7 are taken from Ref.~\cite{Murgui:2019czp}. Uncertainty bands as in Fig.~\ref{fig:dgdw}.  
}  
\label{fig:asjpsi}
\end{figure}

\begin{figure}[tbh]
\includegraphics[height=4.cm]{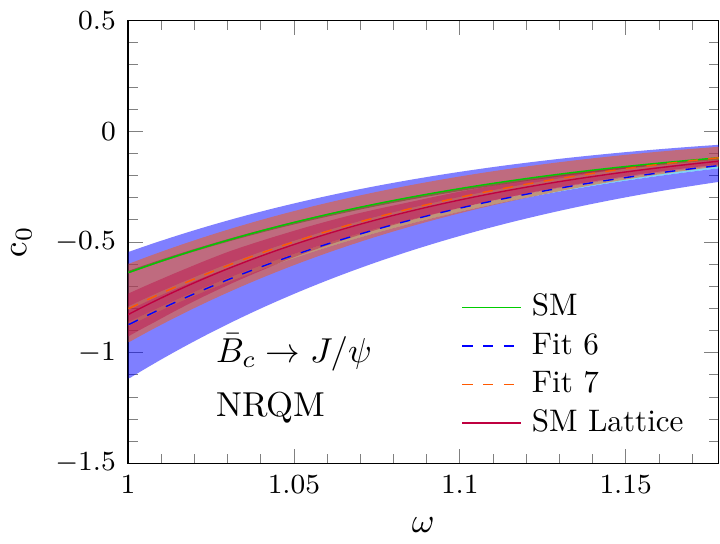} \hspace{.15cm}
\includegraphics[height=4.cm]{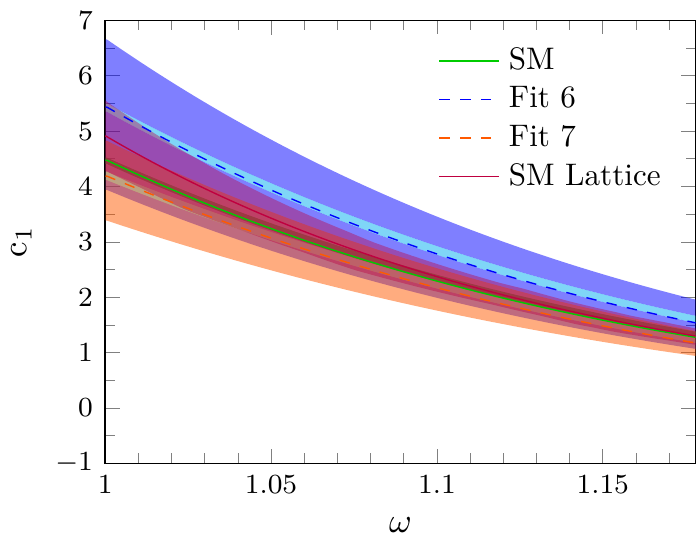}\hspace{.15cm}
\includegraphics[height=4.cm]{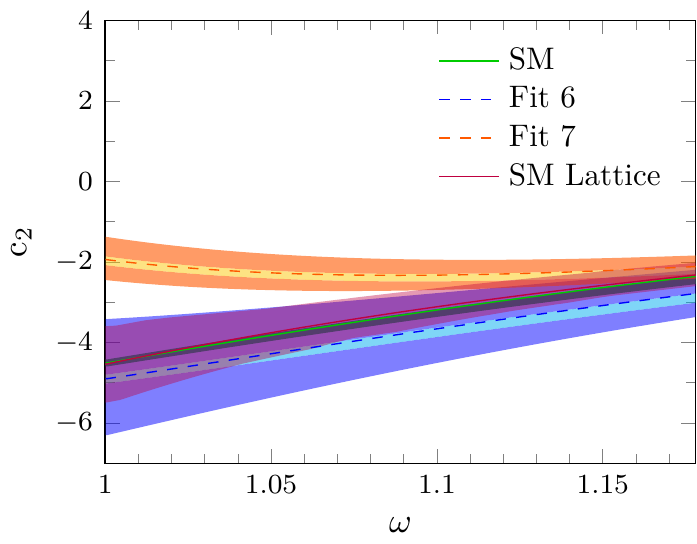}\hspace{.15cm}
\\
\includegraphics[height=4.cm]{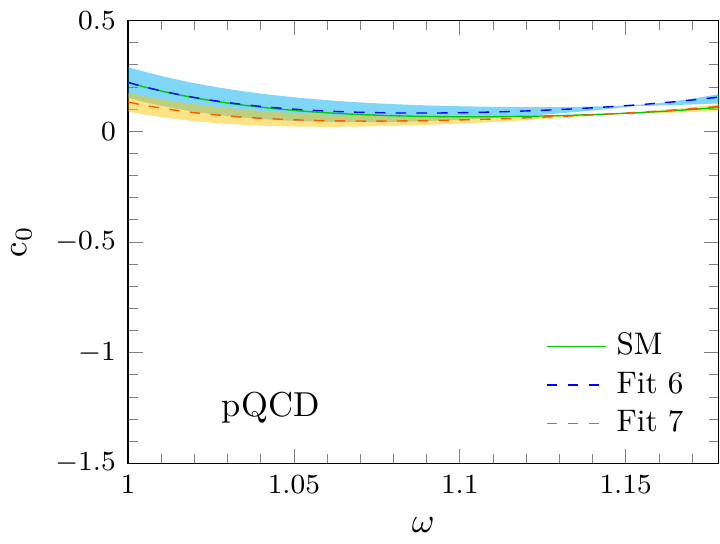} \hspace{.15cm}
\includegraphics[height=4.cm]{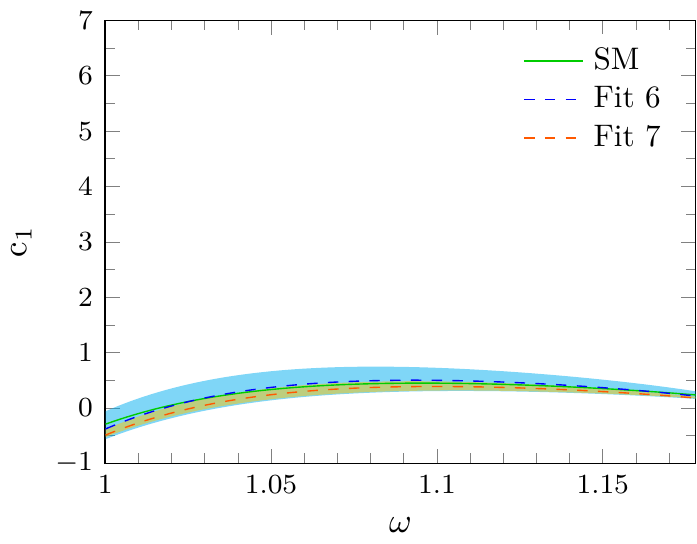}\hspace{.15cm}
\includegraphics[height=4.cm]{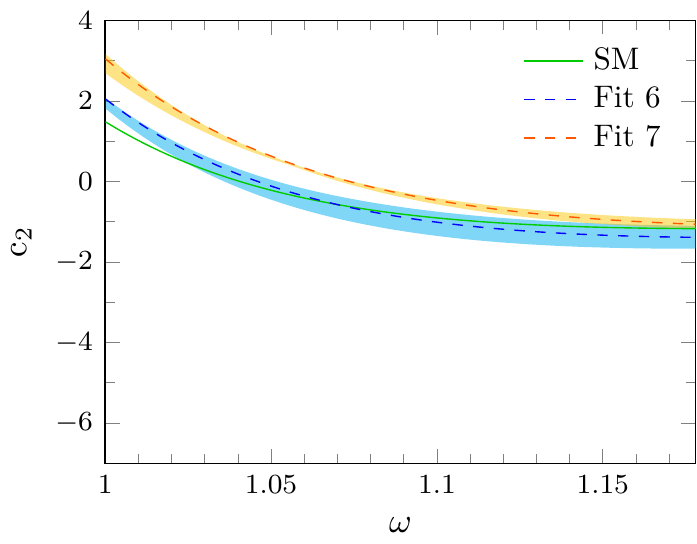}\hspace{.15cm}
\caption{$\bar B_c\to J/\psi\tau\bar\nu_\tau$ decay: Same as in Fig.~\ref{fig:asjpsi} but for the LAB $c_{0,1,2}$  energy expansion coefficients.  
}  
\label{fig:csjpsi}
\end{figure}

Now we discuss results for the   $d^2\Gamma/(d\omega d\cos\theta_\ell)$ 
and $d^2\Gamma/(d\omega dE_\ell)$ double differential distributions. The $a_{0,1,2}$  CM angular   
   and $c_{0,1,2}$ LAB  energy  expansion coefficients are shown in Figs.~\ref{fig:asetac} and \ref{fig:csetac}, and Figs.~\ref{fig:asjpsi} and \ref{fig:csjpsi} for the $\eta_c$ and $J/\psi$ decay modes, respectively. 
 
For  the $\bar B_c\to\eta_c \tau\bar \nu_\tau$ decay,   both sets of form factors lead to qualitatively very similar 
results. As in Ref.~\cite{Penalva:2020xup} for the $\Lambda_b\to\Lambda_c\tau\bar \nu_\tau$ semileptonic 
decay, we find  that, with the exception of $a_0$, all 
the other expansion coefficients serve the purpose of giving a clear distinction between 
NP Fits 6 and 7. Thus, different fits that otherwise give similar  $d\Gamma/d\omega$ decay widths, can be told apart by looking at these
CM angular or LAB energy observables. In addition, we also observe $\omega-$shapes for all  $\bar B_c\to\eta_c$ SM and NP coefficients similar  to those obtained  in Ref.~\cite{Penalva:2020xup} for the $\Lambda_b\to\Lambda_c$ transition, except $a_0$ that here grows with $\omega$ while for the baryon decay it  is a decreasing function of $\omega$. 

\begin{figure}[tbh]
\includegraphics[height=4.cm]{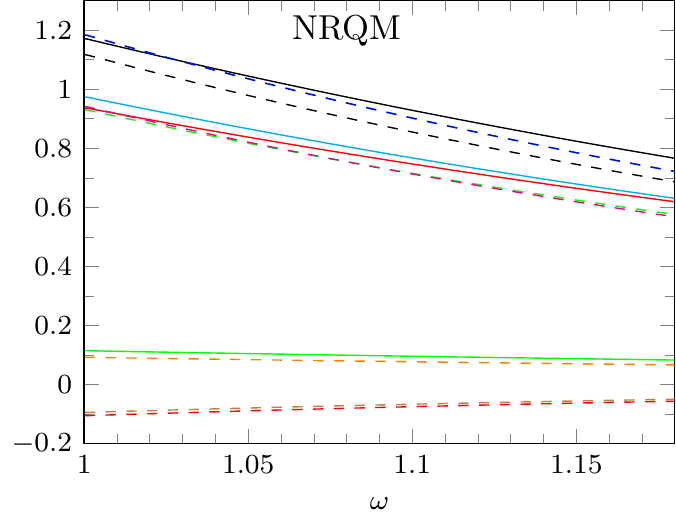} \hspace{.15cm}
\includegraphics[height=4.cm]{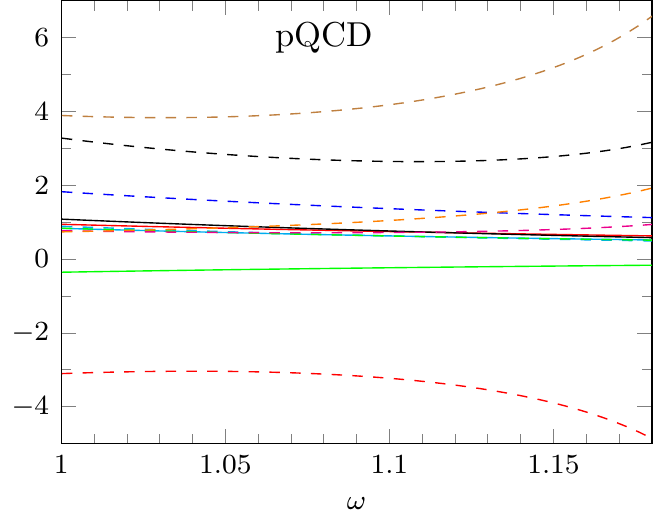}\hspace{.15cm}
\includegraphics[height=4.cm]{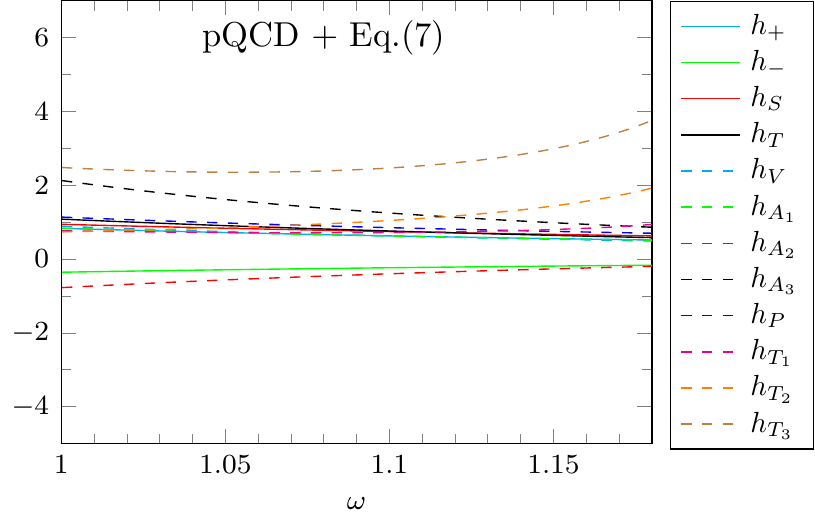}\hspace{.15cm}
\caption{ Different $h_a$ form factors  defined in Eqs.~(\ref{eq.FactoresformaPseudo})
and (\ref{eq.FactoresformaVec}) for $\bar B_c \to \eta_c$  and $\bar B_c \to J/\psi$ semileptonic decays. Left panel: Results obtained with the
NRQM of  Ref.~\cite{Hernandez:2006gt} using the AL1 potential. 
Middle panel: Results obtained from the  form factors of Ref.~\cite{Wen-Fei:2013uea} and the
use of quark-level equations of motion from Ref.~\cite{Sakaki:2014sea}. 
Right panel: Same as the middle panel but forcing Eq.~(\ref{eq:a00}) on $A_0(0)$. 
}  
\label{fig:hh}
\end{figure}

The corresponding results for a decay into $J/\psi$ are shown in 
Figs.~\ref{fig:asjpsi} and \ref{fig:csjpsi}. There are two distinct features 
in this case.
First,  the utility of these observables to distinguish between Fits 6 and 7 
 and, in some cases,
between those NP predictions and the SM results, is not
as good as in the $\eta_c$ case. This happens to be true independent of the form-factor 
set used. Second, the results obtained with the two form-factor sets turn out to be very different
 in most cases, with only $a_0$ and $a_1$ showing a similar qualitative behavior. By looking at SM results alone we find a good qualitative agreement between NRQM and LQCD results while, with the exception of $a_0$ and $a_1$, we find very different shapes for the results obtained  using the pQCD form factors. 
In order to better understand this discrepancy, we show in Fig.~\ref{fig:hh} 
all the  form factors defined in Eqs.~(\ref{eq.FactoresformaPseudo}) and
(\ref{eq.FactoresformaVec}), for decays into both
$\eta_c$ and $J/\psi$. In the left panel we give the results obtained with
the AL1 NRQM of Ref.~\cite{Hernandez:2006gt}. We see the 
results are close to expectations from Eq.~(\ref{eq:hh}) based on HQSS. In the middle panel we give the results
obtained using the form factors from Ref.~\cite{Wen-Fei:2013uea} and the
quark level equations of motion  from  Ref.~\cite{Sakaki:2014sea}. 
Large violations of HQSS
are already seen  for $h_{A_2}$ and $h_{A_3}$ (where no quark level equations of motion are involved), also for $h_{T_3}$ and,
 to a lesser extent, for
$h_P$.
These HQSS violations are related, at least in part,  to the fact that 
the $A_0(q^2), A_1(q^2)$ and $A_2(q^2)$ axial form factors evaluated in
 Ref.~\cite{Wen-Fei:2013uea},  and in terms of which the $h_{A_{1,2,3}}$ ones are determined,
  do not respect the $q^2=0$ constraint
\bea
A_0(0)=\frac{M+M'}{2M'}A_1(0)-\frac{M-M'}{2M'}A_2(0)
\label{eq:a00}
\eea
Even though $q^2\ge m_\tau^2$ for a final $\tau$, taking the wrong values
 of the form factors
at $q^2=0$  affects the determination of the values at larger $q^2$. In the right panel of Fig.~\ref{fig:hh} we see the effect
of imposing the above constraint on $A_0(0)$. The $h_P$ form factor  is now in agreement with HQSS
expectations and things improve for  $h_{A_2},h_{A_3}$ and $h_{T_3}$. 
Note that this restriction also corrects the divergences at $q^2=0$ 
that otherwise  appear for $h_{A_2}$ and $h_{A_3}$  and which signatures  are clearly 
visible in the middle panel at large recoils\footnote{Note that 
when we use the form factors from Ref.~\cite{Wen-Fei:2013uea}, we 
determine $h_{A_{1,2,3}}$ from $A_{0,1,2}$ and  the relation in
Eq.(\ref{eq:a00}) has to be satisfied in order for $h_{A_{2,3}}$ not to diverge
at $q^2=0$.}. Moreover, with this
restriction imposed,
the results for ${\cal R}_{J/\psi}$ would get  smaller, $0.311^{+0.007}_{-0.006}$ for Fit 6 and
$0.303^{+0.005}_{-0.004}$ for Fit 7, and  in much better agreement
with the ones obtained using the NRQM form factors from Ref.~\cite{Hernandez:2006gt} (see 
Table~\ref{tab:ratios}).  Besides, and though they are  
less important numerically, there are 
divergences  in all three $h_{T_{1,2,3}}$ form factors at 
$q^2=0$ when quark-level equations of motion are used to obtain them.  The beginning
of these  
divergences can already be seen in the middle and right panels of Fig.~\ref{fig:hh}. 

\begin{figure}[tbh]
\includegraphics[height=5.cm]{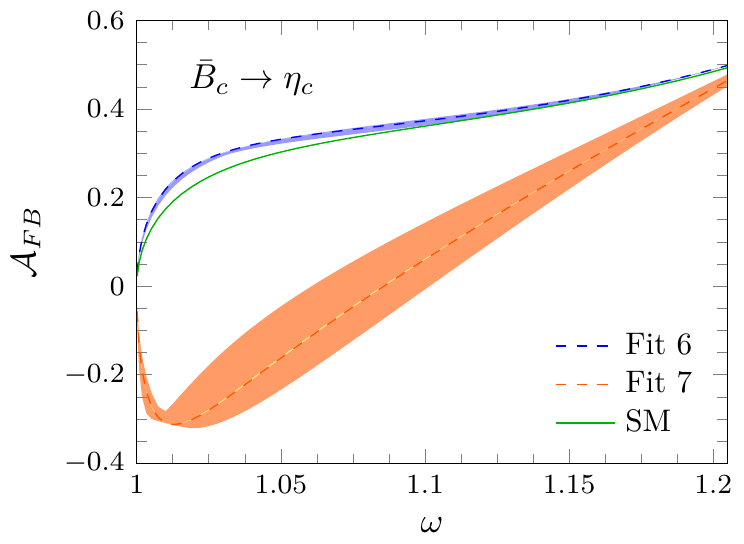} \hspace{.15cm}
\includegraphics[height=5.cm]{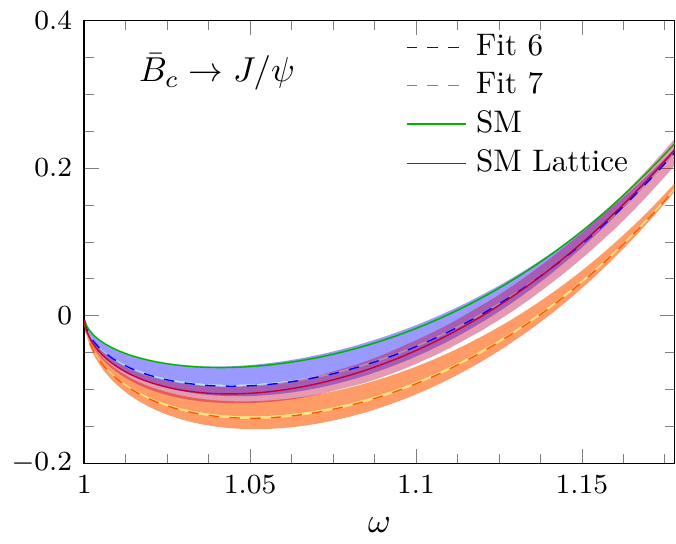}\hspace{.15cm}
\caption{ Forward-backward asymmetry in the CM reference frame
for the $\bar B_c\to\eta_c\tau\bar\nu_\tau$  (left) and 
$\bar B_c\to J/\psi\tau\bar\nu_\tau$ (right) decays. The results have been
obtained with the form factors evaluated with the
NRQM of  Ref.~\cite{Hernandez:2006gt}. For the $\bar B_c\to J/\psi$ case, we also show the SM prediction from the LQCD  results of Ref.~\cite{Harrison:2020nrv}. Uncertainty bands  as in 
Fig.~\ref{fig:dgdw}. 
}  
\label{fig:FB}
\end{figure}

In Fig.~\ref{fig:FB} we now show  the forward-backward asymmetry in the CM 
frame
evaluated with the form factors from the  NRQM of  
Ref.~\cite{Hernandez:2006gt}. This asymmetry is given by the
ratio 
\bea
{\cal A}_{FB}=\frac{a_1(\omega)}{2a_0(\omega)+2a_2(\omega)/3} \label{eq:Afb}
\eea
 For the decay into  $J/\psi$, SM  results for this asymmetry from the LQCD form-factors of Ref.~\cite{Harrison:2020gvo} are also shown. Note that   ${\cal A}_{FB}$ is given in \cite{Harrison:2020nrv} as well. For this decay mode, as it is also the case for other observables, the  SM result falls into the error band of Fit 6. However, for both  $\bar B_c\to \eta_c$ and $\bar B_c\to J/\psi$ transitions, this observable  is also able to distinguish between Fits 6 and 7, in particular for the $\eta_c$ channel. 
 \subsection{Results with a polarized final $\tau$ lepton}
\label{app:ascspol}

\begin{figure}[tbh]
\includegraphics[height=5.5cm]{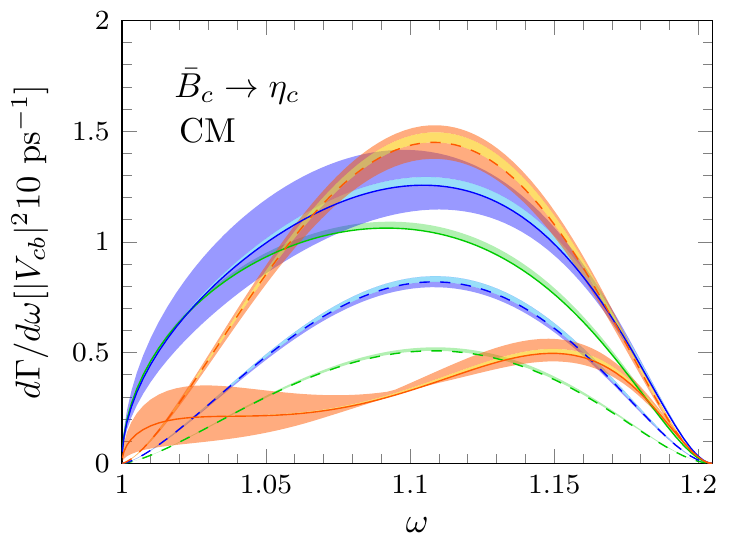} \hspace{.15cm}
\includegraphics[height=5.5cm]{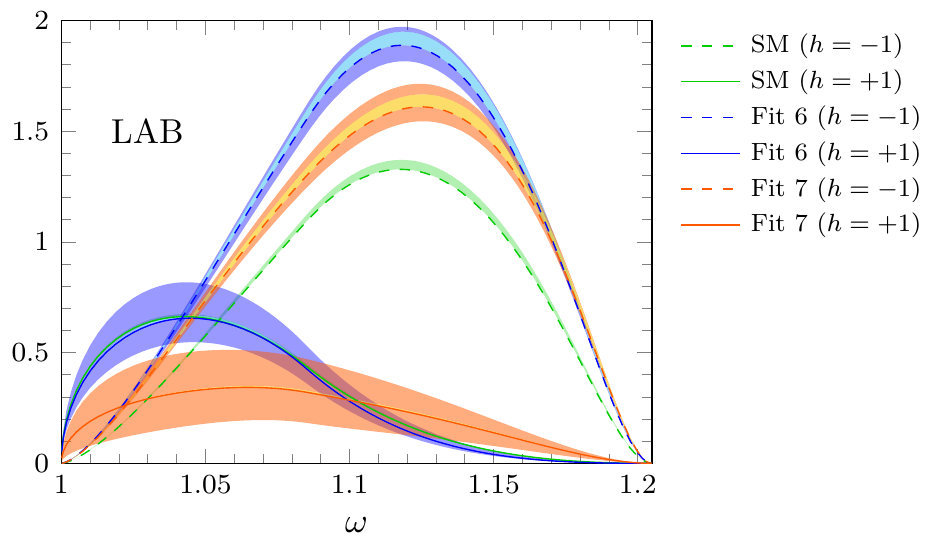}
\caption{ CM (left) and LAB (right) helicity decomposition of the  $d\Gamma(\bar B_c\to\eta_c\tau\bar\nu_\tau)/d\omega$ differential decay width, calculated with the NRQM form factors
from Ref.~\cite{Hernandez:2006gt}. Uncertainty bands  as in 
Fig.~\ref{fig:dgdw}. }  
\label{fig:dgdwpoletac}
\end{figure}

\begin{figure}[tbh]
\includegraphics[height=5cm]{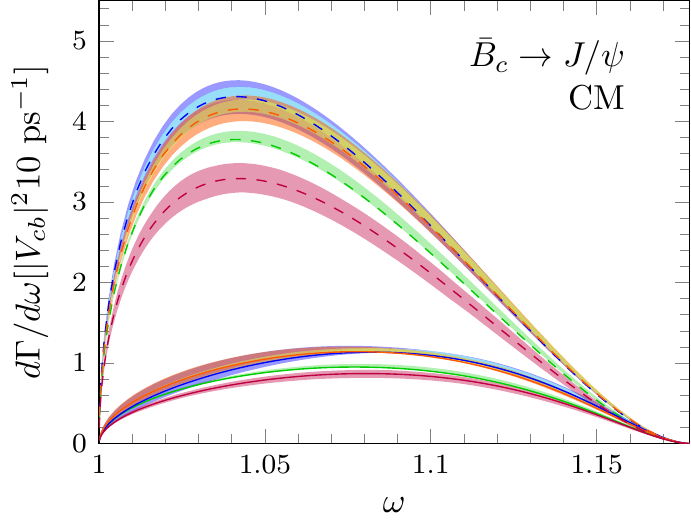} \hspace{.15cm}
\includegraphics[height=5cm]{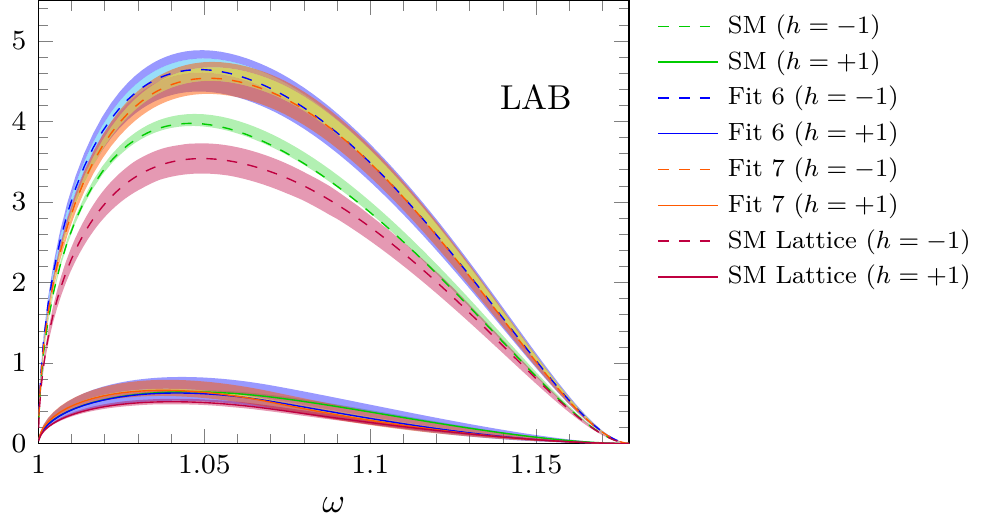}
\caption{ As in Fig.~\ref{fig:dgdwpoletac}, but for the 
$\bar B_c\to J/\psi\tau\bar\nu_\tau$ decay. We also show the SM results obtained with LQCD form factors from Ref.~\cite{Harrison:2020gvo}.
}  
\label{fig:dgdwpoljpsi}
\end{figure}

  In this section we collect the results corresponding to the $\bar B_c\to\eta_c\tau\bar\nu_\tau$ and 
  $\bar B_c\to J/\psi\tau\bar\nu_\tau$ decays 
  with a  polarized $\tau$  (well defined helicity
  $h=\pm1$ in the CM or LAB frames). In this case, and for simplicity, we will only present  results obtained
   with the use of the  NRQM form factors from Ref.~\cite{Hernandez:2006gt}, and, for the $\bar B_c\to J/\psi$ case,  also the SM  results obtained with the LQCD form factors from  Ref.~\cite{Harrison:2020gvo}.
 In Fig.~\ref{fig:dgdwpoletac}  we show the 
 $d\Gamma/d\omega$ differential decay width
 for the $\bar B_c\to\eta_c\tau\bar\nu_\tau$ decay with a final $\tau$
 with well defined helicity in the CM reference frame (left panel) and in the LAB system
 (right panel). The corresponding results for the $\bar B_c\to J/\psi\tau\bar\nu_\tau$
 decay are presented in Fig.~\ref{fig:dgdwpoljpsi}.  The negative helicity 
 contribution is dominant in all cases except for the $\eta_c$ CM distributions obtained both in the SM
 and in Fit 6. This unexpected feature  also occurs for  the polarized $\bar B\to D\tau\bar\nu_\tau$ decay (see Appendix~\ref{app:bddspol}). We see that both CM and LAB  $\tau$ negative-helicity distributions obtained in $\eta_c$ decays clearly discriminate between SM and different 
 NP scenarios. On the other hand, for the  $\bar B_c\to J/\psi\tau\bar\nu_\tau$ decay $d\Gamma/d\omega$ is not an efficient tool for that purpose, even taking into account information on the outgoing $\tau-$polarization.   As we noted in Fig.~\ref{fig:dgdw} for the unpolarized $d\Gamma/d\omega$, 
 the LQCD results~\cite{Harrison:2020gvo} for the SM $\tau$ negative-helicity CM and LAB $\bar B_c$ distributions are around $2\sigma$ below the NRQM predictions, while the shapes turn out to be in excellent agreement.

   As shown   in Ref.~\cite{Penalva:2020xup}, for a polarized final $\tau$ lepton with well defined helicity
  $h=\pm 1$, the CM angular and LAB energy distributions are respectively determined by
  \bea
  \frac{2\overline{\sum}|{\cal M}|^2}{M^2(1-\frac{m_\ell^2}{q^2})}=
  a_0(\omega,h)+a_1(\omega,h)
 \cos\theta_\ell+
 a_2(\omega,h)(\cos\theta_\ell)^2
 \label{eq:cmpol}
  \eea
\begin{eqnarray}
 \frac{2\overline{\sum}|{\cal M}|^2}{M^2}= \frac12\left(c_0+c_1
 \frac{E_\ell}{M}+
c_2 \frac{E^2_\ell}{M^2}\right)-\frac{h}2\frac{M}
{p_\ell}
 \left(\widehat{c}_0+
     \left[c_0+\widehat{c}_1\right] \frac{E_\ell}{M}+
     \left[c_1+\widehat{c}_2\right] \frac{E^2_\ell}{M^2}+
    \left[ c_2+\widehat{c}_3\right] \frac{E^3_\ell}{M^3}\right)
    \label{eq:Chpol}
\end{eqnarray}
In the latter equation, $p_\ell$ is the modulus of the final charged
lepton  three-momentum  measured  in the LAB frame. The general expressions 
of the $a_{0,1,2}(\omega,h)$ and 
$\widehat c_{0,1,2,3}(\omega)$ coefficients in terms of the $\widetilde W$ SFs
can be found in Ref.~\cite{Penalva:2020xup}.
\begin{figure}[tbh]
\includegraphics[height=4cm]{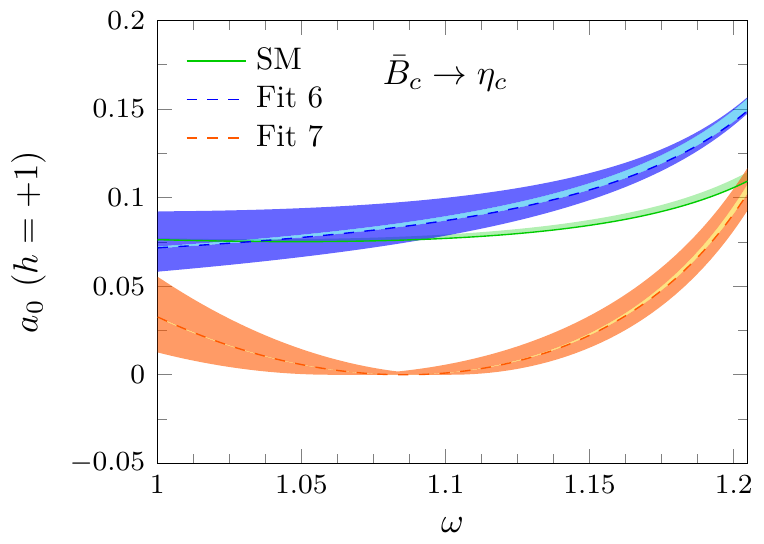} \hspace{.15cm}
\includegraphics[height=4cm]{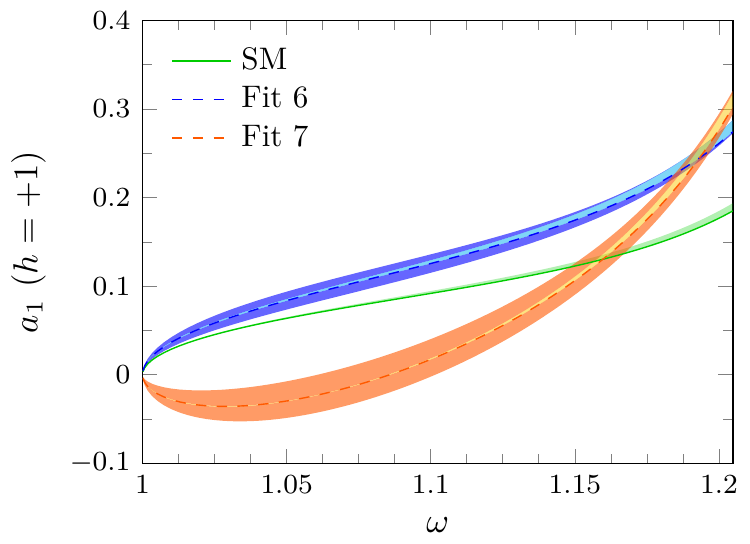}\hspace{.15cm}
\includegraphics[height=4cm]{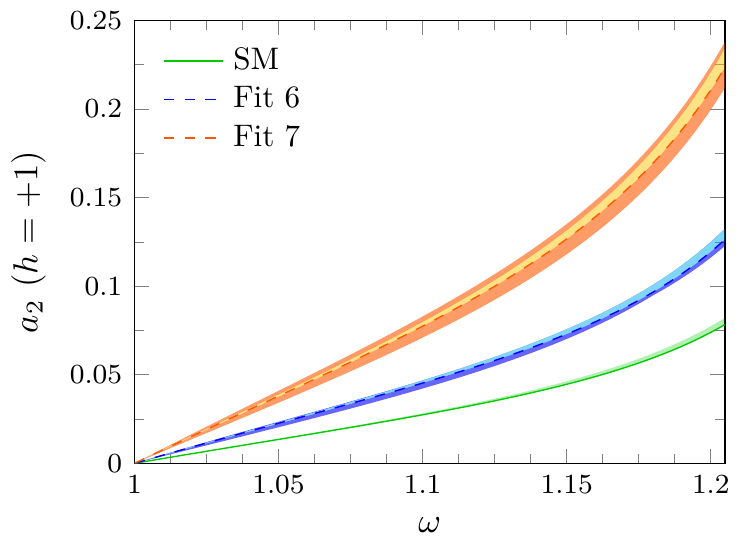}\hspace{.15cm}
\\
\includegraphics[height=4cm]{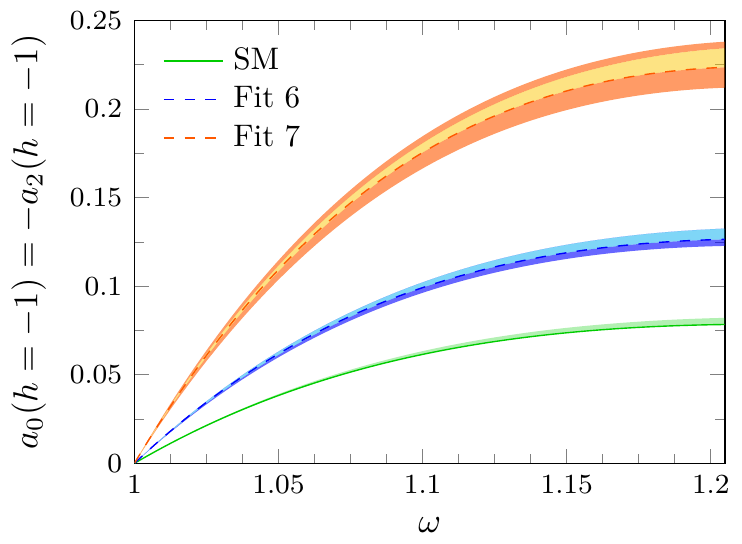} \hspace{.15cm}
\includegraphics[height=4cm]{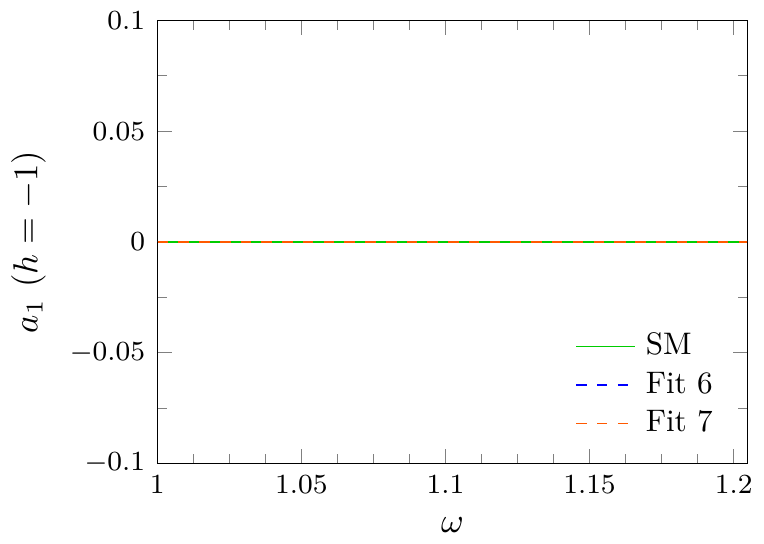}\hspace{.15cm}
\caption{ CM angular expansion coefficients for the 
$\bar B_c\to\eta_c\tau\bar\nu_\tau$ decay with a polarized $\tau$ with positive (upper panels)
and negative (lower panels) helicity.  They have been evaluated with the NRQM form factors
from Ref.~\cite{Hernandez:2006gt}. Uncertainty bands  as in 
Fig.~\ref{fig:dgdw}.
}  
\label{fig:aspoletac}
\end{figure}
  \begin{figure}[tbh]
\includegraphics[height=4cm,width=5.6cm]{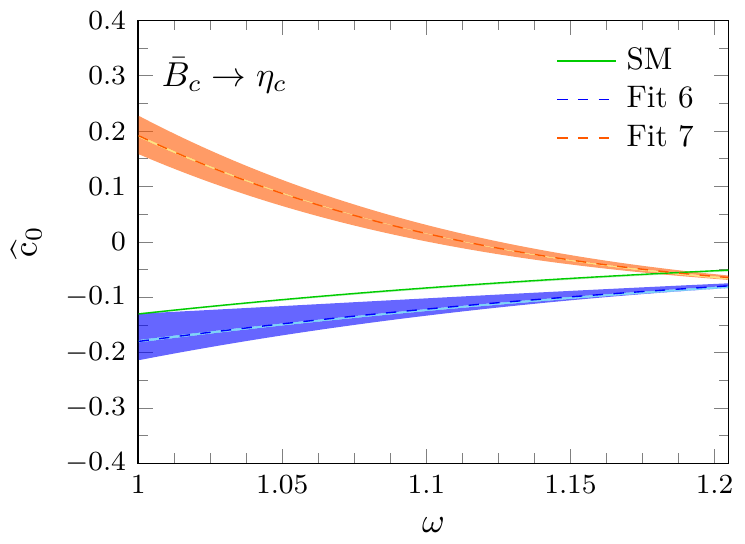}\hspace{.3cm}
\includegraphics[height=4cm,width=5.6cm]{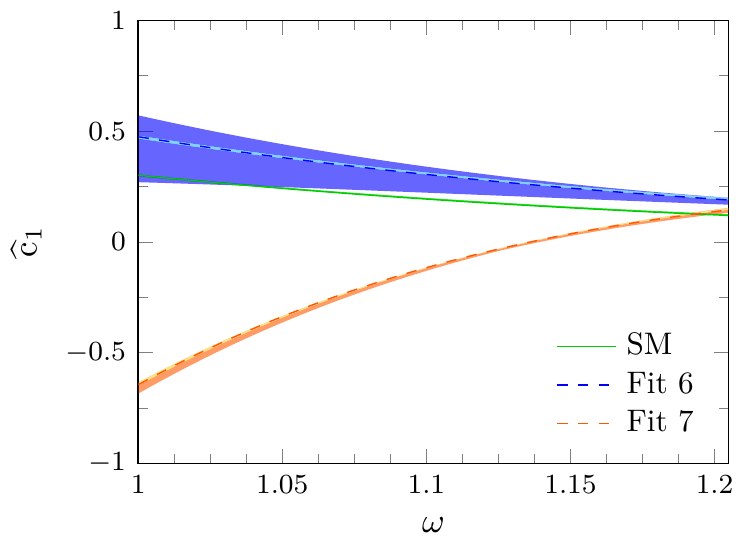}\hspace{.3cm}
\includegraphics[height=4cm,width=5.6cm]{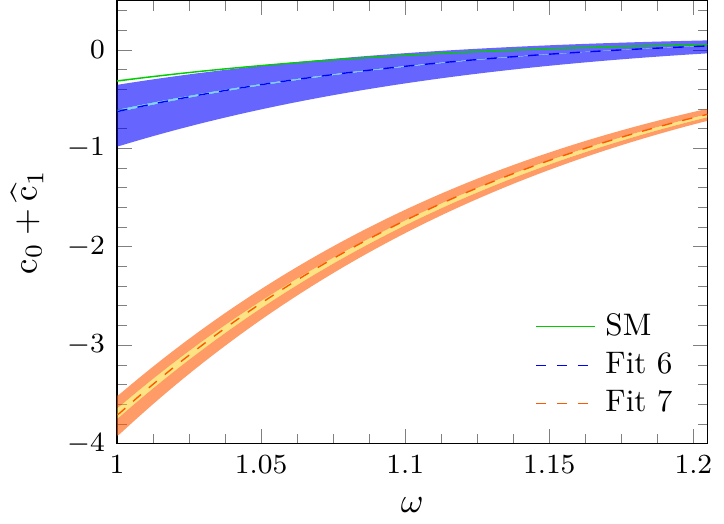} \\\vspace{0.25cm}
\includegraphics[height=4cm,width=4.2cm]{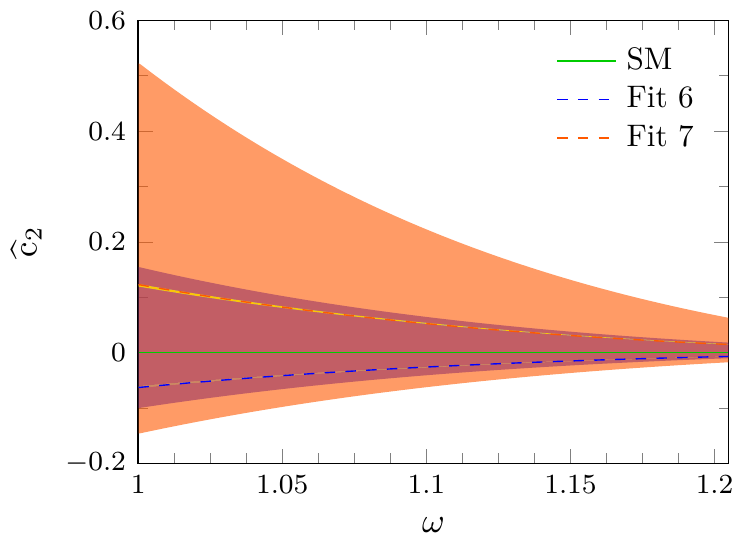}\hspace{.2cm}
\includegraphics[height=4cm,width=4.2cm]{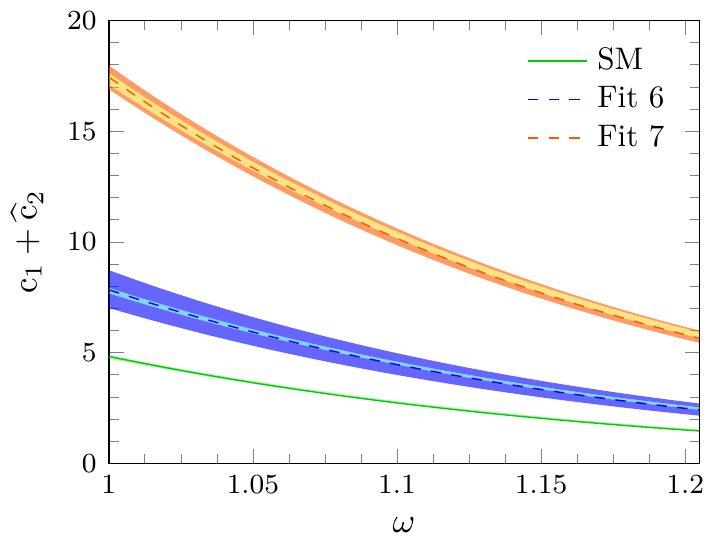} \hspace{0.2cm}
\includegraphics[height=4.2cm,width=4.2cm]{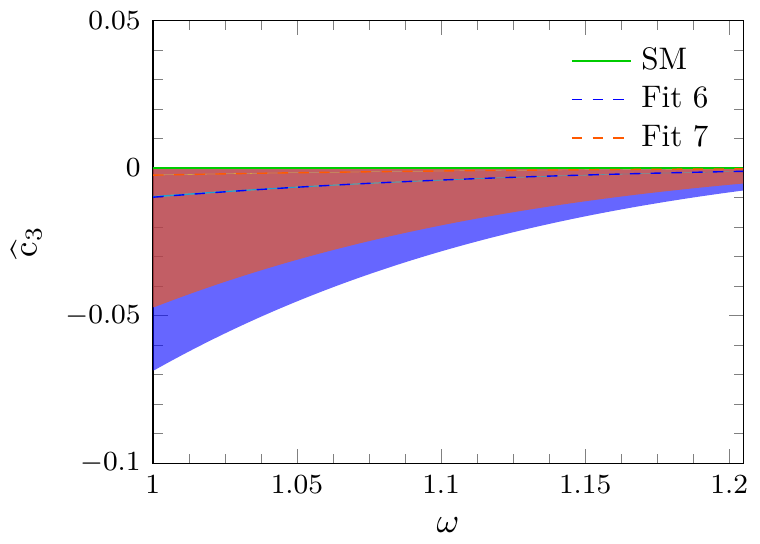}\hspace{0.2cm}
\includegraphics[height=4.2cm,width=4.2cm]{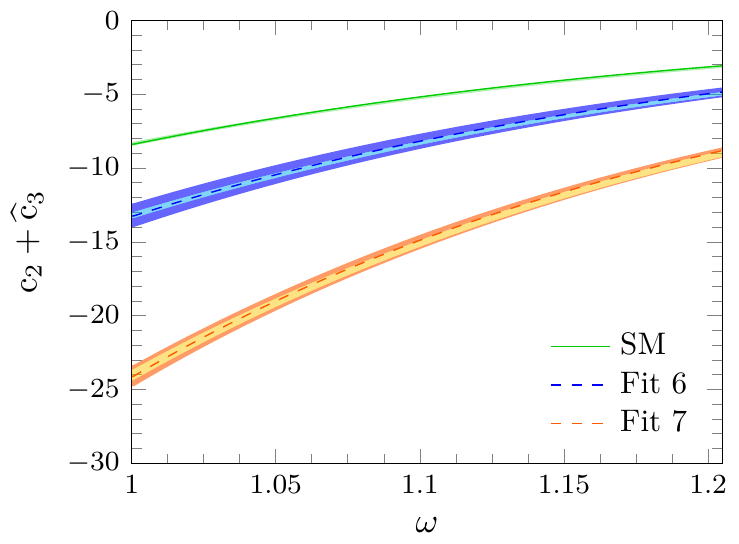}
\caption{ LAB charged lepton energy expansion coefficients $\widehat c_{0,1,2,3}(\omega)$  for the 
polarized $\bar B_c\to\eta_c\tau\bar\nu_\tau$ decay.  We also show the $(c_0+\widehat c_1)$, $(c_1+\widehat c_2)$ and $(c_2+\widehat c_3)$ sums in the third top, second and fourth bottom panels, respectively. All the functions  have been evaluated with the NRQM form factors
from Ref.~\cite{Hernandez:2006gt}. Uncertainty bands  as in 
Fig.~\ref{fig:dgdw}.  }  
\label{fig:cspoletac}
\end{figure}
In Figs.~\ref{fig:aspoletac} and \ref{fig:cspoletac} we present the results for
 the functions $a_{0,1,2}(\omega,h=\pm 1)$ (CM) and  $\widehat c_{0,1,2,3
 }(\omega)$ (LAB) for the polarized $\bar B_c\to\eta_c\tau\bar\nu_\tau$
reaction. 

We see that even taking uncertainties into account, Fits 6 and 7
provide distinct predictions for all non-zero angular coefficients that also
differ from the SM results, with the exception of $a_0(\omega, h=+1)$, for
 which SM and NP Fit 6 results overlap below $\omega\le 1.1$. We also
 observe 
that  for this decay ($\bar B_c\to\eta_c$), the relations 
\begin{equation}
a_0(\omega,h=-1)=-a_2(\omega,h=-1), \qquad a_1(\omega,h=-1)=0 \label{eq:a1hm1}
\end{equation}
are satisfied because of angular momentum conservation.
Since both the initial and final hadrons have zero spin, the virtual particle exchanged  (a $W$ boson in the SM) should have  helicity zero. 
In the CM system this corresponds to a  zero spin projection along the 
 quantization axis defined by its three-momentum in the LAB frame, the same axis that is defined by the
  final hadron LAB (or CM) three-momentum. Thus, in the CM system, the angular
   momentum of the final lepton pair measured along that axis  must be zero. As a consequence,
  the CM helicity   of a
  final $\tau$ lepton emitted along that direction, which corresponds to 
  either $\theta_\ell=0$ or  $\theta_\ell=\pi$, must equal that of the 
  $\bar\nu_\tau$, the latter   being always   positive. This means that a negative helicity   $\tau$ can not be emitted in the CM
   system when $\theta_\ell=0$ or $\pi$. Looking at Eq.~(\ref{eq:cmpol}),
   this implies that $a_0(\omega,h=-1)=-a_2(\omega,h=-1)$ and $a_1(\omega,h=-1)=0$.
   
Besides,  at zero recoil  CM and LAB frames coincide and
angular momentum conservation requires the helicity of the $\tau$ lepton to equal that of the anti-neutrino. 
This implies $a_0(\omega=1,h=-1)=-a_2(\omega=1,h=-1)=0$, and also the cancellation of Eq.~(\ref{eq:Chpol}) at zero recoil  for $h=-1$. 
 In fact,  the LAB $d^2\Gamma(\bar B_c\to\eta_c\tau\bar\nu_\tau)/(d\omega dE_\ell)$ distribution should  cancel for $h=-1$ and any value of $\omega$ 
 when $E_\ell$ equals its maximum value\footnote{The maximum $E_\ell^{+}$ and minimum  $E_\ell^{-}$  energy values allowed to the final charged lepton for a given $\omega$ are
  \begin{equation}
  E_\ell^{\pm}= \frac{(M-M'\omega)(q^2+m^2_\ell)  
  \pm M' \sqrt{\omega^2-1}(q^2-{m^2_\ell})}{2q^2}.
\end{equation}} 
for that particular $\omega$. The reason is that this maximum $E_\ell$ value corresponds necessarily to
$\theta_\ell=\pi$ and in that case the helicity of the $\tau$ is the same in both
CM and LAB frames. Since $h=-1$ is forbidden in the CM  for that specific kinematics it is also forbidden in the
LAB.  Note that any
 violation of these results
will require negative helicity anti-neutrinos which means NP contributions with 
right-handed neutrinos. The possible role of such beyond the SM terms in the explanation of the LFU ratio anomalies have been considered in 
Refs.~\cite{Ligeti:2016npd,Asadi:2018wea,Greljo:2018ogz,Robinson:2018gza,
Azatov:2018kzb,Heeck:2018ntp,Asadi:2018sym,Babu:2018vrl,
Bardhan:2019ljo,Shi:2019gxi,Gomez:2019xfw} and their existence has not
been discarded by the available $\bar B \to D^{(*)}$ data~\cite{Mandal:2020htr}. 

Note also that, as a result of $a_1(\omega,h=-1)$ being zero  for the $P_b \to P_c$ decays, the 
  forward-backward asymmetry in the CM system (${\cal A}_{FB}$ shown in the left panel of 
Fig.~\ref{fig:FB}) can only originate from positive helicity $\tau$'s. For the same reason,
for massless charged leptons ($\ell=e,\mu$), ${\cal A}_{FB}$  vanishes in the SM for transitions between pseudoscalar mesons.

\begin{figure}[tbh]
\includegraphics[height=4cm]{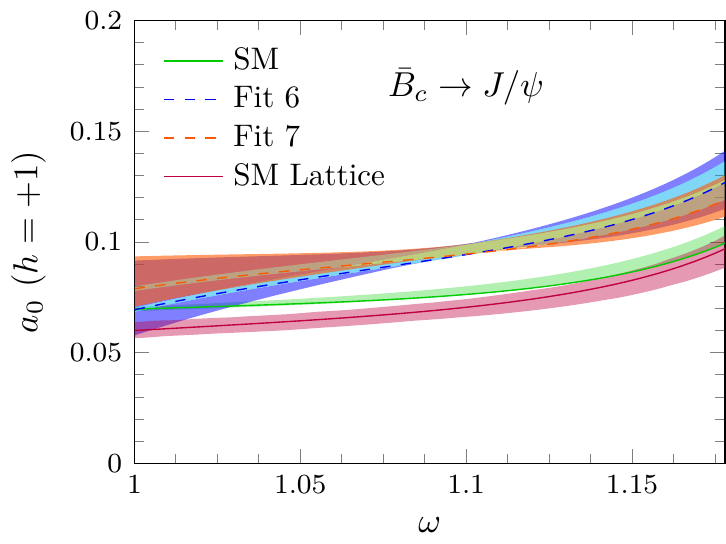} \hspace{.15cm}
\includegraphics[height=4cm]{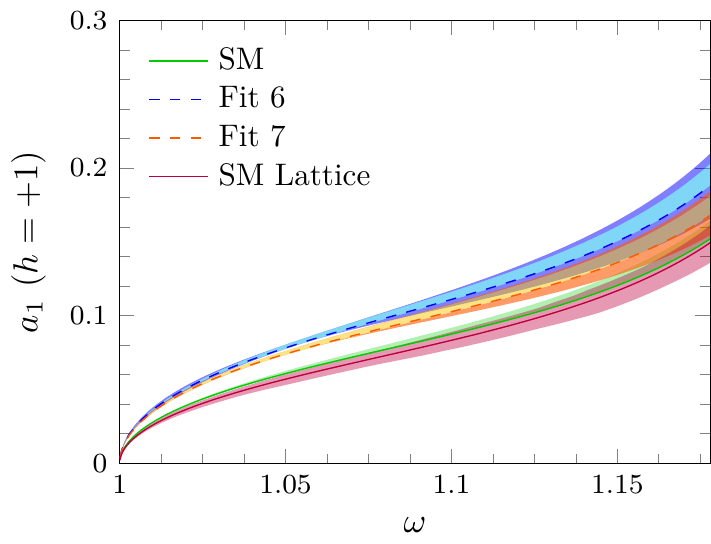}\hspace{.15cm}
\includegraphics[height=4cm]{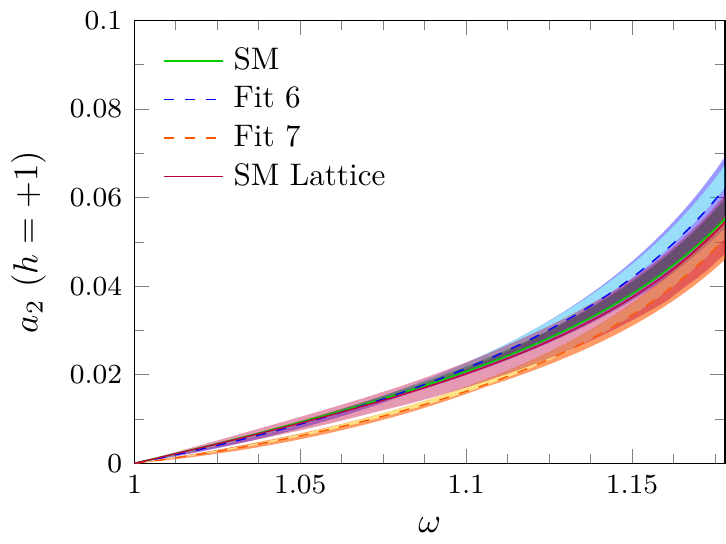}\hspace{.15cm}
\\
\includegraphics[height=4cm]{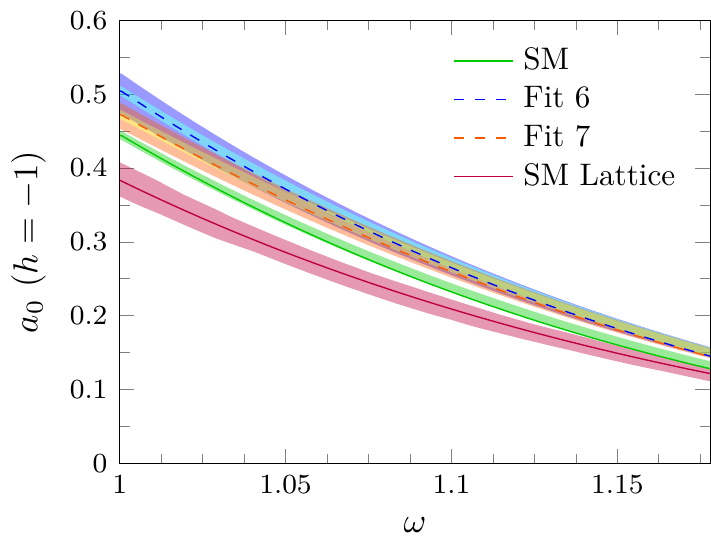} \hspace{.15cm}
\includegraphics[height=4cm]{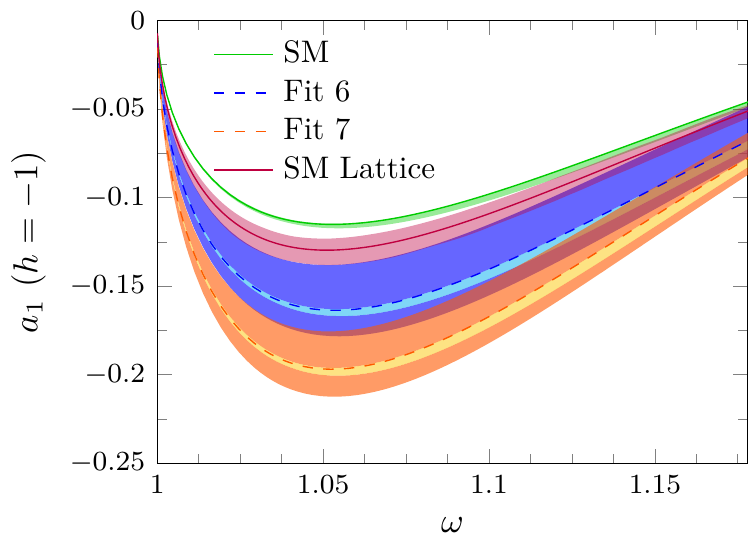}\hspace{.15cm}
\includegraphics[height=4cm]{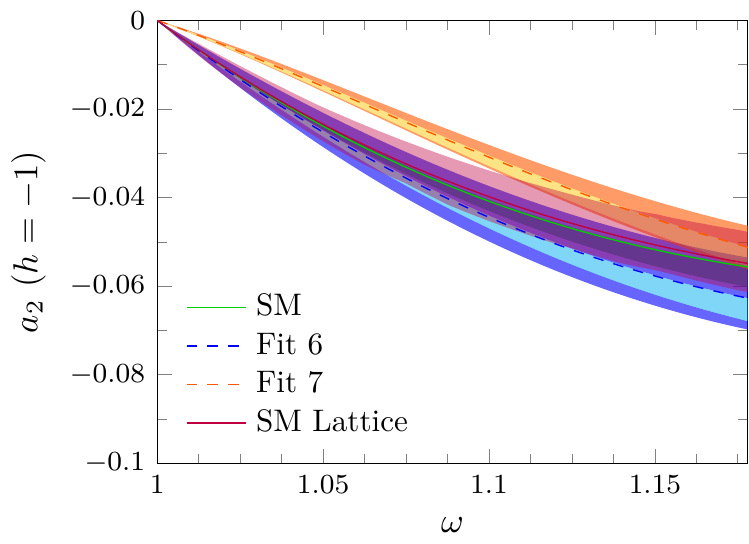}\hspace{.15cm}
\caption{ CM angular expansion coefficients for the 
$\bar B_c\to J/\psi\tau\bar\nu_\tau$ decay with a $\tau$ with positive (upper panels)
and negative (lower panels) helicity. 
They have been evaluated with the NRQM form factors
from Ref.~\cite{Hernandez:2006gt}. We also show the SM results obtained with LQCD form factors from Ref.~\cite{Harrison:2020gvo}. Uncertainty bands  as in 
Fig.~\ref{fig:dgdw}.  
}  
\label{fig:aspoljpsi}
\end{figure}
Looking at positive helicities, one finds that, in the high $\omega$ region, the quantity $a_0(\omega,h=+1)-a_1(\omega,h=+1)+a_2(\omega,h=+1)$ shows a steady decrease,  as $\omega$ increases. In fact, at maximum recoil, one has the approximate result
\bea
a_0(\omega_{\rm max},h=+1)-a_1(\omega_{\rm max},h=+1)+a_2(\omega_{\rm max},h=+1)\approx 0,
\label{eq:a012hp}
\eea
that can be readily inferred from the corresponding figures and
which corresponds to a very small probability of CM positive helicity $\tau$'s emitted at  $\theta_\ell=\pi$ when $\omega=\omega_{\rm max}$. This result can be partially understood taking into account that  our main contribution selects negative chirality
   for the final charged lepton\footnote{Note that only the ${\cal O}_{S_L,S_R}$ and ${\cal O}_T$ NP terms in Eqs.~(\ref{eq:hnp}) and (\ref{eq:hnp2}) select positive chirality for the final charged lepton.}. A $\tau$ lepton emitted with positive
    helicity in the CM frame and with $\theta_\ell=\pi$ will also have positive
     helicity in the LAB frame. However, close to maximum recoil its momentum
 in the    LAB is very large and helicity almost equals chirality, hence the partial cancellation.     
   Note that this result   would be independent of the spin of the
 hadrons involved as long as negative chirality lepton current operators are dominant. This approximate relation in Eq.~(\ref{eq:a012hp})  can already be seen in the polarized results for the $\Lambda_b\to \Lambda_c$ decay shown in Ref.~\cite{Penalva:2020xup}. Besides, and for the same chirality/helicity argument, one expects 
  the  LAB ratio  $\frac{d\Gamma/d\omega(h=+1)}
 {d\Gamma/d\omega(h=-1)}$ to be small in the high $\omega$ region, the reason being that for $\omega$ close to $\omega_{\rm max}$  the charged lepton energies are significantly larger than its mass.

 In  Fig.~\ref{fig:cspoletac} we present the results for the $\widehat c_{0,1,2,3}(\omega)$ coefficients 
 associated to this decay. We observe that $\widehat c_{0}(\omega)$ and $\widehat c_{1}(\omega)$ are able to distinguish between the two NP fits from 
 Ref.~\cite{Murgui:2019czp} considered in the present work. The other two observables $\widehat c_{2}(\omega)$ and $\widehat c_{3}(\omega)$, available from the polarized $d^2\Gamma(\bar B_c\to\eta_c\tau\bar\nu_\tau)/(d\omega dE_\ell)$ distribution, turn out to be very small and negligible when compared with $c_1$ and $c_2$, respectively (see the  plots in Fig.~\ref{fig:cspoletac}). Therefore, these two additional coefficients have little relevance in the discussion of the NP Fits 6 and 7, for which the NP tensor Wilson coefficient $|C_T|\sim 10^{-2}$ is quite small. As discussed in Ref.~\cite{Penalva:2020xup},  
 $\widehat c_2 $ and $\widehat c_3$ are, however, optimal observables to restrict the validity of  NP schemes with larger $|C_T|$ values.
 
\begin{figure}[tbh]
\includegraphics[height=4cm,width=5.6cm]{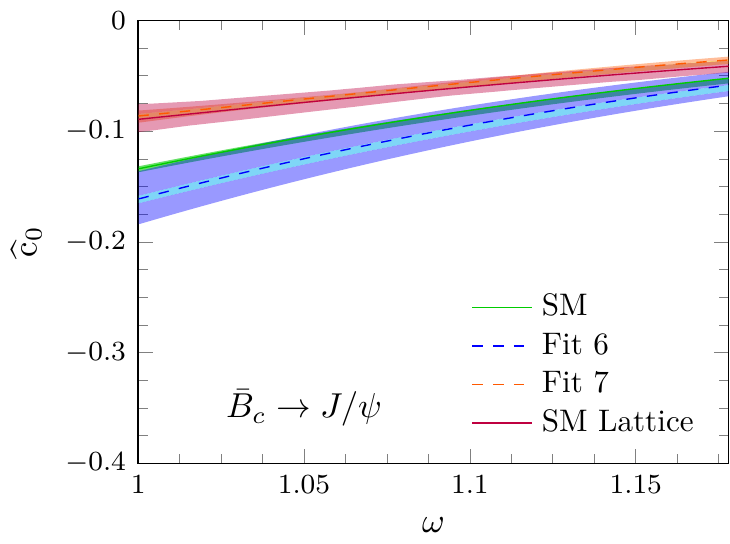}\hspace{.3cm}
\includegraphics[height=4cm,width=5.6cm]{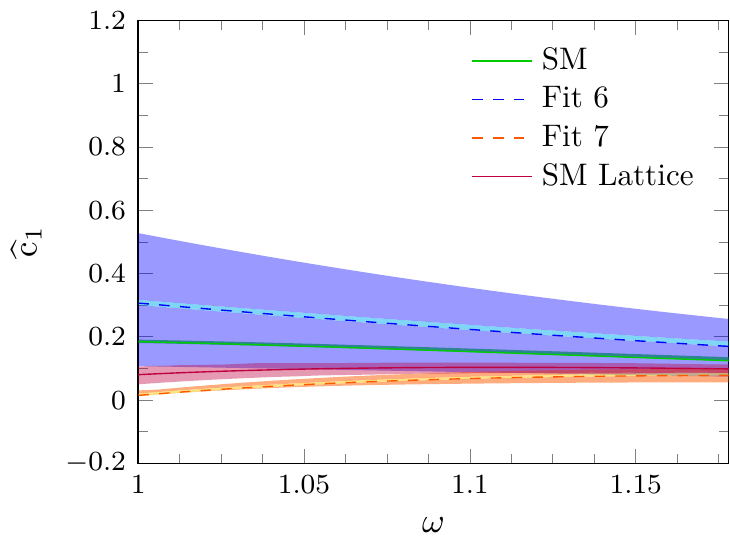}\hspace{.3cm}
\includegraphics[height=4cm,width=5.6cm]{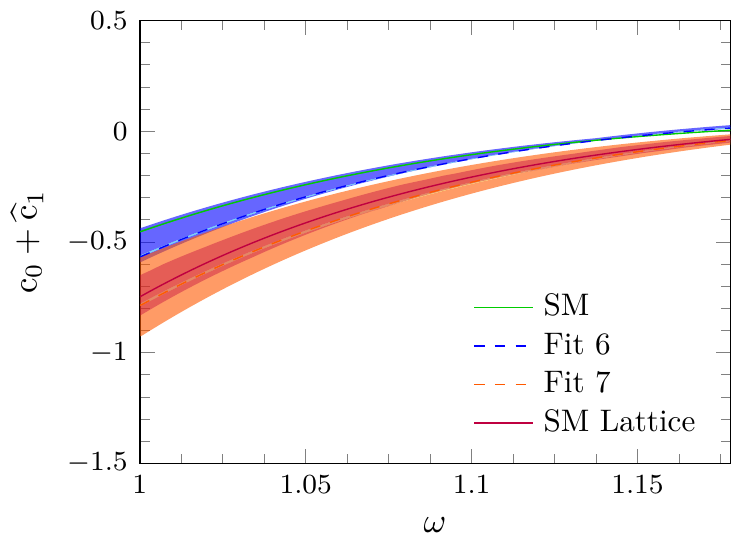} \\\vspace{0.25cm}
\includegraphics[height=4cm,width=4.2cm]{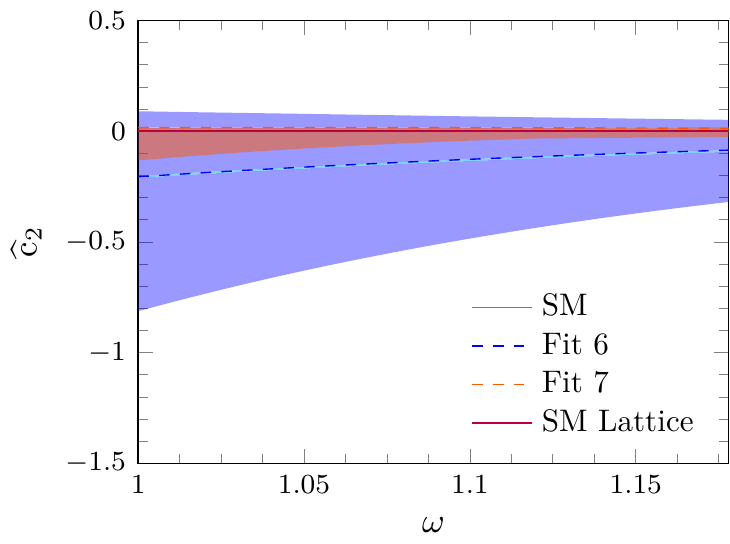}\hspace{.2cm}
\includegraphics[height=4cm,width=4.2cm]{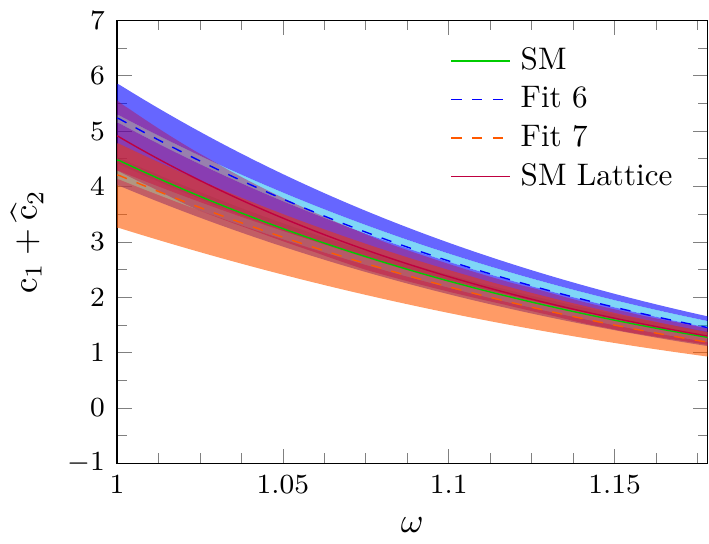} \hspace{0.2cm}
\includegraphics[height=4.2cm,width=4.2cm]{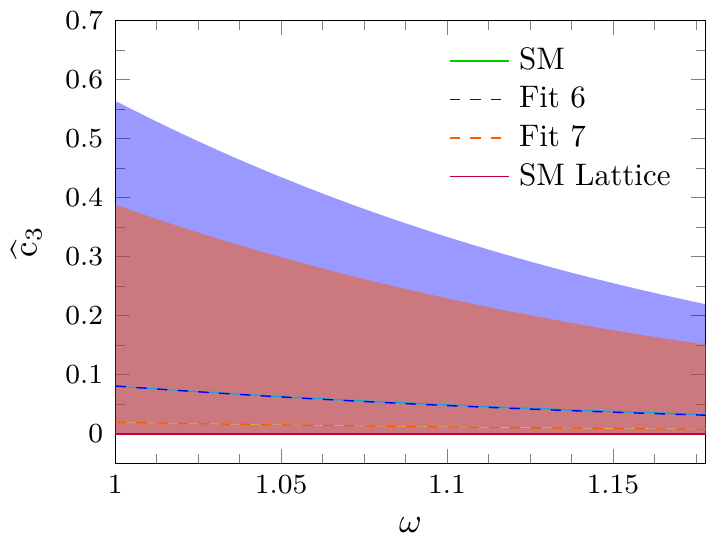}\hspace{0.2cm}
\includegraphics[height=4.2cm,width=4.2cm]{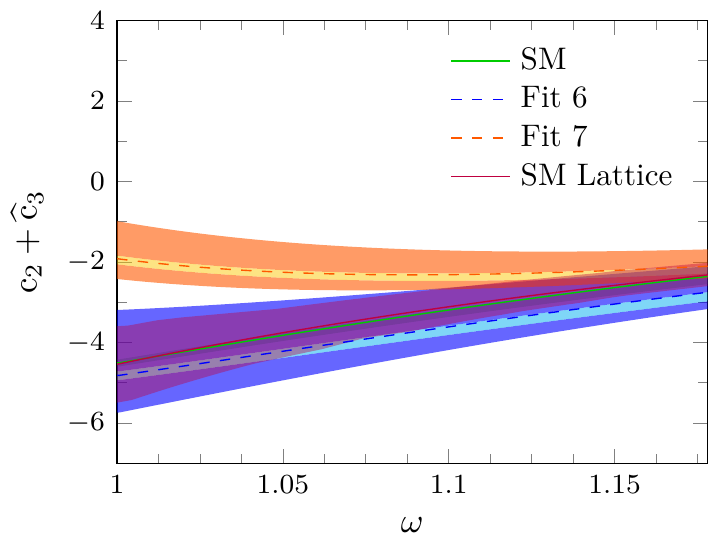}
\caption{ LAB charged lepton energy expansion coefficients $\widehat c_{0,1,2,3}(\omega)$  for the 
polarized $\bar B_c\to J/\psi\tau\bar\nu_\tau$ decay.  We also show the $(c_0+\widehat c_1)$, $(c_1+\widehat c_2)$ and $(c_2+\widehat c_3)$ sums in the third top, second and fourth bottom panels, respectively. All the functions  have been evaluated with the NRQM form factors
from Ref.~\cite{Hernandez:2006gt}.  We also show the SM results obtained with LQCD form factors from Ref.~\cite{Harrison:2020gvo}. Uncertainty bands  as in 
Fig.~\ref{fig:dgdw}.}  
\label{fig:cspoljpsi}
\end{figure}

In Figs.~\ref{fig:aspoljpsi} and \ref{fig:cspoljpsi} we collect the 
corresponding results for the $\bar B_c\to J/\psi\tau\bar\nu_\tau$ decay. In this case, 
no angular momentum related restriction is in place
 for $a_{0,1,2}(\omega,h=-1)$ since the final hadron has spin one and there are
three possible helicity states. However, one can see that the approximate relation in
Eq.~(\ref{eq:a012hp}) is indeed satisfied. Also the discussion above about 
the  LAB ratio  $\frac{d\Gamma/d\omega(h=+1)}
 {d\Gamma/d\omega(h=-1)}$ being small  near maximum recoil also applies for this decay. When looking at SM results alone we find a qualitative agreement of NRQM and LQCD results.

\begin{figure}[tbh]
\includegraphics[height=4.1cm]{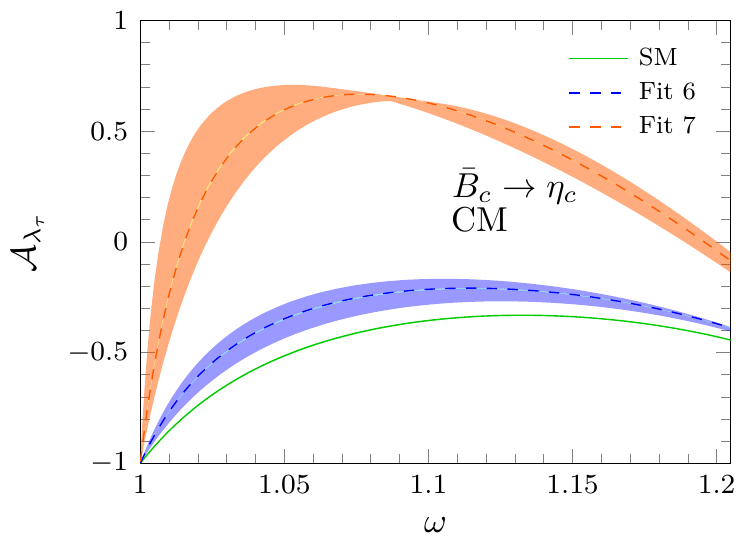} \hspace{.1cm}
\includegraphics[height=4cm]{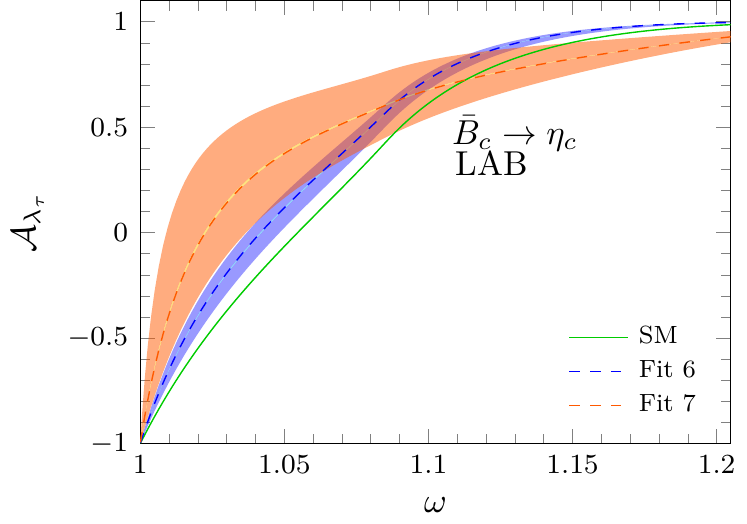}\\

\includegraphics[height=4.1cm]{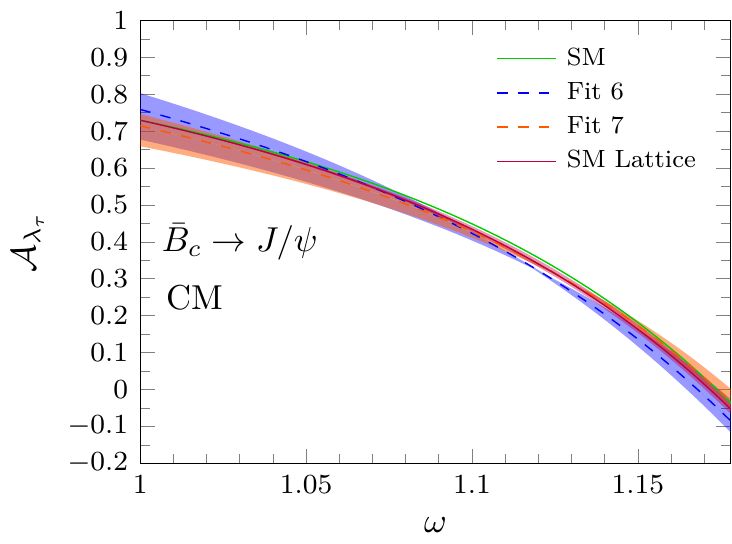} \hspace{.3cm}
\includegraphics[height=4.2cm]{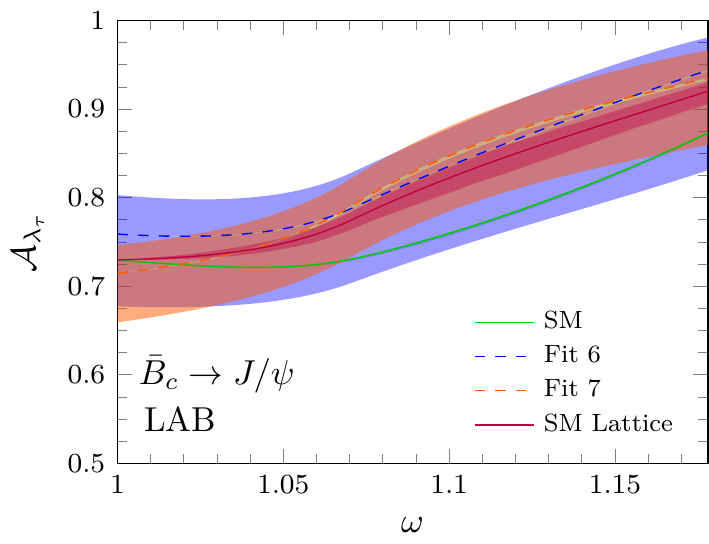}\\
\caption{ $\tau$ polarization asymmetry ${\cal A}_{\lambda_\tau}$, defined in Eq.~(\ref{eq:taupolasy}), for the 
$\bar B_c\to \eta_c$ (upper panels) and $\bar B_c\to J/\psi$ (lower panels) semileptonic decays measured in the CM (left panels) and LAB (right panels) frames. They have been  obtained with  the NRQM form factors
from Ref.~\cite{Hernandez:2006gt}. For the $\bar B_c\to J/\psi$ case we also show the SM results obtained with the LQCD form factors from  Ref.~\cite{Harrison:2020gvo}. Uncertainty bands  as in 
Fig.~\ref{fig:dgdw}.  
}  
\label{fig:poltauasim}
\end{figure}

To conclude this section, in Fig.~\ref{fig:poltauasim}  we present the results for the $\tau$ polarization asymmetry
\bea
{\cal A}_{\lambda_\tau}=\frac{d\Gamma/d\omega(h=-1)-d\Gamma/d\omega(h=+1)}{d\Gamma/d\omega(h=-1)+d\Gamma/d\omega(h=+1)}
\label{eq:taupolasy}
\eea
measured both in the CM and LAB frames.

For the $\bar B_c\to \eta_c$ decay we see that both  polarization asymmetries equal minus one at zero recoil and  that the  LAB one tends to plus one when maximum recoil is approached. The Wilson coefficients of the  charged-lepton positive chirality ${\cal O}_{S_L,S_R}$ operators in the 
 Fit 7 are significantly larger than in the Fit 6, which explains the larger deviations of ${\cal A}_{\lambda_\tau}$ from $+1$ at $\omega=\omega_{\rm max}$ in the first NP scenario. This is in perfect accordance with the discussion above. 
As it is clear from the figure, the CM polarization asymmetry  is a good observable
to distinguish between different NP scenarios. This is not the case however of the LAB one. 
For the $\bar B_c \to J/\psi$ decay the asymmetries are equal at zero recoil, since CM and LAB frames coincide at $\omega=1$, but otherwise they show a very different $\omega$ dependence. None of them is able to distinguish NP results for Fits 6 and 7 among themselves and from the SM. We also observe for these latter decays, a quite good agreement between NRQM and LQCD predictions in particular for the results found in the CM frame, which are also reported in the recent analysis of Ref.~\cite{Harrison:2020nrv}.

Although  the $\bar B_c \to J/\psi$ decay is perhaps easier to measure  experimentally, as a general rule we find the $\bar B_c \to J/\psi$ observables, also in the case of a polarized $\tau$, 
are less optimal for distinguishing between NP Fits 6 and 7 than those discussed above for $\bar B_c \to \eta_c$ decays, or those presented in Ref.~\cite{Penalva:2020xup} for the related $\Lambda_b \to \Lambda_c$ semileptonic decay.

\section{$\bar B\to D$ and $\bar B\to D^*$ semileptonic decay results with an unpolarized final $\tau$ lepton}
\label{sec:B2D}

We present now results for the $\bar B\to D\tau\bar\nu_\tau$ and
$\bar B\to D^*\tau\bar\nu_\tau$ semileptonic decays. As in the previous section, we shall use 
 the Wilson coefficients  and form factors 
corresponding to Fits 6 and 7 in Ref.~\cite{Murgui:2019czp}. 
The form factors are taken from Ref.~\cite{Bernlochner:2017jka}, but in Ref.~\cite{Murgui:2019czp}
not only the Wilson coefficients but also the $1/m_{b,c}$ and $1/m^2_{c}$ corrections to the form factors 
are simultaneously fitted
to experimental data.
To estimate  the
theoretical uncertainties, for each fit, we shall use different sets of Wilson coefficients 
and form factors, selected such that the $\chi^2$ merit function computed in \cite{Murgui:2019czp} changes at most 
by one unit from its value at the fit minimum. With those sets,  for
 each of the  observables that we calculate we determine the maximum deviations 
 above and below  
 their central values. These deviations will give us the $1\sigma$ 
 theoretical uncertainty
 and it will be shown as an error
 band in the figures below.

We start by showing in 
Fig.~\ref{fig:dgdwddstar} the 
$d\Gamma/d\omega$ differential decay width. Both NP fits give similar results that differ
from the SM distribution.
\begin{figure}[tbh]
\includegraphics[height=5.5cm]{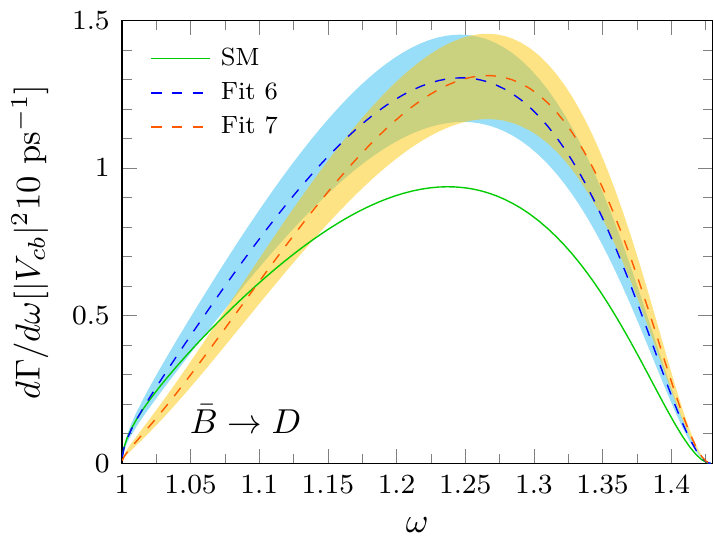} \hspace{1.15cm}
\includegraphics[height=5.5cm]{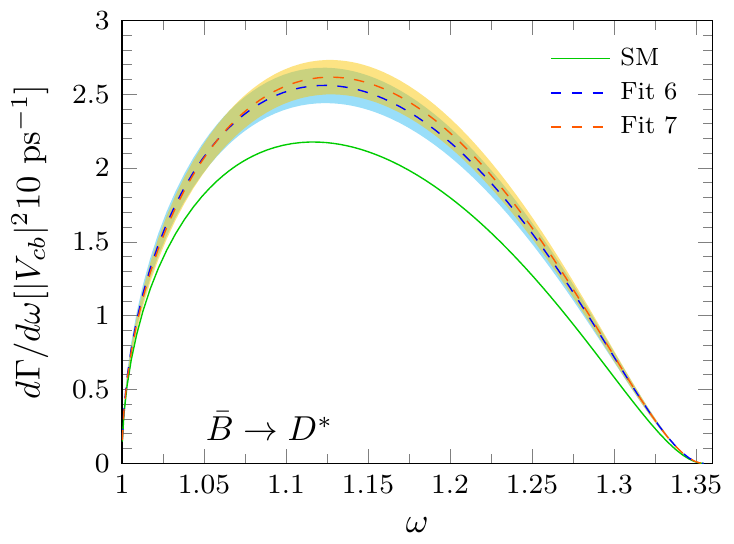}\\
\caption{$d\Gamma(\bar B\to D \tau\bar \nu_\tau)/d\omega$ (left) and
$d\Gamma(\bar B\to D^* \tau\bar \nu_\tau)/d\omega$ (right)
differential decay widths, as a function of $\omega$ 
and in units of $10 |V_{cb}|^2 {\rm ps}^{-1}$. 
We show SM predictions and full NP results obtained using the Wilson coefficients and form factors 
corresponding to
Fits 6  and  7  of Ref.~\cite{Murgui:2019czp}.   Uncertainty bands  obtained 
as explained in the main text. 
}  
\label{fig:dgdwddstar}
\end{figure} 
The corresponding predictions for the ${\cal R}_D$ and  ${\cal R}_{D^*}$ ratios  are given in 
Table~\ref{tab:ratiosd}.
\begin{table}[h!]
\begin{center}
\begin{tabular}{c|c|c|c|}%\cline{2-4}
 \multicolumn{1}{c}{}&\multicolumn{1}{|c|}{SM}&
 \multicolumn{1}{c|}{Fit 6} & 
 \multicolumn{1}{c|}{Fit 7}\\
  \hline
  &&&\\
${\cal R}_{D}=\frac{\Gamma(\bar B\to D \tau\bar \nu_\tau)}
{\tstrut\Gamma(\bar B\to D \ell\bar \nu_\ell)}$ & $0.300 \pm 0.005$&$0.405\pm 0.048$&$0.389\pm 0.045$
\\&&&\\\hline&&&\\
 ${\cal R}_{D^*}=\frac{\Gamma(\bar B\to D^* \tau\bar \nu_\tau)}
{\tstrut\Gamma(\bar B\to D^*\ell\bar \nu_\ell)}$ &$0.251 \pm 0.004$&$0.302\pm0.014$&$0.306\pm0.013$ \\&&& \\\hline
\end{tabular}
\end{center}
\caption{${\cal R}_{D}$ and ${\cal R}_{D^*}$ ratios obtained in the SM and with NP effects from Ref.~\cite{Murgui:2019czp}). }
\label{tab:ratiosd}
\end{table}
The ratios obtained with NP are in agreement with present experimental 
results, though they are located at the high-value corner of the allowed regions, since they were fitted in Ref.~\cite{Murgui:2019czp} to the previous HFLAV world average values quoted in \cite{Amhis:2016xyh}\footnote{In the latest HFLAV average~\cite{Amhis:2019ckw},  a measurement by the BaBar collaboration~\cite{Aubert:2008yv} is omitted, because it does not allow for a separation of the different isospin modes.}. Again, we notice that for these quantities both fits are equivalent within errors and  other observables are needed in order to decide between different NP explanations of the experimental data.
\begin{figure}[tbh]
\includegraphics[height=4cm]{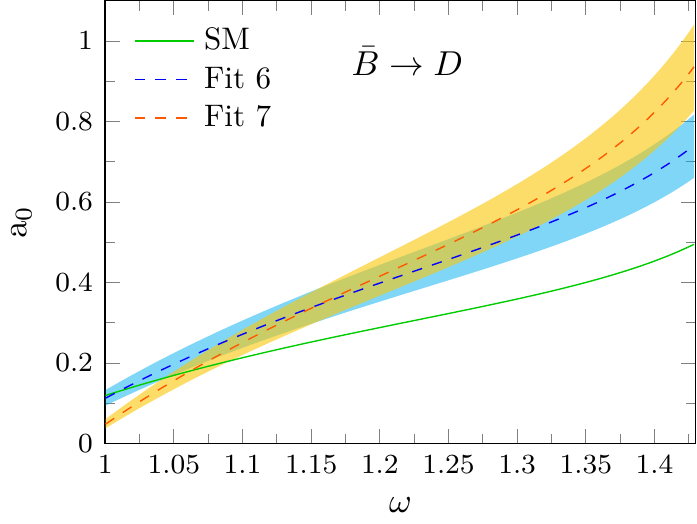} \hspace{.15cm}
\includegraphics[height=4cm]{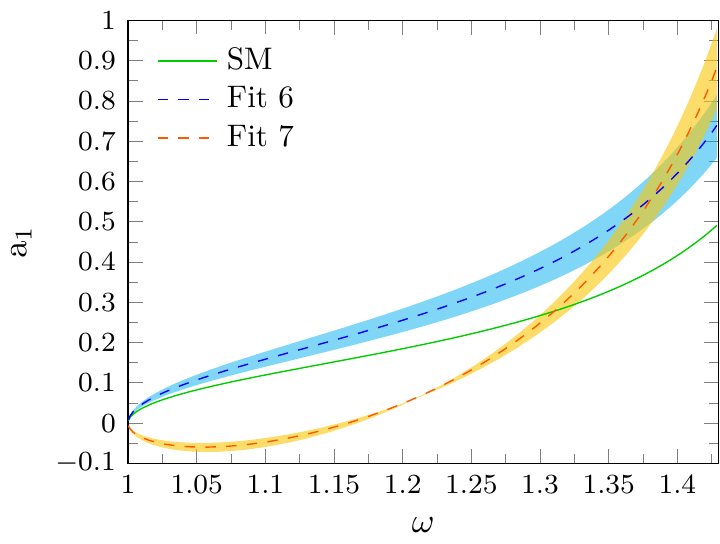}\hspace{.15cm}
\includegraphics[height=4cm]{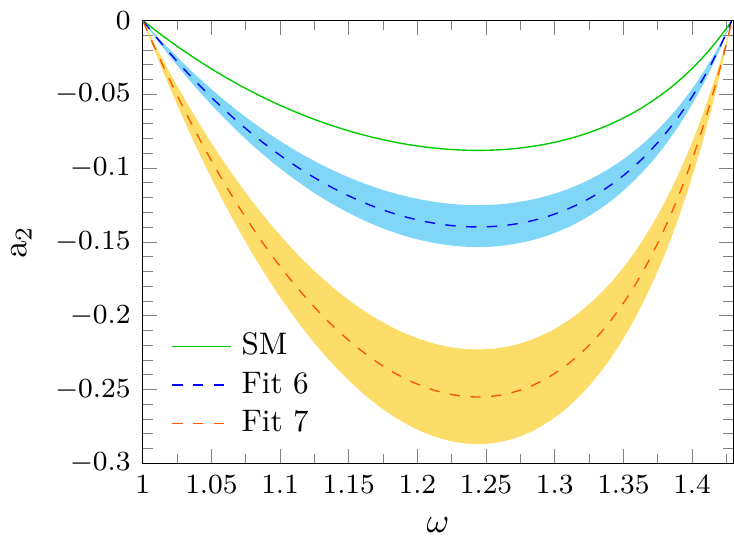}\hspace{.15cm}
\\
\includegraphics[height=4cm]{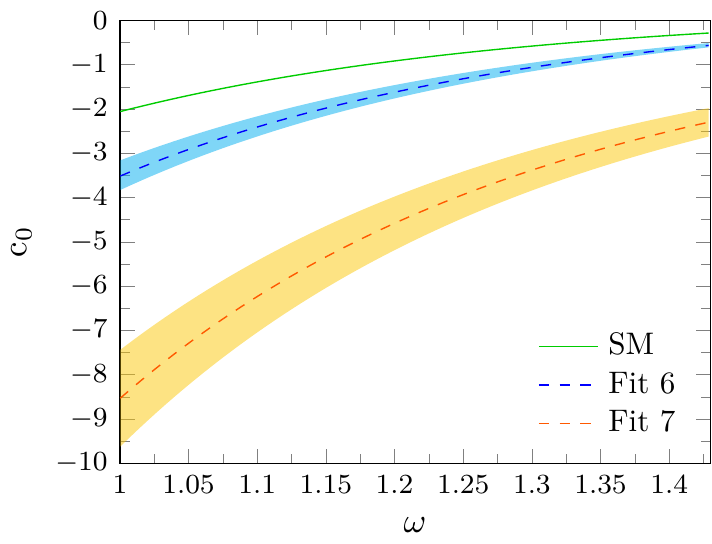} \hspace{.15cm}
\includegraphics[height=4cm]{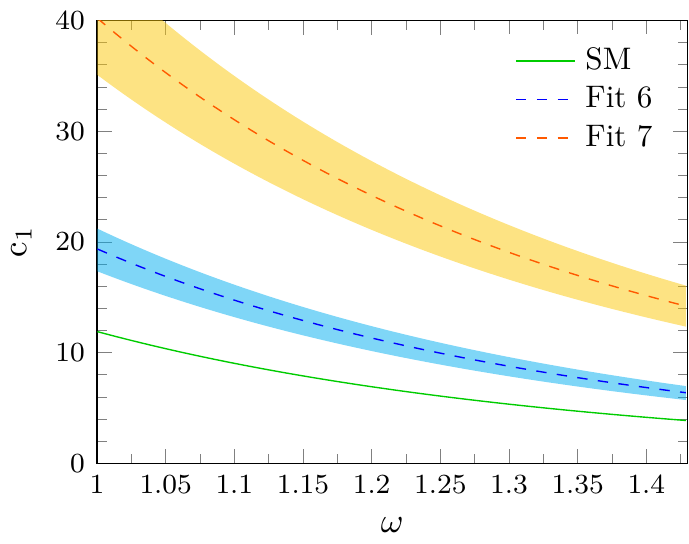}\hspace{.15cm}
\includegraphics[height=4cm]{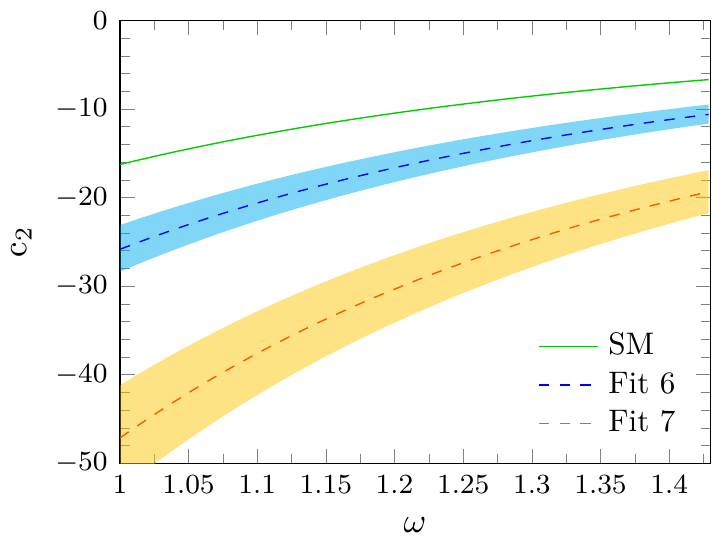}\hspace{.15cm}
\caption{ $\bar B\to D\tau\bar\nu_\tau$ decay: $a_{0,1,2}$ CM angular  and
$c_{0,1,2}$ LAB energy expansion coefficients
as a function of $\omega$.  Uncertainty bands  as in Fig.~\ref{fig:dgdwddstar}. 
}  
\label{fig:ascsd}
\end{figure}
\begin{figure}[tbh]
\includegraphics[height=4cm]{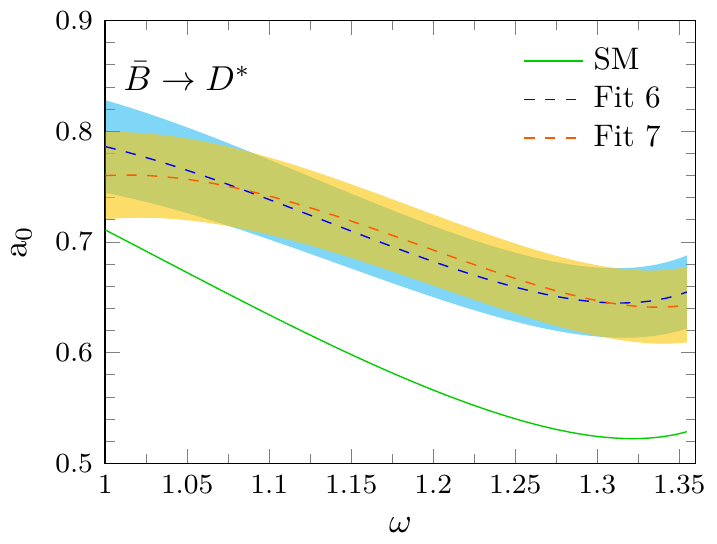} \hspace{.15cm}
\includegraphics[height=4cm]{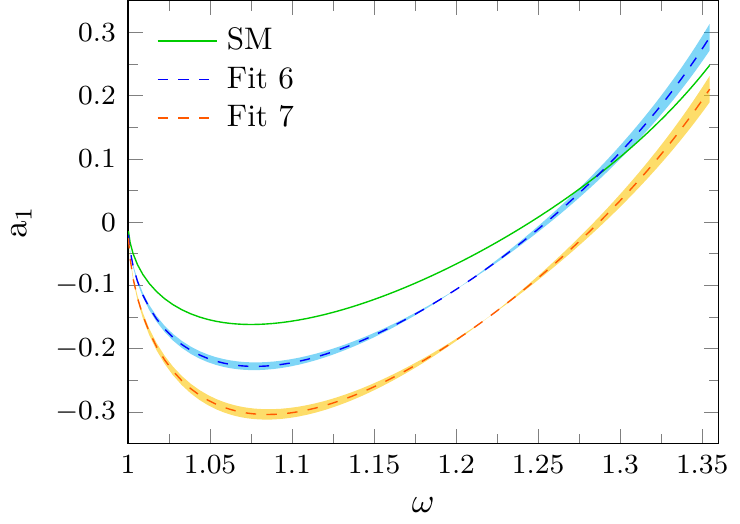}\hspace{.15cm}
\includegraphics[height=4cm]{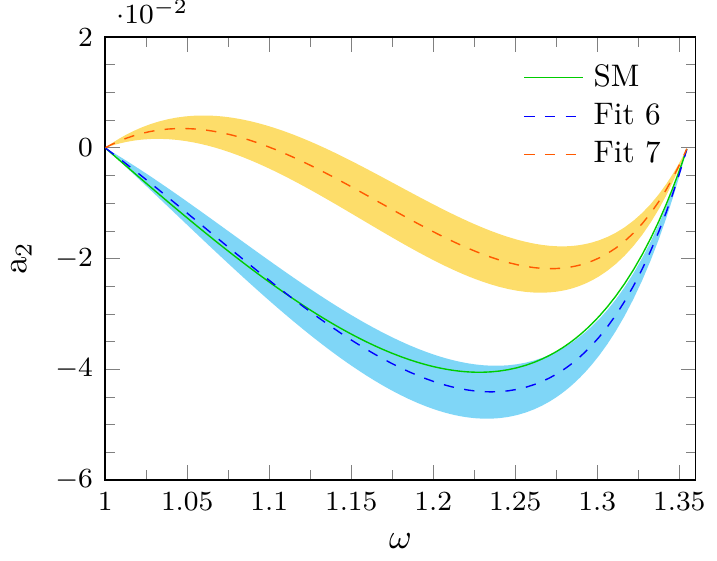}\hspace{.15cm}
%
%\\
\includegraphics[height=4cm]{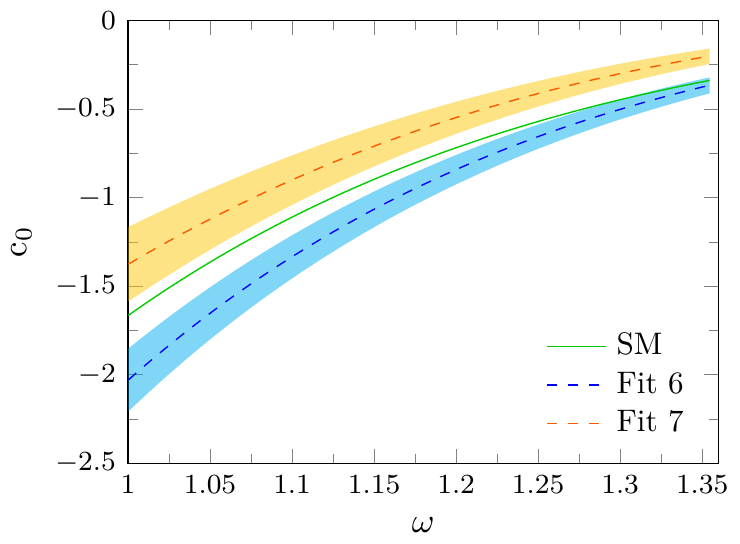} \hspace{.15cm}
\includegraphics[height=4cm]{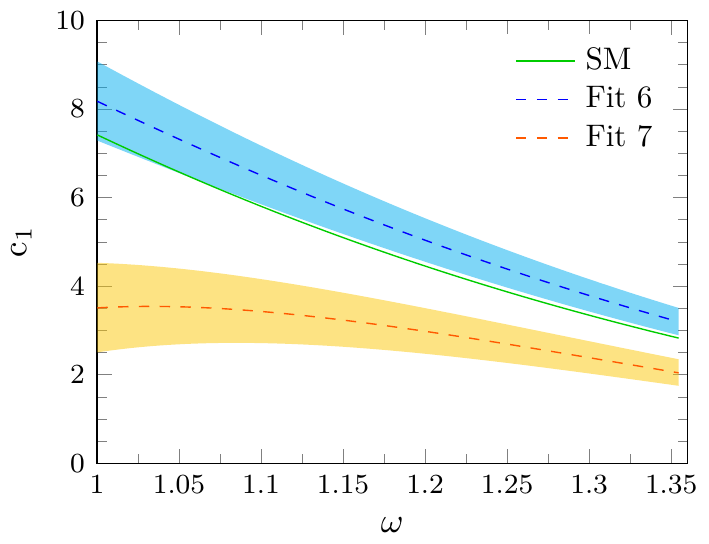}\hspace{.15cm}
\includegraphics[height=4cm]{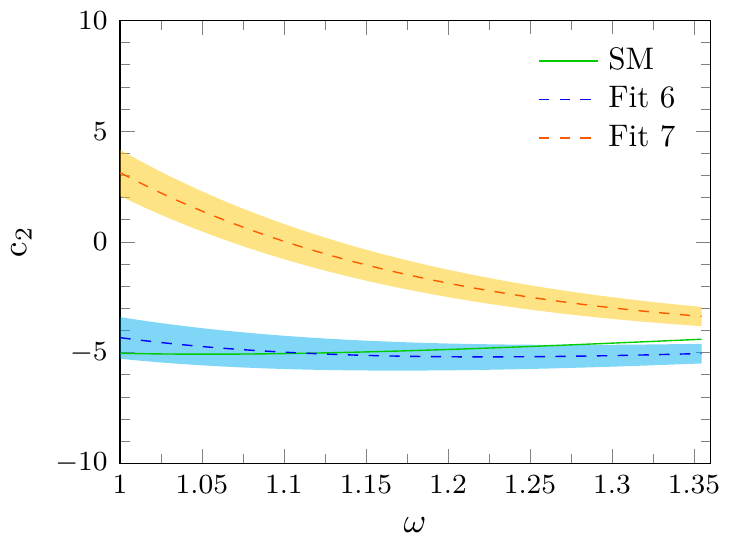}\hspace{.15cm}
\caption{ $\bar B\to D^*\tau\bar\nu_\tau$ decay: $a_{0,1,2}$ CM angular  and
$c_{0,1,2}$ LAB energy expansion coefficients
as a function of $\omega$.  Uncertainty bands  as in Fig.~\ref{fig:dgdwddstar}. 
}  
\label{fig:ascsdstar}
\end{figure}

These observables can be the
$a_{0,1,2}$  and $c_{0,1,2}$  coefficients in the CM angular $d^2\Gamma/(d\omega d\cos\theta_\ell)$ and  LAB energy $d^2\Gamma/(d\omega dE_\ell)$ distributions.  They are shown in Figs.~\ref{fig:ascsd} and \ref{fig:ascsdstar} for the $\bar B \to D\tau\bar\nu_\tau$ and $\bar B \to  D^*\tau\bar\nu_\tau$ decays, respectively. With the only exception of
$a_0$, all of them can be used to distinguish between the two  fits.
However, for the  $\bar B\to D^*$, and similar to what happened for the 
$\bar B_c\to J/\psi$ decay, SM results for 
some of these coefficients fall within the error band of those obtained from  NP Fit 6. In fact the $\omega-$shape patterns exhibited in Figs.~\ref{fig:ascsd} and \ref{fig:ascsdstar} for the  $\bar B  \to D^{(*)}$ reactions are qualitatively similar to those found in Sec.~\ref{sec:Bc2cc} for the $\bar B_c$ decays.  

We stress that the LAB  $d^2\Gamma(\bar B  \to D^{(*)}\tau \bar\nu_\tau  )/(d\omega dE_\ell)$ differential decay widths  are reported for the very  first time in this work.  Though, as shown in \cite{Penalva:2020xup}, CM and LAB  unpolarized distributions provide access to equivalent dynamical information (invariant functions ${\cal A(\omega)}$, ${\cal B(\omega)}$ and ${\cal C(\omega)}$ defined in Eq.~(14) of that reference), it should be explored if the LAB observables could be measured with better precision. 

\begin{figure}[h!]
\includegraphics[height=5.5cm]{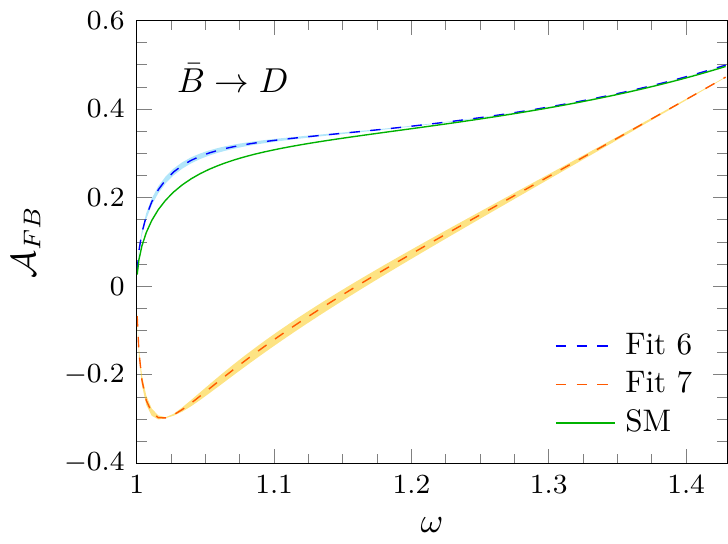} \hspace{.15cm}
\includegraphics[height=5.5cm]{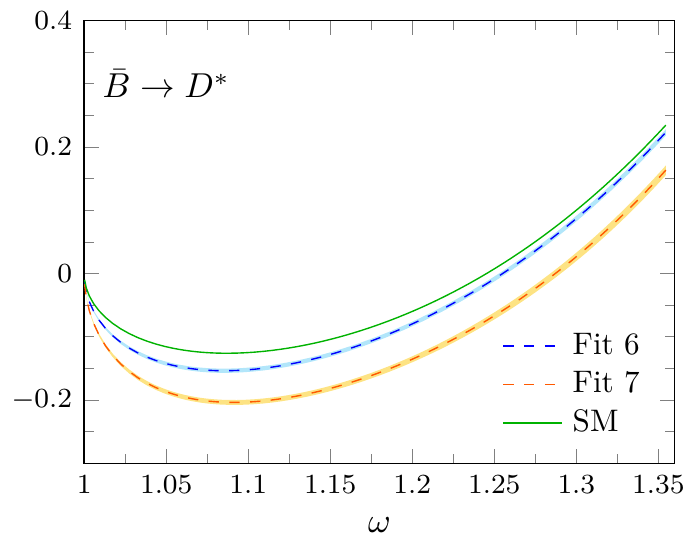}\hspace{.15cm}
\caption{ Forward-backward asymmetry in the CM reference frame
for the $\bar B\to D\tau\bar\nu_\tau$  (left) and 
$\bar B\to D^*\tau\bar\nu_\tau$ (right) decays. 
 Uncertainty bands    as in Fig.~\ref{fig:dgdwddstar}. 
}  
\label{fig:FBD}
\end{figure}

%\newpage
In Fig.~\ref{fig:FBD} we show the CM forward-backward asymmetry (Eq.~\eqref{eq:Afb}). The
shape in each case is very similar to what we obtained respectively for 
$\bar B_c\to\eta_c$ and $\bar B_c\to J/\psi$ decays, see Fig.~\ref{fig:FB},  with very close values at maximum recoil 
and significantly smaller errors.

\begin{figure}[tbh]
\includegraphics[height=4cm]{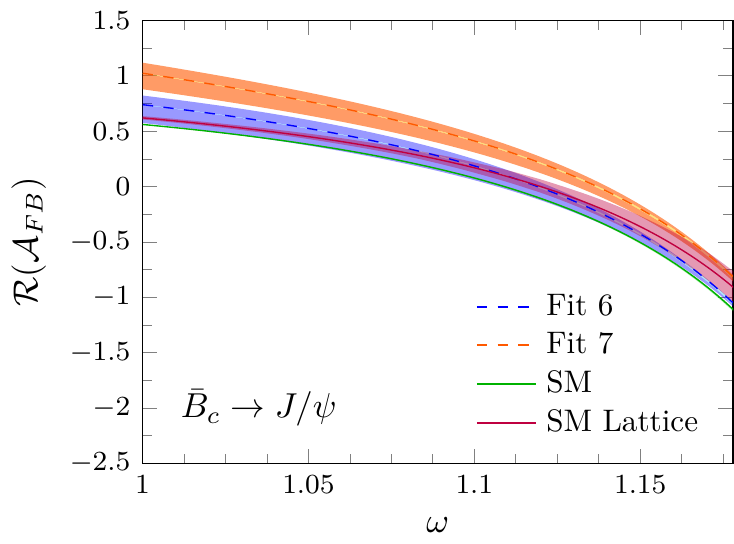} \hspace{.15cm}
\includegraphics[height=4cm]{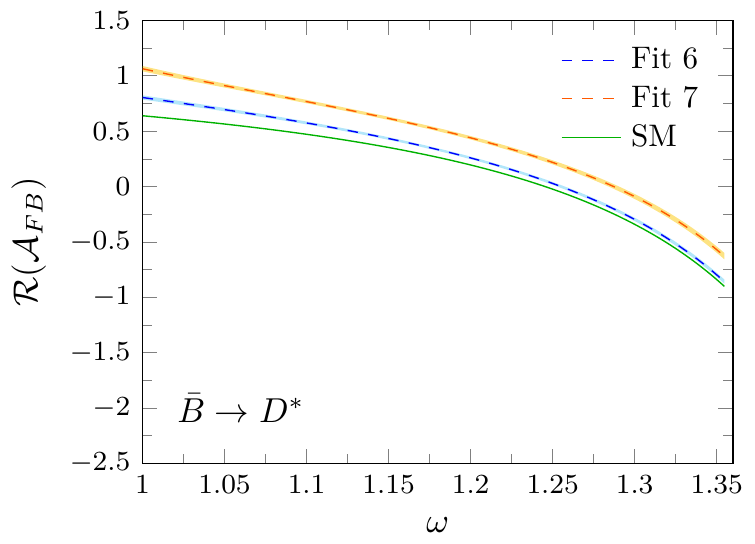}\hspace{.15cm}
\includegraphics[height=4cm]{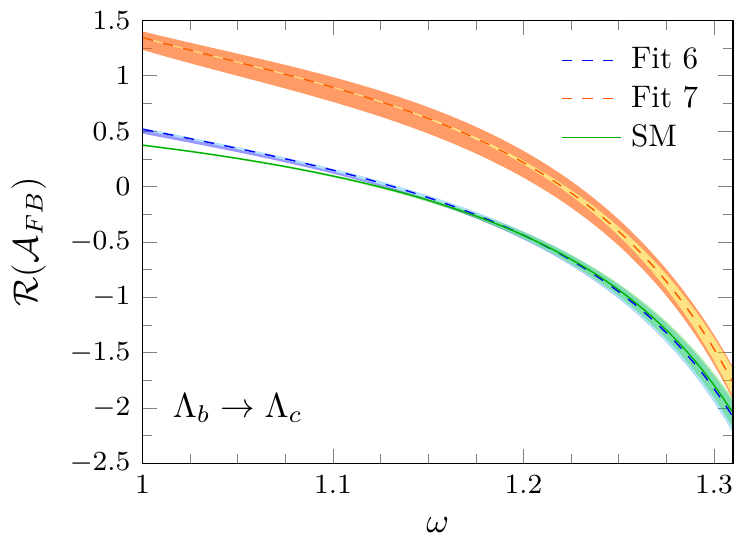}\hspace{.15cm}
\caption{${\cal R}({\cal A}_{FB})$ ratios defined in Eq.~\eqref{eq:ratiosFB} for the $\bar B_c \to J/\psi$, $\bar B \to D^*$ and $\Lambda_b \to \Lambda_c$ semileptonic decays, as a function of $\omega$.  Errors bands have been calculated as in Figs.~\ref{fig:dgdw} and \ref{fig:dgdwddstar} for the $\bar B_c$ and $\bar B$ meson decays. The $\Lambda_b$ plot has been taken from Fig.4 of Ref.~\cite{Penalva:2020xup}, while for the 
for the $\bar B_c \to J/\psi$ decay, we also show the SM results obtained with LQCD form factors from Ref.~\cite{Harrison:2020gvo}.}  
\label{fig:ratioFB}
\end{figure}

To minimize experimental and theoretical uncertainties, it was proposed in Ref.~\cite{Penalva:2020xup} to pay attention to the ratio ${\cal R}({\cal A_{FB}})$, defined as
\begin{equation}
{\cal R}({\cal A}_{FB}) = \frac{({\cal A}_{FB})_\tau}{({\cal A}_{FB})_{\ell=e,\mu}^{\rm SM}\tstrut}= 
\frac{ \Big[ \frac{a_1 \tstrut}{2 a_0 +2a_2/3} \Big]_\tau\tstrut}
 {\tstrut \Big[ \frac{a_1 \tstrut}{2 a_0 +2a_2/3} \Big]_{\ell=e,\mu}^{\rm SM} } \label{eq:ratiosFB}
\end{equation}
In Fig.~\ref{fig:ratioFB}, we show the theoretical predictions for 
${\cal R}({\cal A}_{FB})$ for the $\bar B_c \to J/\psi$, $\bar B \to D^*$ and $\Lambda_b \to \Lambda_c$
semileptonic decays, with the latter taken from Ref.~\cite{Penalva:2020xup} where details of the LQCD form factors used in the calculation 
can be found. In addition, for $\bar B_c \to J/\psi$, we also display the SM results obtained with LQCD form factors from Ref.~\cite{Harrison:2020gvo}, which agree remarkably well with the NRQM distribution. Note that for $\bar B_c \to \eta_c$ and $\bar B \to D$ decays, the denominator in Eq.~\eqref{eq:ratiosFB} vanishes in the massless lepton limit ($m_\ell\to 0)$, since $a_1(\omega)=a_1(\omega, h=-1)+ a_1(\omega, h=+1)$, and the negative helicity contribution is zero (Eq.~\eqref{eq:a1hm1}), while the positive helicity one is proportional to $m_\ell$.

The ratio ${\cal R}({\cal A}_{FB})$ can be measured by subtracting the number of events seen for $\theta_\ell \in [0,\pi/2]$ and  
for  $\theta_\ell \in [\pi/2, \pi]$ and dividing by the total sum of observed events, in each of 
the $H_b\to H_c \tau\bar \nu_\tau$ and $H_b\to H_c e(\mu)\bar \nu_{e(\mu)}$ reactions.  We expect that this observable should be free  
of a good part of experimental normalization errors.  On the theoretical side, we see in Fig.~\ref{fig:ratioFB} that  predictions for this ratio have indeed small uncertainties, and that this quantity has the potential to establish the validity of the NP scenarios associated 
to Fit 7, even more if all  three reactions shown in Fig.~\ref{fig:ratioFB} are simultaneously confronted with experiment.

For completeness, $\bar B \to D^{(*)}$ results with a polarized final $\tau$ are given
 in Appendix~\ref{app:bddspol}. Roughly, the same qualitative features that we have discussed for the polarized $\bar B_c\to\eta_c$ and $\bar B_c\to J/\psi$ semileptonic decays are also found in this case.
%
%
%
%
%
%
%\newpage
\section{Conclusions}
\label{sec:conclusions}
We have shown the relevant role  that the $a_{0,1,2}(\omega)$ CM and $c_{0,1,2}(\omega)$ LAB scalar functions, in terms of which the CM $d^2\Gamma/(d\omega d\cos\theta_\ell)$ and LAB $d^2\Gamma/(d\omega dE_\ell)$ differential decay widths are expanded, could play in order to separate between different NP scenarios that otherwise
give rise to the same ${\cal R}_{D^{(*)}}, {\cal R}_{\eta_c,J/\psi}$ ratios. The scheme we have used is the one originally developed in Ref.~\cite{Penalva:2020xup}, and applied there to the analysis of the $\Lambda_b\to \Lambda_c \tau \bar\nu_\tau$ decay, that we have extended in this work to the study of the  $\bar B\to D, \bar B\to D^*, \bar B_c\to \eta_c$ and  $\bar B_c\to J/\psi$  meson  reactions.

  For the $\bar B_c\to \eta_c,J/\psi$ transitions we have obtained results from a NRQM scheme, consistent with the expected breaking pattern of HQSS from $\bar B \to D^{(*)}$ decays~\cite{Neubert:1993mb},  estimating the  systematic uncertainties caused by the use of different inter-quark potentials.  Besides, and since SM LQCD vector and axial form factors  for $\bar B_c\to J/\psi$ have recently  been reported by the HPQCD collaboration~\cite{Harrison:2020gvo}, 
  we have made systematic comparisons with the SM observables computed with the LQCD input. In general, though there appear some overall normalization inconsistencies, we find quite good agreements for $\omega-$shapes, which become much better for observables constructed out of ratios of distributions,  like the $\tau-$forward-backward [${\cal A}_{FB}$] and $\tau-$polarization [${\cal A}_{\lambda_\tau}$] asymmetries, as well as the  ratios between predictions obtained in $\tau$ and ($e/\mu$) modes like
${\cal R}_{J/\psi}$ (Table~\ref{tab:ratios}) or  ${\cal R}({\cal A}_{FB})$ (Fig.~\ref{fig:ratioFB}).  This further supports  the reliability of our results for the LAB or the
  $\bar B \to \eta_c$ distributions, not yet studied.

As a general rule,  the $\bar B_c \to J/\psi$ observables, even involving $\tau$ polarization, 
are less optimal for distinguishing between NP scenarios than those obtained from $\bar B_c \to \eta_c$ decays, or those discussed in Ref.~\cite{Penalva:2020xup} for the related $\Lambda_b \to \Lambda_c$ semileptonic decay. We have also found qualitative similar behaviors for $\bar B\to D$ and  $\bar B_c\to \eta_c$, and $\bar B\to D^*$ and  $\bar B_c\to J/\psi$ decay observables.

We have also drawn the attention to the ratio ${\cal R}({\cal A_{FB}})$, defined in Eq.~\eqref{eq:ratiosFB} and shown in Fig.~\ref{fig:ratioFB} for $\bar B_c \to J/\psi$, $\bar B \to D^*$ and $\Lambda_b \to \Lambda_c$ decays, as a promising quantity, both from the experimental and theoretical points of view, to shed light into details of  different NP scenarios in $b\to c \tau \bar\nu_\tau$ transitions.  

One should notice however that the effective Hamiltonian of Eq.~\eqref{eq:hnp}, despite  excluding right-handed neutrino terms,  contains five, complex in general, NP Wilson  coefficients. While one of them can always be taken to be real,  that still leaves  nine free parameters to be determined from data. Even assuming that the form factors were known, and therefore the genuinely hadronic part ($W$) of the $\widetilde W$ SFs, it would be difficult to determine all NP parameters  just from the study of a  unique reaction.  As shown in Ref.~\cite{Penalva:2020xup}, for decays with an unpolarized final $\tau$ lepton, the 
CM $d^2\Gamma/(d\omega d\cos\theta_\ell)$ and LAB $d^2\Gamma/(d\omega dE_\ell)$ differential decay widths
 are completely determined by only three independent functions  which are linear combinations of the $\widetilde W$ SFs, the latter depending on  the NP Wilson coefficients. 
 This means that $a_{0,1,2}(\omega)$ and $c_{0,1,2}(\omega)$ contain the same information. For the case of polarized final $\tau$'s, the CM $d^2\Gamma/(d\omega d\cos\theta_\ell)$ and LAB   and $d^2\Gamma/(d\omega dE_\ell)$ distributions provide complementary information giving access  to another five independent linear combinations of  the $\widetilde W$'s~\cite{Penalva:2020xup}. But in this case it is the  experimental measurement of the required   polarized decay that could become a  very difficult task.  We think it is therefore more convenient to analyze data from various 
types of semileptonic decays simultaneously (e.g. $\bar B\to D, \bar B\to D^*$, $\Lambda_b\to \Lambda_c, \bar B_c\to \eta_c, \bar B_c\to J/\psi$...), 
 considering both the $e/\mu$ and $\tau$ modes. The scheme presented in  \cite{Penalva:2020xup} is a powerful tool to achieve this objective.

\section*{Acknowledgements}
We warmly thank C. Murgui, A. Pe\~nuelas  and A. Pich for useful
discussions.
This research has been supported  by the Spanish Ministerio de
Econom\'ia y Competitividad (MINECO) and the European Regional
Development Fund (ERDF) under contracts FIS2017-84038-C2-1-P, 
FPA2016-77177-C2-2-P and PID2019-105439G-C22, by Generalitat Valenciana under contract PROMETEO/2020/023 
and by the EU STRONG-2020 project under the program 
H2020-INFRAIA-2018-1, grant agreement no. 824093. 
\appendix

\section{Form Factors for $P_b( 0^-)\to P_c(0^-)$ and $P_b( 0^-)\to P_c^*(1^-)$ transitions}
\label{app:ff}
For these two transitions we use the standard definitions of the form factors  taken from
Ref.~\cite{Bernlochner:2017jka}\footnote{Note however that within the conventions 
of Ref.~\cite{Penalva:2020xup}, that we follow here, our hadronic matrix elements are 
dimensionless and they should be compared to those given 
in \cite{Bernlochner:2017jka}  divided by a  $\sqrt{2M}\sqrt{2M'}$ 
factor.},
\begin{itemize}
 \item $P_b\to  P_c$
 \begin{eqnarray}
 \langle P_c;\vec{p}^{\,\prime}| \bar c(0)b(0)|P_b;\vec{p}\,\rangle &=& 
 \frac{1+\omega}{2}h_S(\omega) \nonumber \\
\langle P_c;\vec{p}^{\,\prime}| \bar c(0)\gamma_5 b(0)|P_b;\vec{p}\,\rangle &=& \langle P_c;\vec{p}^{\,\prime}| \bar c(0)\gamma^\alpha \gamma_5 b(0)|P_b;\vec{p}\,\rangle 
 = 0  \nonumber \\
 \langle P_c;\vec{p}^{\,\prime}| \bar c(0)\gamma^\alpha b(0)|P_b;\vec{p}\,\rangle 
 &=& \frac12 \left(v^\alpha+v^{\prime \alpha}\right) \,h_+(\omega)+ \frac12 \left(v^\alpha-v^{\prime \alpha}\right)\,h_-(\omega)\nonumber \\
\langle P_c;\vec{p}^{\,\prime}| \bar c(0)\sigma^{\alpha\beta}  b(0)|P_b;\vec{p}\,\rangle 
 &=& \frac{i}{2} \left(v^{\prime \alpha} v^\beta-v^{\prime \beta} v^\alpha\right)h_T(\omega)\nonumber \\
 \langle P_c;\vec{p}^{\,\prime}| \bar c(0)\sigma^{\alpha\beta}\gamma_5  b(0)|P_b;\vec{p}\,\rangle 
 &=& \frac{1}{2} \epsilon^{\alpha\beta\delta \eta} v^{\prime}_\delta v_\eta h_T(\omega)
\label{eq.FactoresformaPseudo}
\end{eqnarray}
with $v^\alpha=p^\alpha/M$ and $v^{\prime \alpha}=p^{\prime\alpha}/M'=
(p^\alpha-q^\alpha)/M'$, the quadrivelocities of the initial and final hadrons, which have masses $M$ and $M'$, respectively, $\omega=\left(v\cdot v'\right)$ and $\epsilon_{0123}=+1$.

\item $P_b\to  P_c^*$
\begin{eqnarray}
 \langle P_c^*;\vec{p}^{\,\prime}, r\,| \bar c(0)b(0)|P_b;\vec{p}\,\rangle &=& 0 \nonumber \\
 \langle P_c^*;\vec{p}^{\,\prime}, r\,| \bar c(0)\gamma_5 b(0)|P_b;\vec{p}\,\rangle &=& -\frac12 \left(\epsilon_r^*\cdot v\right)h_P(\omega)\nonumber \\
 \langle P_c^*;\vec{p}^{\,\prime}, r\,| \bar c(0)\gamma^\alpha b(0)|P_b;\vec{p}\,\rangle &=&  \frac{i}{2}\epsilon^{\alpha}_{\, \, \delta \eta\gamma}
\epsilon^{*\delta}_r v^{\prime\eta} v^\gamma  h_V(\omega) \nonumber \\
 \langle P_c^*;\vec{p}^{\,\prime}, r\,| \bar c(0)\gamma^\alpha \gamma_5 b(0)|P_b;\vec{p}\,\rangle &=& \frac{\omega+1}{2}\epsilon^{*\alpha}_r h_{A_1}(\omega)
 -\frac{\left(\epsilon^{*}_r\cdot v\right)}{2}\Big[v^\alpha h_{A_2}(\omega)+
v^{\prime \alpha} h_{A_3}(\omega)\Big] \nonumber \\
\langle P_c^*;\vec{p}^{\,\prime}, r\,| \bar c(0) \sigma^{\alpha\beta} b(0)|P_b;\vec{p}\,\rangle &=& -\frac12\epsilon^{\alpha\beta}_{\, \,\,\,\,\, \, \delta \eta}\Big\{ \epsilon^{*\delta}_r\left[\left(v^\eta+v^{\prime \eta}\right) \,h_{T_1}(\omega)+ \left(v^\eta-v^{\prime \eta}\right)\,h_{T_2}(\omega)\right]+ v^\delta v^{\prime\eta}\left(\epsilon_r^*\cdot v\right)\,h_{T_3}(\omega)\Big\}  \nonumber \\
\langle P_c^*;\vec{p}^{\,\prime}, r\,| \bar c(0) \sigma^{\alpha\beta} \gamma_5 b(0)|P_b;\vec{p}\,\rangle &=& -\frac{i}{2}\Bigg\{\epsilon^{*\alpha}_r\left[\left(v^\beta+v^{\prime \beta}\right) \,h_{T_1}(\omega)+ \left(v^\beta-v^{\prime \beta}\right)\,h_{T_2}(\omega)\right]\nonumber \\
&&-\epsilon^{*\beta}_r\left[\left(v^\alpha+v^{\prime \alpha}\right) \,h_{T_1}(\omega)+ \left(v^\alpha-v^{\prime \alpha}\right)\,h_{T_2}(\omega)\right]\nonumber \\
&& +\left(v^\alpha v^{\prime \beta} -v^\beta v^{\prime \alpha}\right)
\left(\epsilon_r^*\cdot v\right)\,h_{T_3}(\omega)\Bigg\}
\label{eq.FactoresformaVec}
\end{eqnarray}
where $r$ is the helicity of the final vector meson, with $\epsilon_r$ its corresponding polarization vector. In short, 
\be
\langle P_c^*; \vec{p}^{\,\prime}, r\, | \bar c(0) \Gamma^{(\alpha\beta)} b(0) | P_b; \vec{p}\,\rangle = T^{(\alpha\beta)}_\mu \epsilon_r^{\mu *}
\ee
with  $\Gamma^{(\alpha\beta)}= 1,\gamma_5, \gamma^\alpha,\gamma^\alpha\gamma_5,\sigma^{\alpha\beta}$ and $\sigma^{\alpha\beta}\gamma_5$ and $T^{(\alpha\beta)}_\mu$ read from Eq.~\eqref{eq.FactoresformaVec}. 

\end{itemize}

The form factors are real functions of $\omega$ greatly
constrained by HQSS near zero recoil ($\omega=1$)~\cite{Neubert:1993mb,Bernlochner:2017jka}. Indeed, 
all factors in Eqs.~\eqref{eq.FactoresformaPseudo} and 
\eqref{eq.FactoresformaVec} have been chosen such that in the
heavy quark limit each form factor either vanishes or equals
the leading-order Isgur-Wise function\footnote{These relations trivially 
follow from
\begin{equation}
\langle P_c^{(*)};\vec{p}^{\,\prime}, (r)\,| \bar c(0)\Gamma
b(0)|P_b;\vec{p}\,\rangle = -\frac12\xi(\omega) {\rm Tr} \left[\bar
H_{v'}^{(c)} \Gamma H_{v}^{(b)}\right],
\end{equation}
where the pseudoscalar and vector mesons are represented by
a super-field, which has the right transformation properties
under heavy quark and Lorentz
symmetry~\cite{Neubert:1993mb,Bernlochner:2017jka}
\begin{equation}
  H_{v}^{(Q)} = \frac{1+\slashed{v}}{2} \left(V_v^{(Q)}
\slashed{\epsilon}- P_v^{(Q)}\gamma_5\right)
\end{equation}
and $\bar H_{v}^{(Q)} = \gamma^0 H_{v}^{(Q)\dagger}\gamma^0$. For $\bar B_c
\to \eta_c, J/\psi$ transitions, the appropriate $4\times 4$ field 
accounts also for the heavy anticharm quark both in the initial and
final mesons~\cite{Jenkins:1992nb}
\begin{equation}
  H_{v}^{(Q\bar c)} = \frac{1+\slashed{v}}{2} \left(V_v^{(Q\bar c)}
\slashed{\epsilon}- P_v^{(Q\bar c)}\gamma_5\right)\frac{1-\slashed{v}}{2}
\end{equation}
}
\begin{eqnarray}
h_-= h_{A_2}= h_{T_2} = h_{T_3}= 0, \quad h_+ =  h_V = h_{A_1} = 
h_{A_3}= h_S = h_P = h_T = h_{T_1} =\xi
\label{eq:hh}
\end{eqnarray}

The hadron tensors and $\widetilde W$ SFs introduced in 
Ref.~\cite{Penalva:2020xup} are straightforwardly  obtained from Eq.~\eqref{eq.FactoresformaPseudo} in the case of $P_b\to P_c$ transitions, while for decays into vector mesons, we use  
\be
 \sum_r \langle P_c^*; \vec{p}^{\,\prime}, r\, | \bar c(0) \Gamma^{(\alpha\beta)} b(0) | P_b; \vec{p}\,\rangle 
 \langle P_c^*; \vec{p}^{\,\prime}, r\, | \bar c(0) \Gamma^{(\rho\lambda)} b(0) | P_b; \vec{p}\,\rangle ^* 
= T^{(\alpha \beta)}_\mu T^{(\rho \lambda)}_\nu \left( -g^{\mu\nu}+ v^{\prime \mu}v^{\prime \nu}\right)  \label{eq:wmunu-baryons}
\ee
The explicit expressions for the $\widetilde W$ SFs in 
terms of the above 
form factors and the Wilson coefficients are given in the following  appendix.
%in  Appendices~\ref{app:ff1} and  
%\ref{app:ff2} for decays into pseudoscalar and vector mesons, respectively. 
%We finally obtain the $d\Gamma/(d\omega dE_\ell)$ and 
%$d\Gamma/(d\omega d\cos\theta_\ell$) distributions from 
%the general results derived in Ref.~\cite{Penalva:2020xup} 
%
%

  \section{Hadron tensor $\widetilde W$ SFs for  the $P_b\to  P_c\ell^-\bar\nu_\ell$ and $P_b\to  P_c^*\ell^-\bar\nu_\ell$ decays}

We compile here the $\widetilde W$ SFs introduced  in Ref.~\cite{Penalva:2020xup} for the particular meson decays studied in this work. As shown in that reference,  these $\widetilde W$ SFs determine the LAB $d^2\Gamma/(d\omega dE_\ell)$ and CM $d^2\Gamma/(d\omega d\cos\theta_\ell$) differential decay widths, for the full set of NP operators in Eq.~\eqref{eq:hnp}, for generally complex Wilson coefficients, and for the case where the final charged lepton has a well defined helicity in either reference frame. In the equations below, we use $C_{V,A}=1+C_{V_L}\pm C_{V_R}$ and $C_{S,P}=C_{S_L}\pm C_{S_R}$.

\subsection{$P_b\to  P_c\ell^-\bar\nu_\ell$}
\label{app:ff1}

In this case, the SFs related to the SM currents are
\begin{eqnarray}
\widetilde W_1&=& \widetilde W_3=0,\quad \widetilde W_2=  \frac{|C_V|^2}{r}F_+^2, \quad \widetilde W_4=  \frac{|C_V|^2}{4r}(F_+-F_-)^2, \quad
\widetilde W_5= \frac{|C_V|^2}{r}\, F_+\left(F_--F_+\right)\nonumber\\
\end{eqnarray}
where
\begin{eqnarray}
F_+ &=&\frac1{R}\left(h_+-\frac{1-r}{1+r}h_-\right),\quad
F_- = \frac1{R}\left(h_- -\frac{1-r}{1+r}h_+\right)= \frac{1-r^2}{1+r^2-2r\omega}\left(F_0-F_+\right) \nonumber \\
F_0 &=& \frac{2r(1+\omega)}{R(1+r)^2}\Big[h_+-\frac{1+r}{1-r}\frac{\omega-1}{\omega+1}\,h_-\Big]
\label{eq:fpm0}
\end{eqnarray}
with $r= M'/M$ and $R= 2\sqrt{r}/(1+r)$, and we have also introduced the $F_0$ form-factor 
in the definition of $F_-$, as commonly done in this type of calculations. In  addition, 
\begin{eqnarray}
\widetilde W_{SP}&=& |C_S|^2 \left(\frac{1+\omega}{2}\right)^2  h_S^2, \quad \widetilde W_{I1}=  C_V C_S^*\, \frac{1+\omega}{\sqrt{r}}h_S\,F_+,\quad \widetilde W_{I2}=  C_V C_S^*\, \frac{1+\omega}{2\sqrt{r}}h_S\,\left(F_--F_+\right)\nonumber \\
\widetilde W_{I3}&=&-C_T^* C_S \, \frac{1+\omega}{2r}h_S\,h_T, \quad
\widetilde W_{I4}=-C_T^* C_V \frac{h_T F_+}{r^{3/2}}, \quad
\widetilde W_{I5}=C_T^* C_V  \frac{h_T \left(F_+-F-\right)}{2r^{3/2}}, \quad \widetilde W_{I6}=\widetilde W_{I7}=0 \nonumber \\
\widetilde W_{1}^T&=&\frac{|C_T|^2}{4}\left(\omega^2-1\right)h_T^2,\quad \widetilde W_{2}^T = \frac{|C_T|^2}{4r^2}\left(1+r^2-2r\omega\right)h_T^2,\quad \widetilde W_{3}^T = \frac{|C_T|^2}{4r^2}h_T^2,\nonumber \\
\widetilde W_4^T &=& -\frac{|C_T|^2}{4r^2}\left(1-r\omega\right)h_T^2, \quad \widetilde W_5^T=0. 
\label{eq:ws1}
\end{eqnarray}
As derived in Ref.~\cite{Penalva:2020xup}, the tensor $\widetilde W$ SFs accomplish:
\begin{equation}
 2 \widetilde W_1^T +\widetilde W_2^T + (1 - 2 r \omega + 
    r^2)\widetilde W_3^T + 2 (1 - r \omega) \widetilde W_4^T = 0 \label{eq:test}
\end{equation}

\subsection{$P_b\to  P_c^*\ell^-\bar\nu_\ell$}
\label{app:ff2}
In this case, the $\widetilde W$ SFs related to the SM currents are
\begin{eqnarray}
\widetilde W_1&=& \frac{|C_V|^2}{4} \left(\omega^2-1\right)h_V^2+\frac{|C_A|^2}{4} \left(\omega+1\right)^2h_{A_1}^2 \nonumber \\
\widetilde W_2&=&-\frac{|C_V|^2}{4r^2} \left(1+r^2-2r\omega\right)h_V^2+\frac{|C_A|^2}{4r^2}\left(\omega+1 \right)^2\left(h_{A_1}^2
-2\, \frac{\omega-r}{\omega+1}h_{A_1}\left(r\, h_{A_2}+h_{A_3}\right)+ \frac{\omega-1}{\omega+1}\left(r\, h_{A_2}+h_{A_3}\right)^2\right) \nonumber \\
\widetilde W_3&=&\frac{{\rm Re}[C_V C_A^*]}{r}(\omega+1)h_Vh_{A_1} \nonumber \\
\widetilde W_4&=&-\frac{|C_V|^2}{4r^2} h_V^2+\frac{|C_A|^2}{4r^2}\left(\omega+1 \right)^2\left(h_{A_1}-h_{A_3}\right)
\left(h_{A_1}+ \frac{1-\omega}{1+\omega}\,h_{A_3}\right) \nonumber \\
\widetilde W_5&=&\frac{|C_V|^2}{2r^2}\left(1-r\omega \right) h_V^2-\frac{|C_A|^2}{2r^2}\Big\{\left(1+\omega\right)\left(h_{A_1}-r\, h_{A_2}-h_{A_3}\right)\left[\left(1+\omega\right)h_{A_1}- (\omega-r)h_{A_3}\right] \nonumber \\
&&+ \left(1+\omega\right)\left(r\, h_{A_2}+h_{A_3}\right)\left(h_{A_1}-(1-r)h_{A_3}\right)\Big\}.
\end{eqnarray}
The rest of NP $\widetilde W$ SFs  are
\begin{eqnarray}
\widetilde W_{SP}&=& \frac{|C_P|^2}{4}\left(\omega^2-1\right)h_P^2 \nonumber \\
\widetilde W_{I1}&=& \frac{C_A C_P^*}{2  r}\left(\omega^2-1\right)\left[r\, h_{A_2}+h_{A_3}-\frac{\omega-r}{\omega-1}\,h_{A_1}\right]h_P\nonumber \\
\widetilde W_{I2}&=& \frac{C_A C_P^*}{2  r}\left(\omega+1\right)\left[\omega h_{A_1}+\left(1-\omega\right)h_{A_3}\right]h_P\nonumber \\
\widetilde W_{I3}&=& -\frac{C_P C_T^*}{2  r}  (\omega^2 - 1) h_P T_1\nonumber \\
\widetilde W_{I4}&=& \frac{C_V C_T^*}{2 r^2}\left [ (1- r \omega)T_2+(\omega-r) T_3\right]h_V 
- \frac{C_A C_T^*}{2r^2}\Big \{(\omega + 1)  h_{A_1} \left[ (r-\omega)T_1 +r T_2 +T_3 \right] 
+(\omega^2-1)(r h_{A_2}  +  h_{A_3})T_1\Big\} \nonumber \\
\widetilde W_{I5}&=& -\frac{C_V C_T^*}{2 r^2}\left (T_2+\omega T_3 \right)h_V  
+ \frac{C_A C_T^*}{2r^2}\Big \{ (\omega^2-1)h_{A_3} T_1 -
   (\omega + 1) h_{A_1}\left[\omega T_1-T_3 \right]\Big\} \nonumber \\
\widetilde W_{I6}&=& \frac{C_V C_T^*}{2  r}(\omega^2-1)h_V\left(r T_2+T_3\right)
-\frac{C_A C_T^*}{2  r}(\omega+1)h_{A_1}\left[(1-r \omega)T_2 +(\omega-r)T_3 \right]\nonumber \\
\widetilde W_{I7}&=&-\frac{C_V C_T^*}{2  r} (\omega^2-1) h_V T_3 +\frac{C_A C_T^*}{2  r}(\omega+1)h_{A_1}\left [T_2+\omega T_3 \right]  \nonumber \\
\widetilde W_1^T&=&\frac{|C_T|^2}{4}(\omega^2 - 1)^2 T_1^2\nonumber \\
\widetilde W_2^T&=&\frac{|C_T|^2}{4 r^2}\left[(r^2-2 r \omega+1) (\omega^2-1)T_1^2 + (1-r^2)(T_3^2-T_2^2)+2r(1-r\omega)(\omega T_2+T_3)T_2+2(r-\omega)(T_2+\omega T_3)T_3\right]\nonumber \\
\widetilde W_3^T&=&\frac{|C_T|^2}{4 r^2}\left[(\omega^2-1)(T_1^2-T_3^2)-(T_2+\omega T_3)^2\right]\nonumber \\
 W_4^T&=&\frac{|C_T|^2}{4 r^2}\left[ (1 - r \omega) (T_2^2 + T_3^2)+ 
 2  (\omega - r) T_2 T_3 + (\omega^2-1)\left(2T_3^2-(1 - r \omega)T_1^2  \right)\right] \nonumber \\
 \widetilde W_5^T&=& 0
\end{eqnarray}
with the tensor $\widetilde W_{1,2,3,4}^T$ SFs satisfying Eq.~\eqref{eq:test}, and
\begin{equation}
 T_1 = \frac{ (\omega+1)h_{T_1}+ (\omega-1)h_{T_2}}{\omega^2-1}-h_{T_3},\quad T_2 =- \frac{ (\omega+1)h_{T_1}+ (\omega-1)h_{T_2}}{\omega^2-1},\quad T_3 = \frac{ (\omega+1)h_{T_1}- (\omega-1)h_{T_2}}{\omega^2-1}
 \label{eq:htt}
\end{equation}
Although  $T_1$, $T_2$ and $T_3$  behave as $\pm 1/(\omega-1)$ in the heavy 
quark limit, the corresponding $\widetilde W_{1,2,3,4}^T$ SFs 
are finite at zero recoil, as they  should be, with their values being 
 given by
$|C_T|^2\Big [1$, $-\frac{r^2+6r +1}{4r^2}$, 
$-\frac1{4r^2}$, $\frac{3 r  +1}{4r^2}\Big]$, respectively.

%their values close to zero recoil
% are given by
%$|C_T|^2\Big [(\omega+1)^2/4$, $-(r^2+2 r \omega+4r +1)/4r^2$, 
%$-1/4r^2$, $(2 r +r\omega +1)/4r^2\Big]$, respectively.

%
%
%
%
%

\section{Evaluation of the  $\bar B_c\to \eta_c$ and $\bar B_c\to J/\psi$ semileptonic decay form factors within the NRQM of Ref.~\cite{Hernandez:2006gt}}
\label{app:NRQM}
Within the NRQM calculation of Ref.~\cite{Hernandez:2006gt}, and with the global phases used in the present work, we obtain the following expressions
for the different form factors.
\subsection{$\bar B_c\to \eta_c$}
For the pseudoscalar-pseudoscalar $\bar B_c\to \eta_c$  transition we have
\bea
\label{eq:fpm0-}
F_+=\frac{1}{2M}\,\Big(
V^0+{V^3}\frac{E'-M}{|\vec{q}\,|}\Big)\ &,& \
F_-=\frac{1}{2M}\,\Big(
V^0+{V^3}\frac{E'+M}{|\vec{q}\,|}\Big)
,\nonumber\\
h_S=\frac{S}{(\omega+1)\sqrt{MM'}}\ &,& \
h_T=-\sqrt{\frac{M'}{M}}\,i\frac{T^{03}}{|\vec q\,|},
\eea
with $F_\pm$ defined in Eq.~(\ref{eq:fpm0}), and where $V^\mu$, $S$ and
$T^{\mu\nu}$ stand for the NRQM matrix elements  of the vector, scalar  and tensor   $b\to c$ transition currents, respectively. 
In addition, $E'=\sqrt{M^{\prime 2}+\vec{q}\,^2}$ is the energy of the final meson that has three-momentum $-\vec q$ in the LAB frame, with $\vec q$  the three-momentum transferred in the LAB frame and that for the purpose of calculation we take it along the positive $Z$ axis. For the matrix elements one has the results
\bea
V^0&=&\sqrt{2M2E'}\ \int\,d^3p\ \frac{1}{4\pi}
\big[\hat{\phi}^{(\eta_c)}(|\vec{p}\,|)\big]^*\,
\hat{\phi}^{(\bar B_c)}\big(\big|\,\vec{p}-\frac12
\vec{q}
\, \big|\big)\ \sqrt{\frac{\widehat{E}_{c}\widehat{E}_{b}}{4E_{c}
E_{b}}}
%\nonumber\\
\left(
1+\frac{(-\frac12\,\vec{q}-\vec{p}\,)
\cdot(\frac12\,\vec{q}-\vec{p}\,)}{\widehat{E}_{c}\widehat{E}_{b}}
\right), \nonumber\\
V^3&=&\sqrt{2M2E'}\ \int\,d^3p\ \frac{1}{4\pi}
\big[\hat{\phi}^{(\eta_c)}(|\vec{p}\,|)\big]^*\,
 \hat{\phi}^{(\bar B_c)}\big(\big|\,\vec{p}-\frac12
\vec{q}\,
\big|\big)
 \ \sqrt{\frac{\widehat{E}_{c}\widehat{E}_{b}}{4E_{c}
E_{b}}}%\nonumber\\
\left(\frac{\frac12\,|\vec{q}\,|-p_z}{\widehat{E}_{b}}+
\frac{-\frac12\,|\vec{q}\,|-p_z}{\widehat{E}_{c}}
\right), \nonumber\\
S&=&\sqrt{2M2E'}\ \int\,d^3p\ \frac{1}{4\pi}
\big[\hat{\phi}^{(\eta_c)}(|\vec{p}\,|)\big]^*\,
\hat{\phi}^{(\bar B_c)}\big(\big|\,\vec{p}-\frac12
\vec{q}\, \big|\big)\ \sqrt{\frac{\widehat{E}_{c}\widehat{E}_{b}}{4E_{c}
E_{b}}}
%\nonumber\\
\left(
1-\frac{(-\frac12\,\vec{q}-\vec{p}\,)
\cdot(\frac12\,\vec{q}-\vec{p}\,)}{\widehat{E}_{c}\widehat{E}_{b}}
\right), \nonumber\\
T^{03}&=&i\sqrt{2M2E'}\ \int\,d^3p\ \frac{1}{4\pi}
\big[\hat{\phi}^{(\eta_c)}(|\vec{p}\,|)\big]^*\,
 \hat{\phi}^{(\bar B_c)}\big(\big|\,\vec{p}-\frac12
\vec{q}\,
 \big|\big)
\ \sqrt{\frac{\widehat{E}_{c}\widehat{E}_{b}}{4E_{c}
E_{b}}}%\nonumber\\
\left(\frac{\frac12\,|\vec{q}\,|-p_z}
{\widehat{E}_{b}}-
\frac{-\frac12\,|\vec{q}\,|-p_z}{\widehat{E}_{c}}
\right). 
\eea
Here, $\hat\phi$ stands for  the orbital part of the meson wave functions in momentum space and $\hat E_f=E_f+m_f$, with
$m_f, E_f$ the mass and relativistic energy of the quark with flavor $f$. The corresponding  three-momenta are $\frac12\,\vec q-\vec p$ for the  quark $b$ and $-\frac12\,\vec q-\vec p$ for the quark $c$.

\subsection{$\bar B_c\to J/\psi$}
For the pseudoscalar-vector $\bar B_c\to J/\psi$ transition we now have
\bea
\label{eq:fpm0-}
h_V&=&\frac{M'\sqrt2}{|\vec{q}\,|} \frac{V^1_{\lambda=-1}}{\sqrt{MM'}},\ \ h_P=\frac{M'}{|\vec{q}\,|}\frac{P_{\lambda=0}}{\sqrt{MM'}},\nonumber\\
h_{A_1}&=&\frac{\sqrt2}{\omega+1}\,\frac{A^1_{\lambda=-1}}{\sqrt{MM'}},\ \
h_{A_2}=-\frac{M'}{|\vec{q}\,|}\Big[
-\frac{A^0_{\lambda=0}}{\sqrt{MM'}}-\frac{E'}{|\vec{q}\,|}\,\frac{A^3_{\lambda=0}}{\sqrt{MM'}}
+\sqrt2\,\frac{M'}{|\vec{q}\,|}\,\frac{A^1_{\lambda=-1}}{\sqrt{MM'}}\Big],\ \nonumber \\
h_{A_3}&=&-\frac{M^{\prime 2}}{|\vec{q}\,|^2}\Big(
\frac{A^3_{\lambda=0}}{\sqrt{MM'}}-\sqrt2
\frac{E'}{M'}\frac{A^1_{\lambda=-1}}{\sqrt{MM'}}\Big),\nonumber\\
T_1&=&\frac{M^{\prime 2}}{|\vec{q}\,|^2}\frac{T^{12}_{\lambda=0}}{\sqrt{MM'}},\ \
T_2=\sqrt2\,\frac{M'}{|\vec{q}\,|}\Big(
i\,\frac{T^{01}_{\lambda=-1}}{\sqrt{MM'}}-
\frac{E'}{|\vec{q}\,|}\frac{T^{23}_{\lambda=-1}}{\sqrt{MM'}}\Big),\ \
T_3=\sqrt2\,\frac{M^{\prime 2}}{|\vec{q}\,|^2}
\,\frac{T^{23}_{\lambda=-1}}{\sqrt{MM'}},
\eea
with $T_{1,2,3}$ defined in Eq.~(\ref{eq:htt}) and 
$V^\mu_\lambda$, $A^\mu_\lambda$, $P_\lambda$ and $T^{\mu\nu}_\lambda$  the NRQM  matrix elements of the vector, axial, pseudoscalar and tensor $b\to c$ transition currents, respectively.  Here, $\lambda$ is the polarization of the final $J/\psi$ meson. We use states that have well defined spin  in the $Z$ direction in the $J/\psi$ rest frame. Since the $J/\psi$  three-momentum  equals $-\vec q$ (which is directed along the  negative $Z$ axis), $\lambda$ coincides with minus the helicity, the latter being the same in  the CM and LAB frames.
We obtain the following expressions for the matrix elements
\bea
V^{1}_{\lambda=-1}&=&\frac{-1}{\sqrt2}
\sqrt{2M2E'}\ \int\,d^3p\ \frac{1}{4\pi}
\big[\hat{\phi}^{(J/\psi)}(|\vec{p}\,|)\big]^*\,
\hat{\phi}^{(\bar B_c)}\big(\big|\,\vec{p}-\frac12
\vec{q}\, \big|\big)\ \sqrt{\frac{\widehat{E}_{c}\widehat{E}_{b}}{4E_{c}
E_{b}}}
%\nonumber\\
\left(-\frac{\frac12\,|\vec{q}\,|-p_z}{\widehat{E}_{b}}+
\frac{-\frac12\,|\vec{q}\,|-p_z}{\widehat{E}_{c}}
\right),\nonumber\\
A^{0}_{\lambda=0}&=&
\sqrt{2M2E'}\ \int\,d^3p\ \frac{1}{4\pi}
\big[\hat{\phi}^{(J/\psi)}(|\vec{p}\,|)\big]^*\,
\hat{\phi}^{(\bar B_c)}\big(\big|\,\vec{p}-\frac12
\vec{q}\, \big|\big)\ \sqrt{\frac{\widehat{E}_{c}\widehat{E}_{b}}{4E_{c}
E_{b}}}
%\nonumber\\
\left(\frac{\frac12\,|\vec{q}\,|-p_z}{\widehat{E}_{b}}+
\frac{-\frac12\,|\vec{q}\,|-p_z}{\widehat{E}_{c}}
\right)\nonumber\\
A^{1}_{\lambda=-1}&=&\frac{1}{\sqrt2}
\sqrt{2M2E'}\ \int\,d^3p\ \frac{1}{4\pi}
\big[\hat{\phi}^{(J/\psi)}(|\vec{p}\,|)\big]^*\,
\hat{\phi}^{(\bar B_c)}\big(\big|\,\vec{p}-\frac12
\vec{q}\, \big|\big)\ \sqrt{\frac{\widehat{E}_{c}\widehat{E}_{b}}{4E_{c}
E_{b}}}
\left(1+\frac{2p_x^2-(-\frac12\,\vec{q}-\vec{p}\,)
\cdot(\frac12\,\vec{q}-\vec{p}\,)}
{\widehat{E}_{c}\widehat{E}_{b}}
\right)
\nonumber\\
A^{3}_{\lambda=0}&=&
\sqrt{2M2E'}\ \int\,d^3p\ \frac{1}{4\pi}
\big[\hat{\phi}^{(J/\psi)}(|\vec{p}\,|)\big]^*\,
\hat{\phi}^{(\bar B_c)}\big(\big|\,\vec{p}-\frac12
\vec{q}\, \big|\big)\ \sqrt{\frac{\widehat{E}_{c}\widehat{E}_{b}}{4E_{c}
E_{b}}}
\ \left(
1+\frac{2(-\frac12\,|\vec{q}\,|-p_z\,)
\,(\frac12\,|\vec{q}\,|-p_z\,)}
{\widehat{E}_{c}\widehat{E}_{b}}\right.\nonumber\\
&&\left.-\frac{(-\frac12\,\vec{q}-\vec{p}\,)
\cdot(\frac12\,\vec{q}-\vec{p}\,)}
{\widehat{E}_{c}\widehat{E}_{b}}
\right),\nonumber
\eea
\bea
P_{\lambda=0}&=&
\sqrt{2M2E'}\ \int\,d^3p\ \frac{1}{4\pi}
\big[\hat{\phi}^{(J/\psi)}(|\vec{p}\,|)\big]^*\,
\hat{\phi}^{(\bar B_c)}\big(\big|\,\vec{p}-\frac12
\vec{q}\, \big|\big)\ \sqrt{\frac{\widehat{E}_{c}\widehat{E}_{b}}{4E_{c}
E_{b}}}
%\nonumber\\
\left(\frac{\frac12\,|\vec{q}\,|-p_z}
{\widehat{E}_{b}}-
\frac{-\frac12\,|\vec{q}\,|-p_z}{\widehat{E}_{c}}
\right)\nonumber\\
T^{12}_{\lambda=0}&=&
\sqrt{2M2E'}\ \int\,d^3p\ \frac{1}{4\pi}
\big[\hat{\phi}^{(J/\psi)}(|\vec{p}\,|)\big]^*\,
\hat{\phi}^{(\bar B_c)}\big(\big|\,\vec{p}-\frac12
\vec{q}\, \big|\big)\ \sqrt{\frac{\widehat{E}_{c}\widehat{E}_{b}}{4E_{c}
E_{b}}}
\ \left(
1-\frac{2(-\frac12\,|\vec{q}\,|-p_z\,)
\,(\frac12\,|\vec{q}\,|-p_z\,)}
{\widehat{E}_{c}\widehat{E}_{b}}\right.\nonumber\\
&&\left.+\frac{(-\frac12\,\vec{q}-\vec{p}\,)
\cdot(\frac12\,\vec{q}-\vec{p}\,)}
{\widehat{E}_{c}\widehat{E}_{b}}
\right),\nonumber\\
T^{23}_{\lambda=-1}&=&\frac{1}{\sqrt2}
\sqrt{2M2E'}\ \int\,d^3p\ \frac{1}{4\pi}
\big[\hat{\phi}^{(J/\psi)}(|\vec{p}\,|)\big]^*\,
\hat{\phi}^{(\bar B_c)}\big(\big|\,\vec{p}-\frac12
\vec{q}\, \big|\big)\ \sqrt{\frac{\widehat{E}_{c}\widehat{E}_{b}}{4E_{c}
E_{b}}}
\left(1-\frac{2p_x^2-(-\frac12\,\vec{q}-\vec{p}\,)
\cdot(\frac12\,\vec{q}\, -\vec{p}\,)}
{\widehat{E}_{c}\widehat{E}_{b}}
\right),\nonumber\\
T^{01}_{\lambda=-1}&=&\frac{-i}{\sqrt2}
\sqrt{2M2E'}\ \int\,d^3p\ \frac{1}{4\pi}
\big[\hat{\phi}^{(J/\psi)}(|\vec{p}\,|)\big]^*\,
\hat{\phi}^{(\bar B_c)}\big(\big|\,\vec{p}-\frac12
\vec{q}\, \big|\big)\ \sqrt{\frac{\widehat{E}_{c}\widehat{E}_{b}}{4E_{c}
E_{b}}}
%\nonumber\\
\left(-\frac{\frac12\,|\vec{q}\,|-p_z}
{\widehat{E}_{b}}-
\frac{-\frac12\,|\vec{q}\,|-p_z}{\widehat{E}_{c}}
\right).
\nonumber\\
\eea

\section{Results for the $\bar B\to D\tau\bar\nu_\tau$ and
$\bar B\to D^*\tau\bar\nu_\tau$ decays for the case of a polarized final $\tau$}
\label{app:bddspol}

\begin{figure}[tbh]
\includegraphics[height=5.5cm]{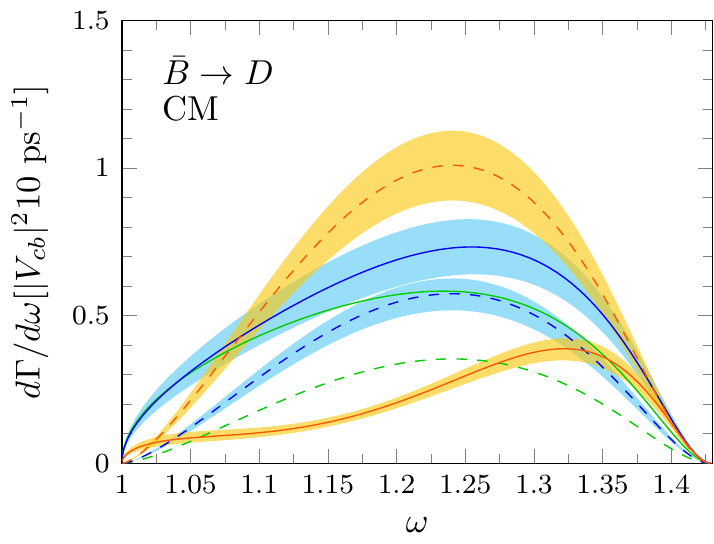} \hspace{.15cm}
\includegraphics[height=5.5cm]{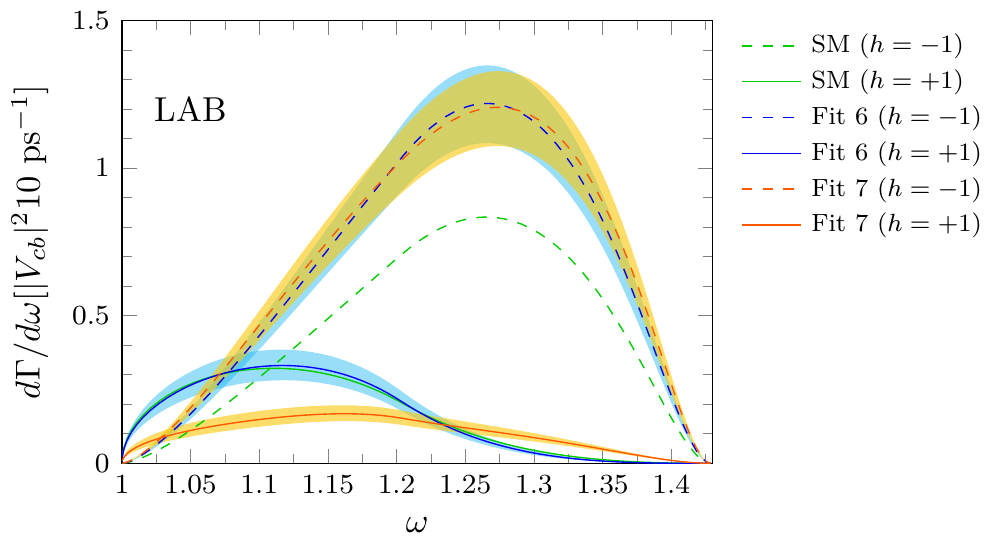} \\
\includegraphics[height=5.5cm]{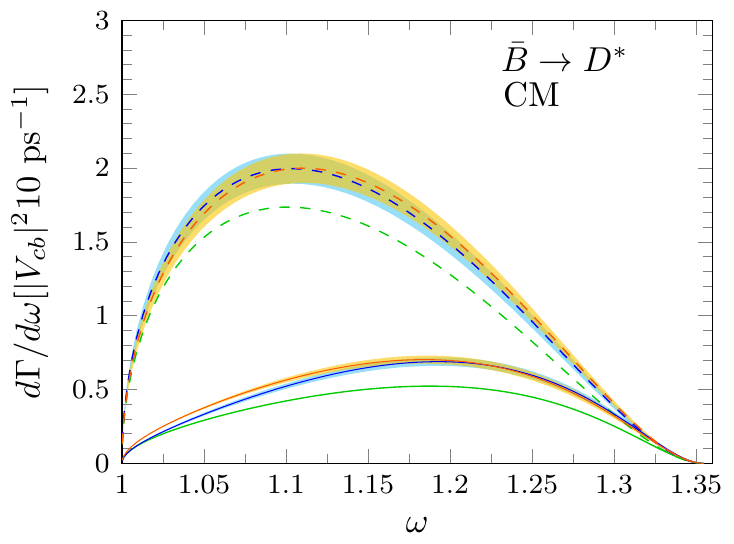} \hspace{.15cm}
\includegraphics[height=5.5cm]{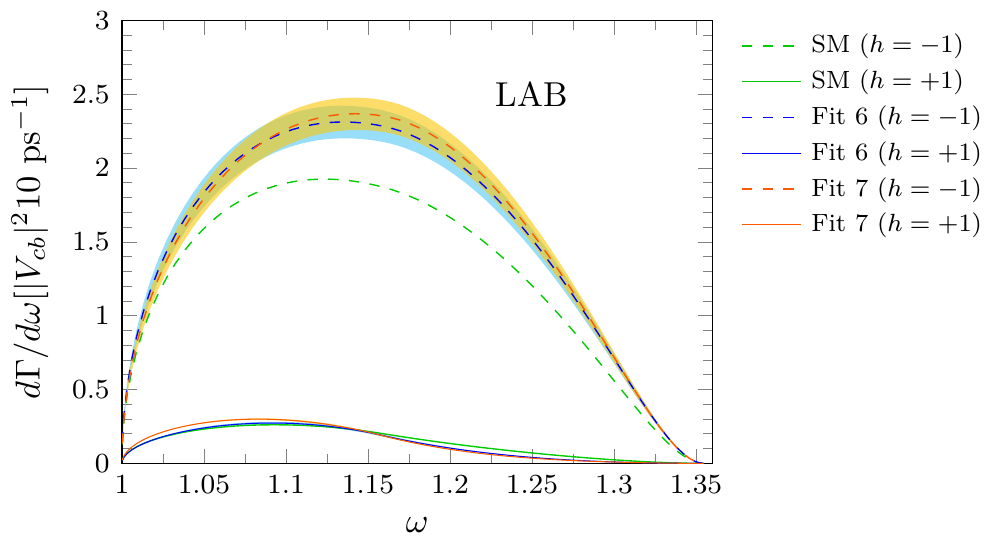}
\caption{ CM (left) and LAB (right) helicity decomposition of the $d\Gamma/d\omega$ differential decay width with a polarized $\tau$. We show distributions for $\bar B\to D\tau\bar\nu_\tau$ (top) and  $\bar B\to D^*\tau\bar\nu_\tau$ reactions (bottom), which have been  evaluated with  
Wilson coefficients and form factors
from Ref.~\cite{Murgui:2019czp}. Uncertainty bands  as in 
Fig.~\ref{fig:dgdwddstar}. }  
\label{fig:dgdwpoleD}
\end{figure}

\begin{figure}[tbh]
\includegraphics[height=4.cm]{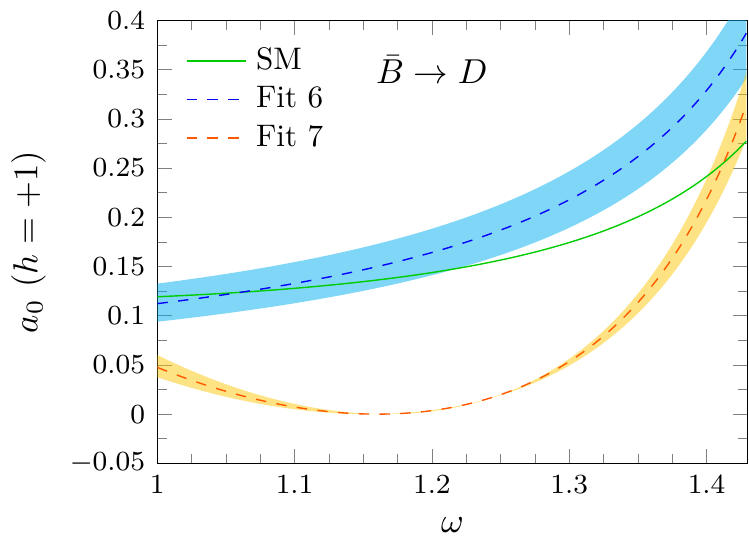} \hspace{.15cm}
\includegraphics[height=4.cm]{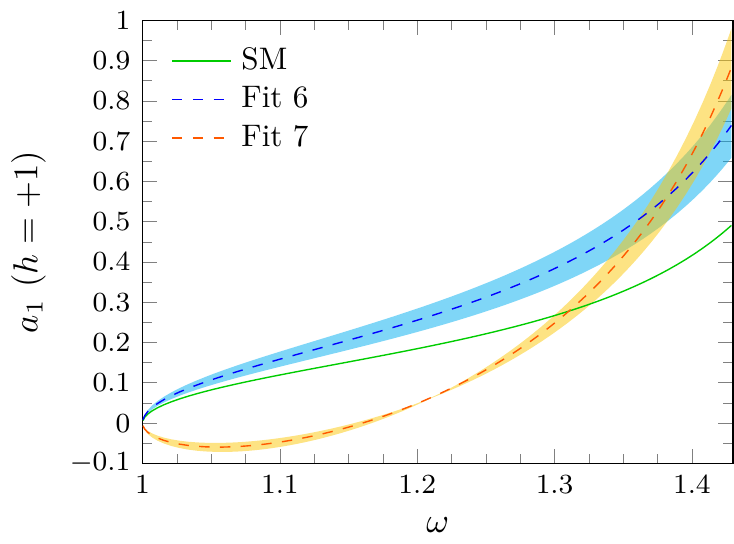}\hspace{.15cm}
\includegraphics[height=4.cm]{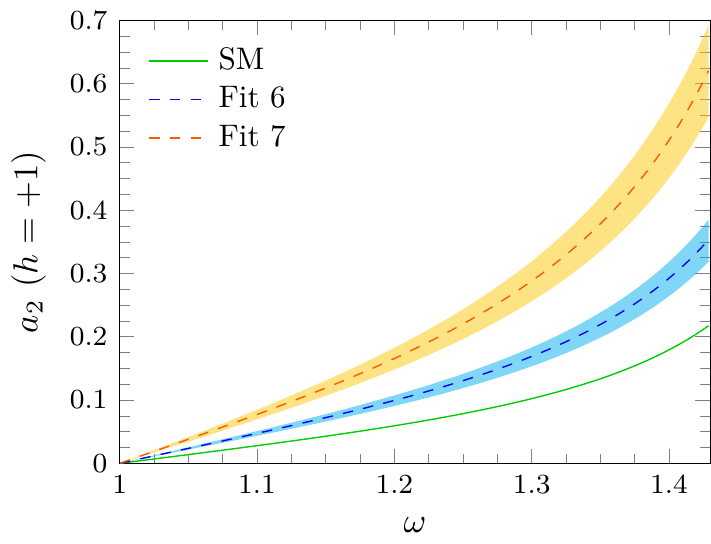}\hspace{.15cm}
\\
\includegraphics[height=4.cm]{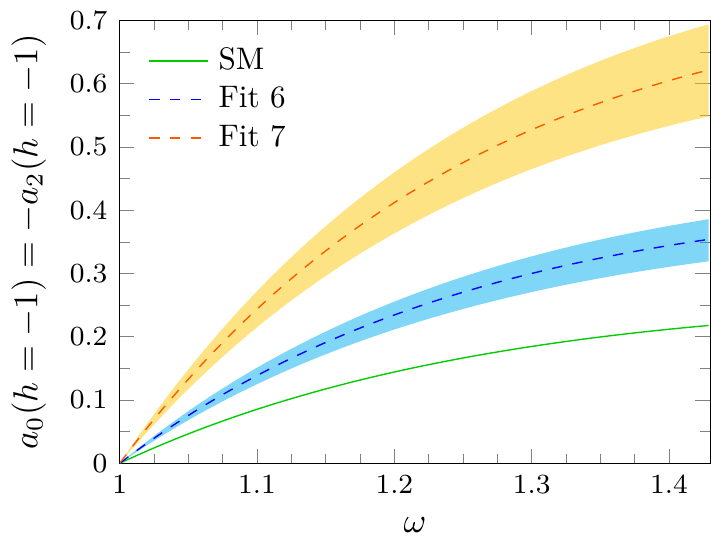} \hspace{.15cm}
\includegraphics[height=4.cm]{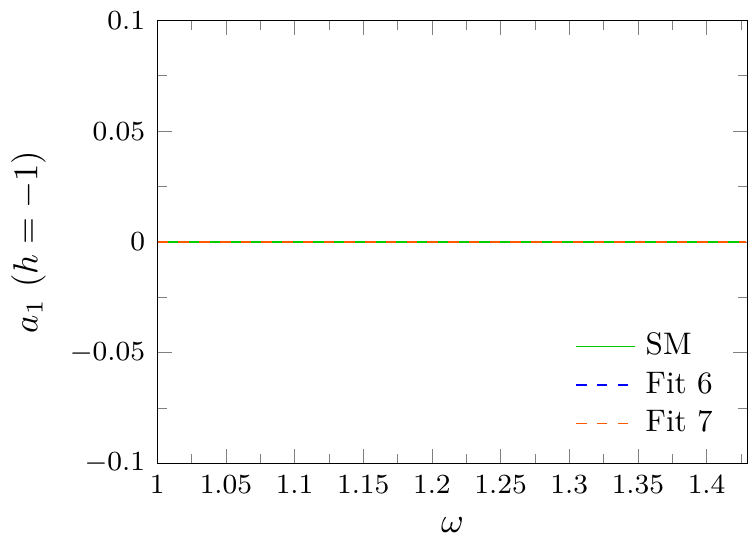}\hspace{.15cm}
\caption{ CM angular expansion coefficients for the 
$\bar B\to D\tau\bar\nu_\tau$ decay with a polarized $\tau$ with positive (upper panels)
and negative (lower panels) helicity.  They have been evaluated  with the Wilson coefficients and form factors
from Ref.~\cite{Murgui:2019czp}. Uncertainty bands  as in 
Fig.~\ref{fig:dgdwddstar}.
}  
\label{fig:aspoleD}
\end{figure}
\begin{figure}[tbh]
\includegraphics[height=4cm,width=5.6cm]{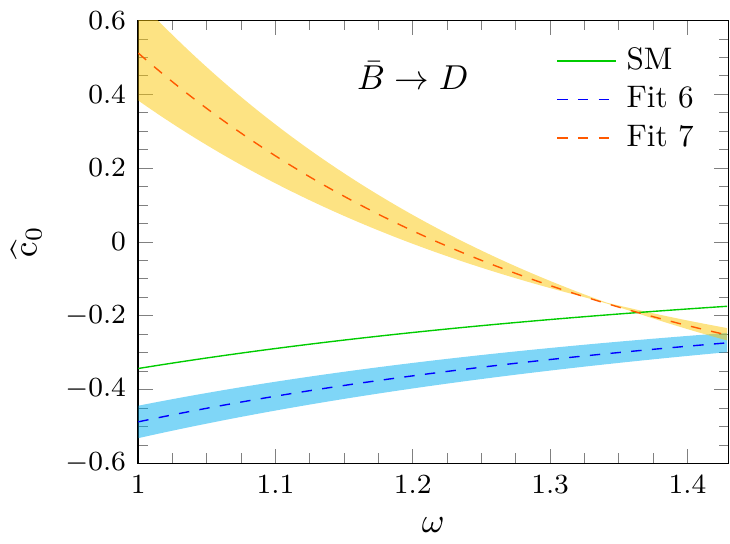}\hspace{.3cm}
\includegraphics[height=4cm,width=5.6cm]{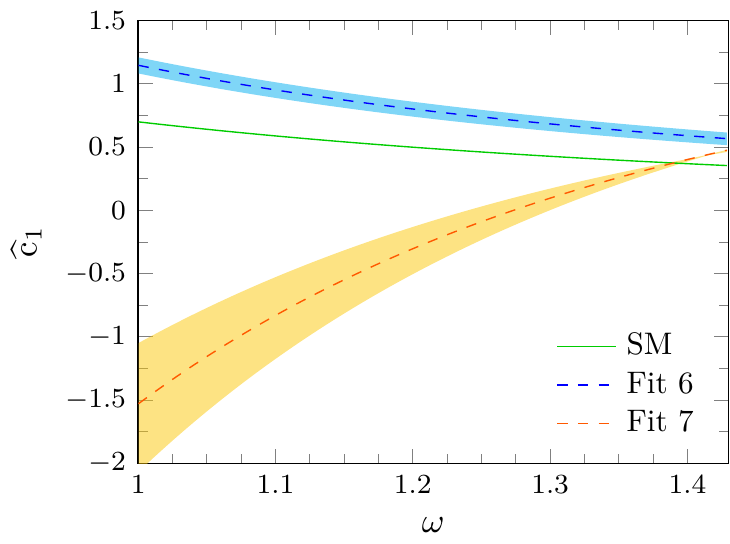}\hspace{.3cm}
\includegraphics[height=4cm,width=5.6cm]{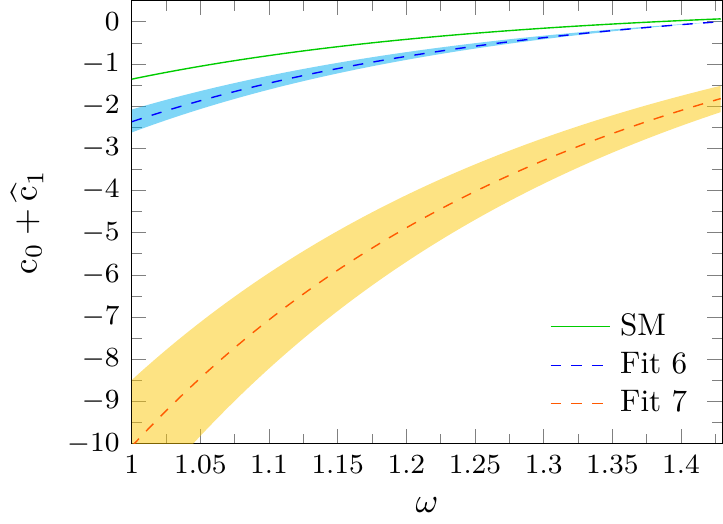} \\\vspace{0.25cm}
\includegraphics[height=4cm,width=4.2cm]{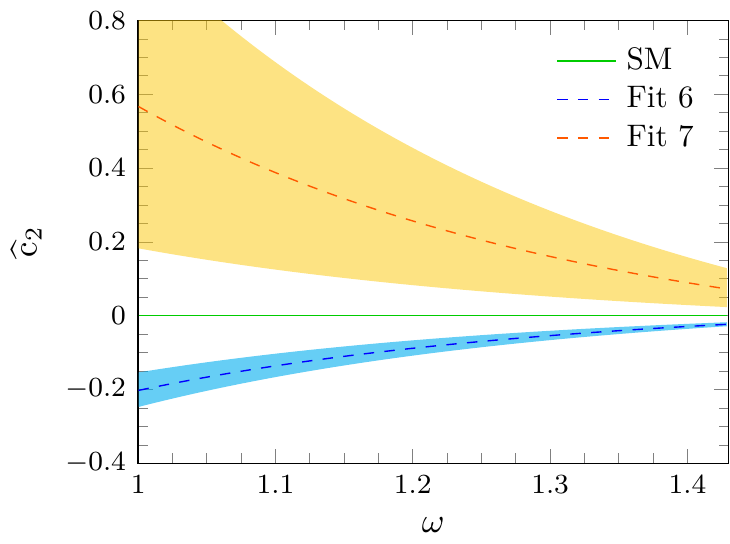}\hspace{.2cm}
\includegraphics[height=4cm,width=4.2cm]{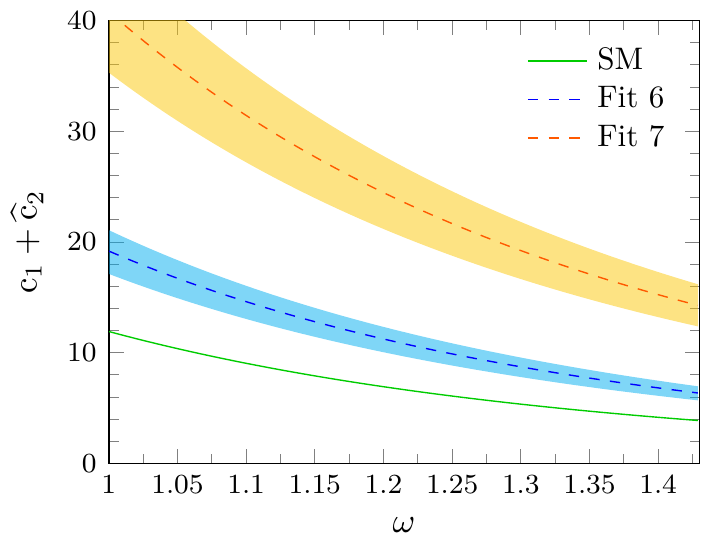} \hspace{0.2cm}
\includegraphics[height=4.2cm,width=4.2cm]{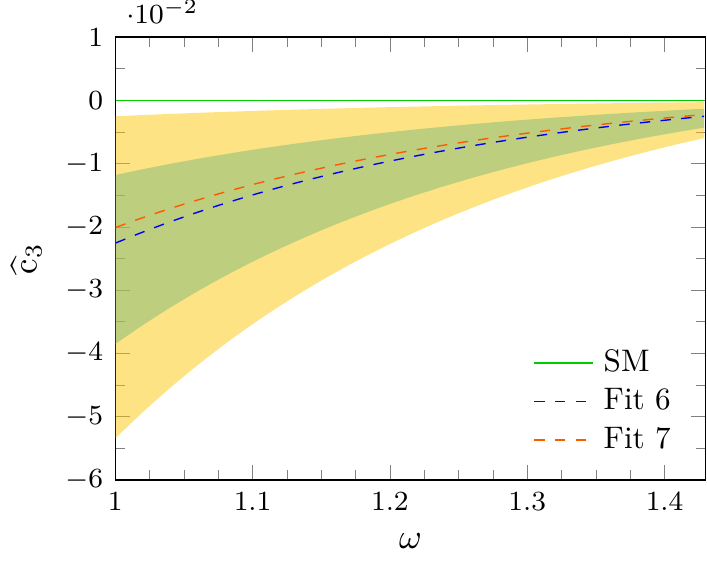}\hspace{0.2cm}
\includegraphics[height=4.2cm,width=4.2cm]{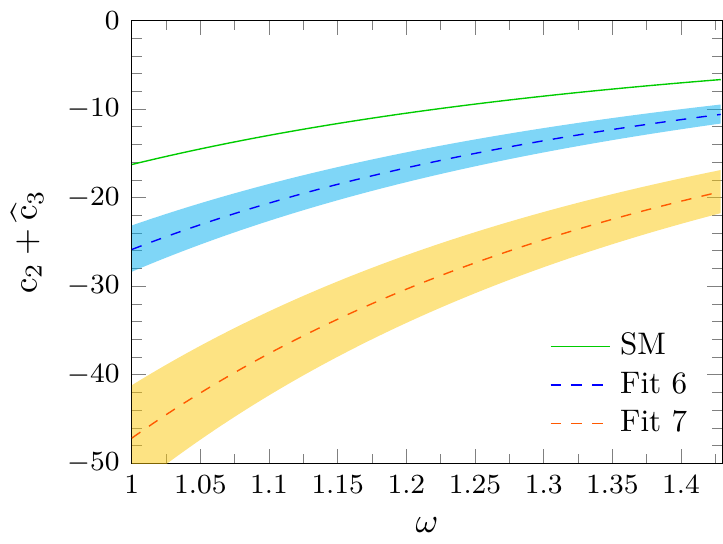}
\caption{ LAB  charged lepton energy expansion coefficients $\widehat c_{0,1,2,3}(\omega)$  for the 
polarized $\bar B\to D\tau\bar\nu_\tau$ decay.  We also show the $(c_0+\widehat c_1)$, $(c_1+\widehat c_2)$ and $(c_2+\widehat c_3)$ sums in the third top, second and fourth bottom panels, respectively. All quantities have been evaluated  with the Wilson coefficients and form factors
from Ref.~\cite{Murgui:2019czp}. Uncertainty bands  as in 
Fig.~\ref{fig:dgdwddstar}.   }
\label{fig:cspoleD}
\end{figure}
\begin{figure}[tbh]
\includegraphics[height=4.cm]{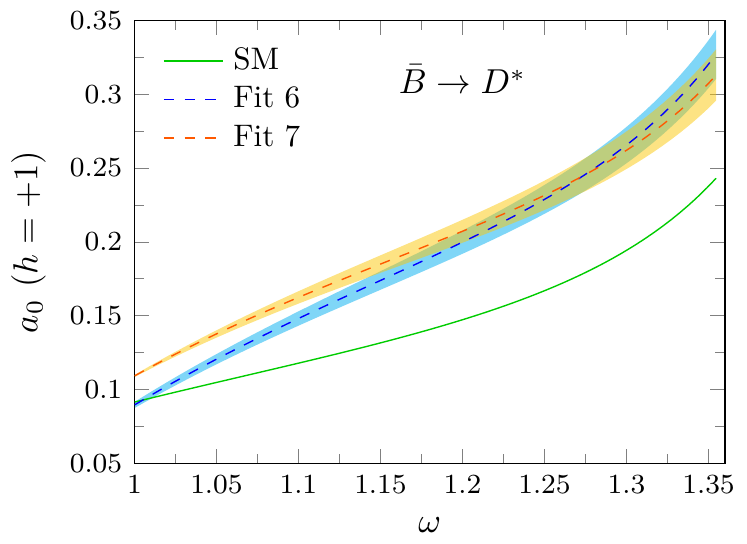} \hspace{.15cm}
\includegraphics[height=4.cm]{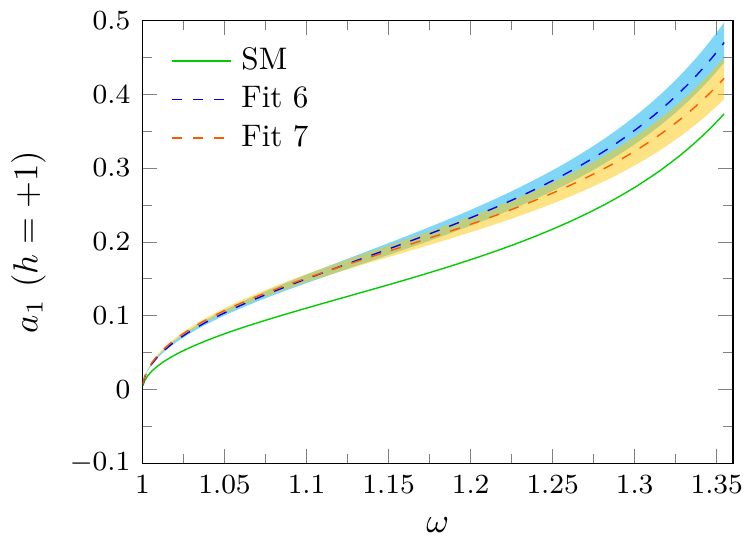}\hspace{.15cm}
\includegraphics[height=4.cm]{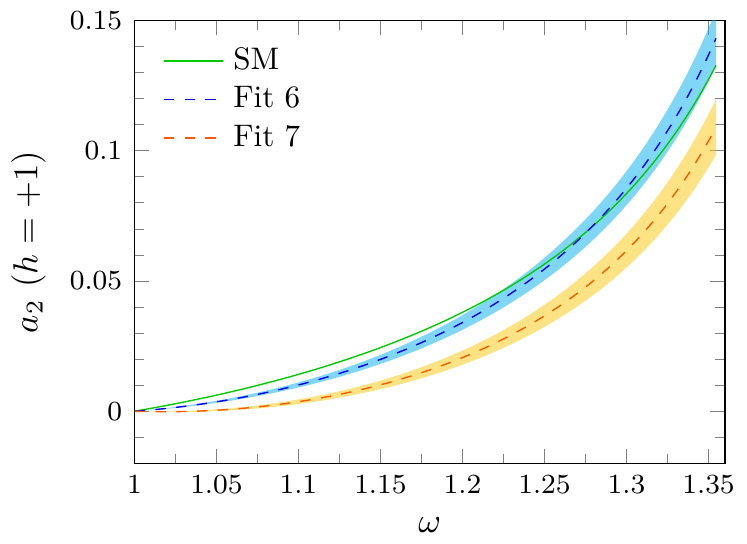}\hspace{.15cm}
\\
\includegraphics[height=4.cm]{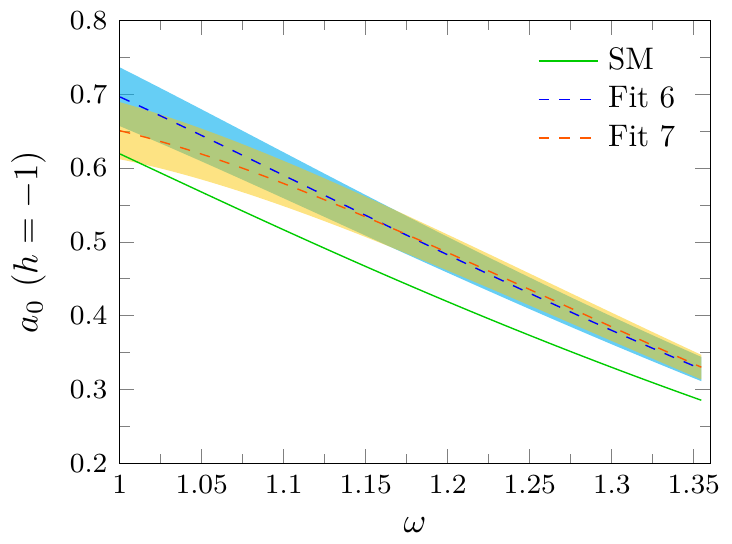} \hspace{.15cm}
\includegraphics[height=4.cm]{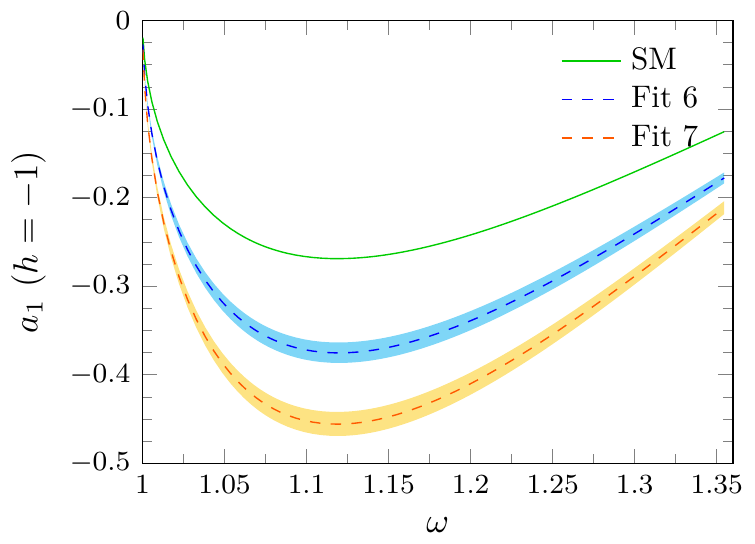}\hspace{.15cm}
\includegraphics[height=4.cm]{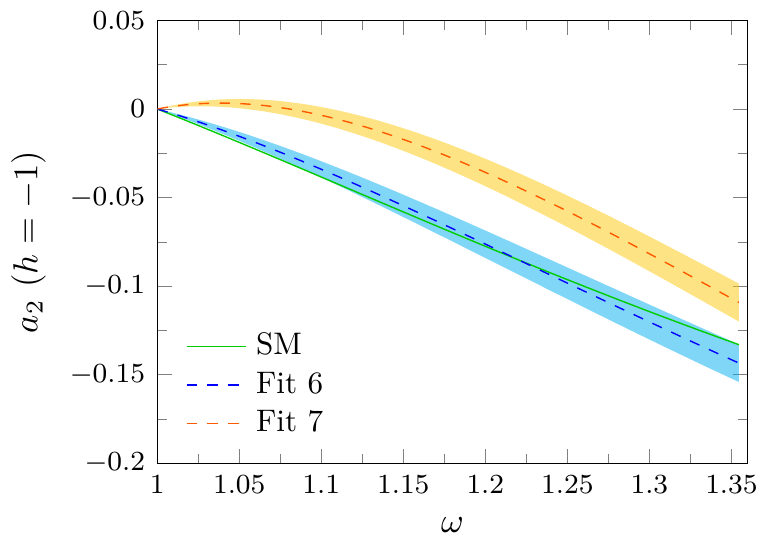}\hspace{.15cm}
\caption{ CM angular expansion coefficients for the 
$\bar B\to D^*\tau\bar\nu_\tau$ decay with a $\tau$ with positive (upper panels)
and negative (lower panels) helicity. 
They have been evaluated  with the Wilson coefficients and form factors
from Ref.~\cite{Murgui:2019czp}. Uncertainty bands  as in 
Fig.~\ref{fig:dgdwddstar}.  
}  
\label{fig:aspolDstar}
\end{figure}
\begin{figure}[tbh]
\includegraphics[height=4cm,width=5.6cm]{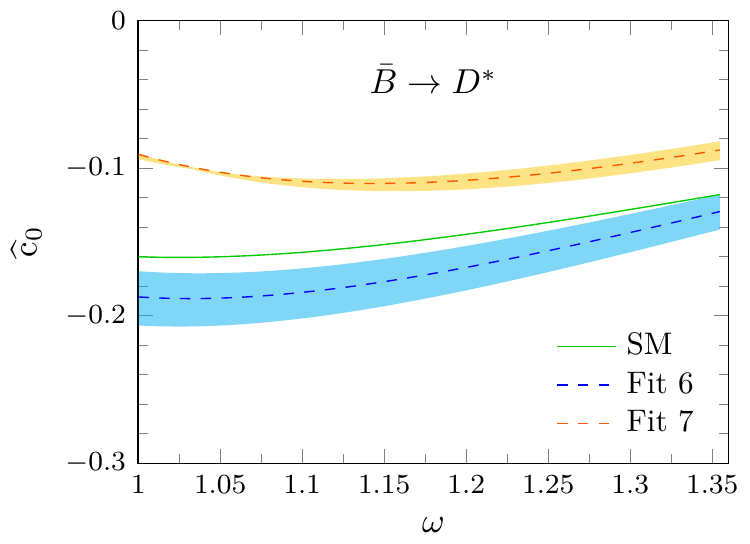}\hspace{.3cm}
\includegraphics[height=4cm,width=5.6cm]{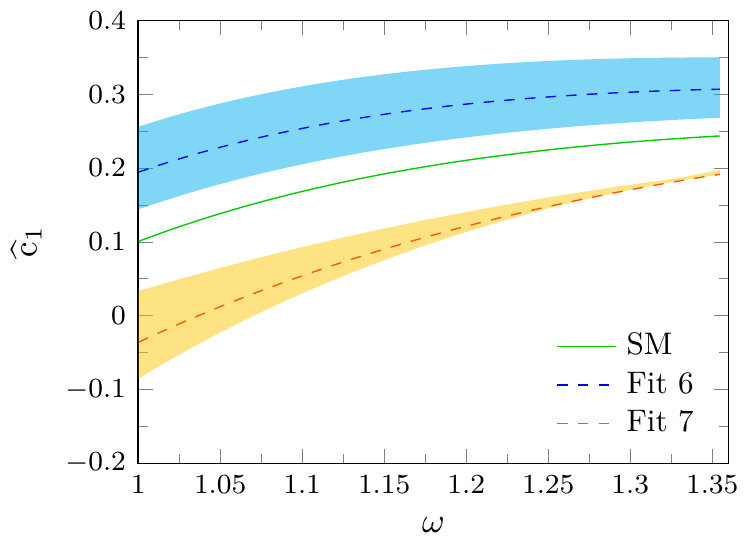}\hspace{.3cm}
\includegraphics[height=4cm,width=5.6cm]{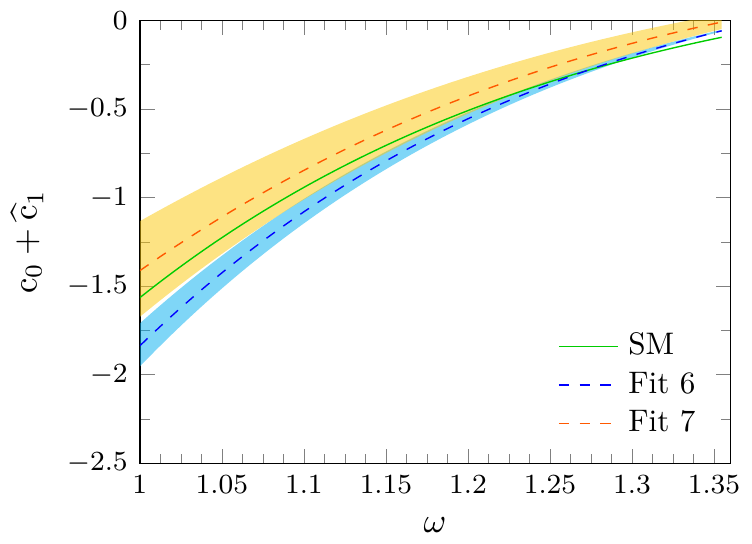} \\\vspace{0.25cm}
\includegraphics[height=4cm,width=4.2cm]{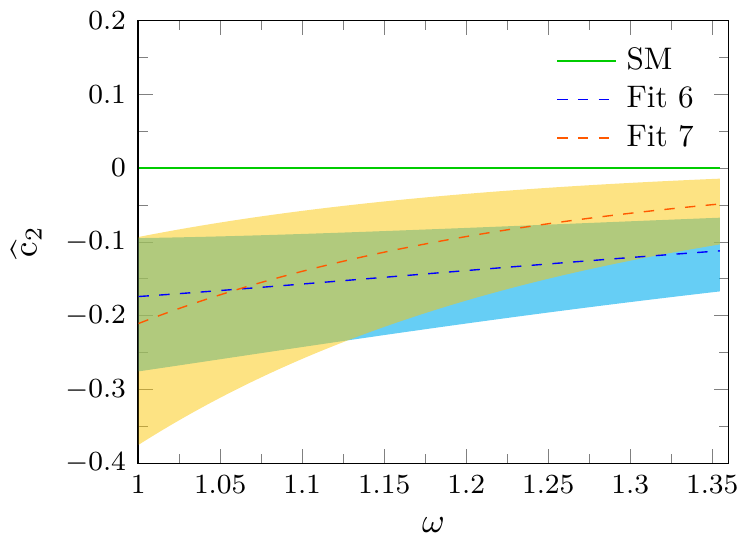}\hspace{.2cm}
\includegraphics[height=4cm,width=4.2cm]{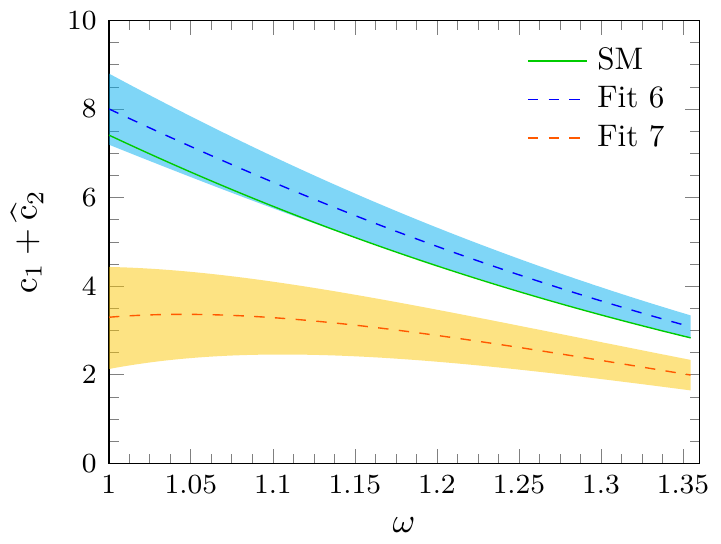} \hspace{0.2cm}
\includegraphics[height=4.2cm,width=4.2cm]{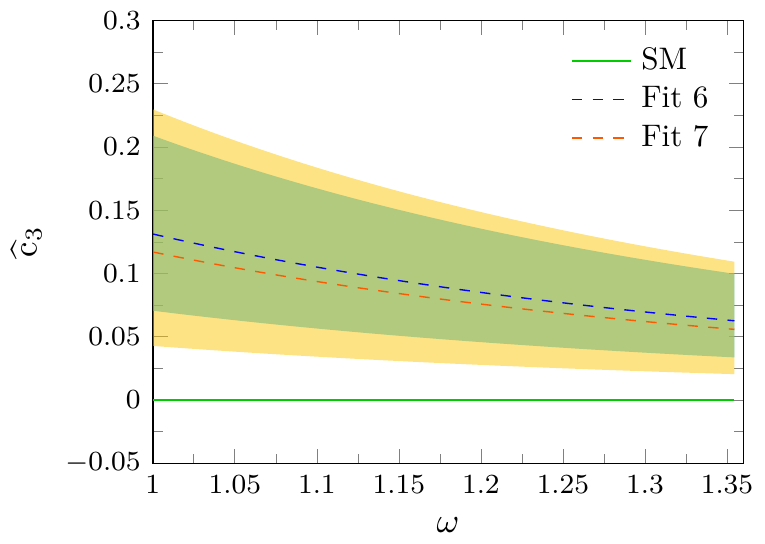}\hspace{0.2cm}
\includegraphics[height=4.2cm,width=4.2cm]{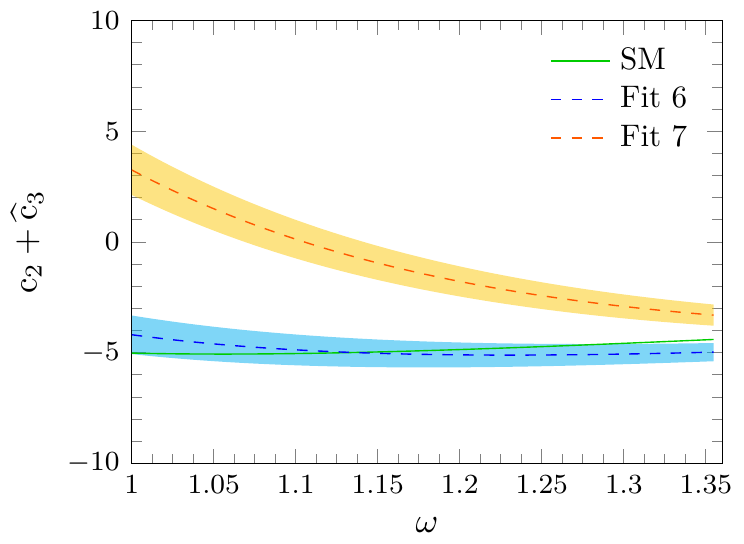}
\caption{ LAB  charged lepton energy expansion coefficients $\widehat c_{0,1,2,3}(\omega)$  for the 
polarized $\bar B\to D^*\tau\bar\nu_\tau$ decay.  We also show the $(c_0+\widehat c_1)$, $(c_1+\widehat c_2)$ and $(c_2+\widehat c_3)$ sums in the third top, second and fourth bottom panels, respectively. All quantities have been evaluated  with the Wilson coefficients and form factors
from Ref.~\cite{Murgui:2019czp}. Uncertainty bands  as in 
Fig.~\ref{fig:dgdwddstar}.   
}  
\label{fig:cspolDstar}
\end{figure}
\begin{figure}[tbh]
\includegraphics[height=4cm]{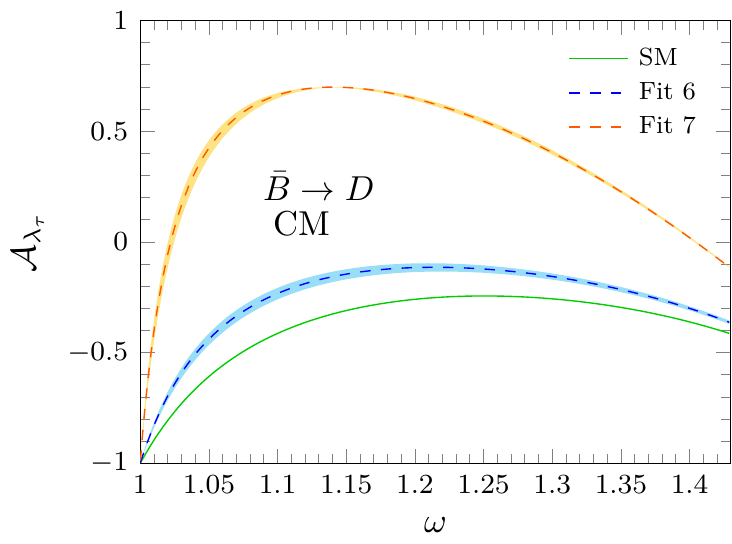} \hspace{.3cm}
\includegraphics[height=4cm]{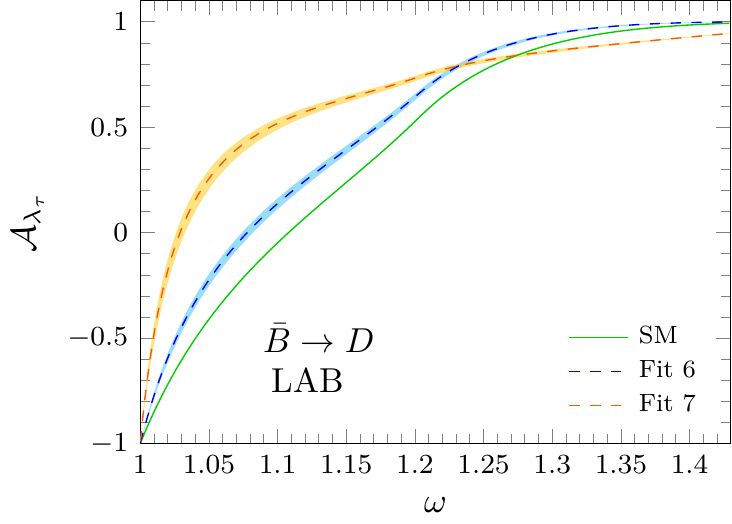}\\

\includegraphics[height=4cm]{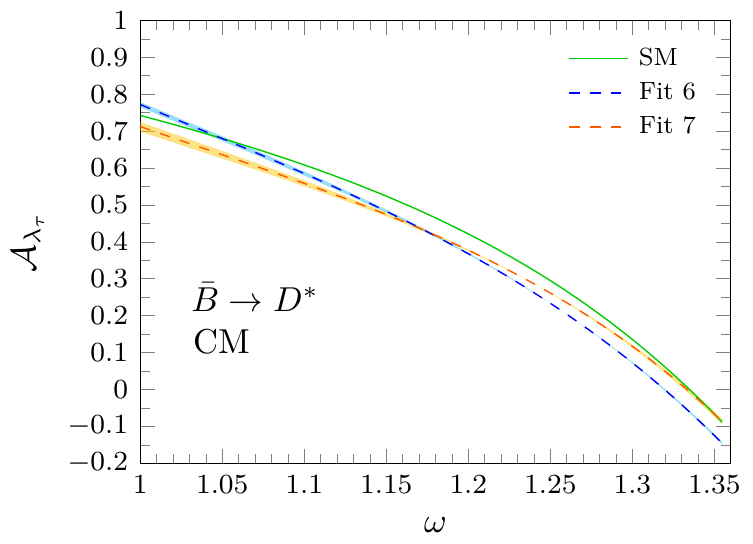} \hspace{.4cm}
\includegraphics[height=4.1cm]{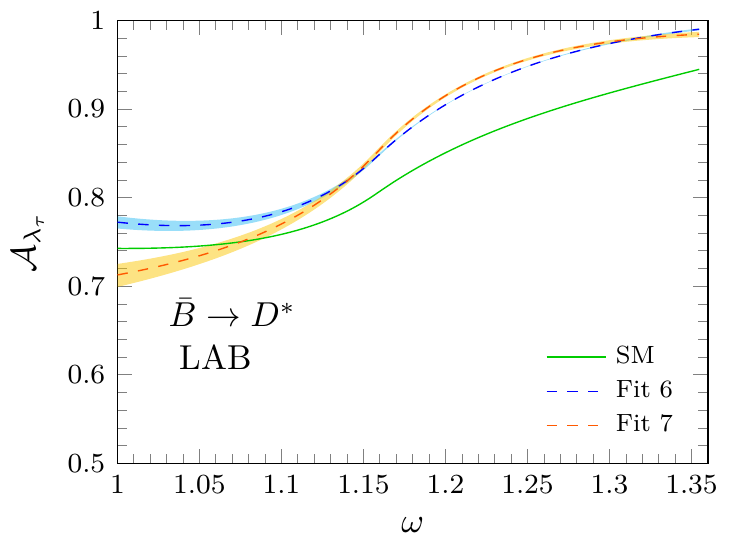}
\caption{$\tau$ polarization asymmetry ${\cal A}_{\lambda_\tau}$ for the 
$\bar B \to D$ (upper panels) and $\bar B \to D^*$ (lower panels) semileptonic decays measured in the CM (left panels) and LAB (right panels) frames. 
 All quantities have been evaluated  with the Wilson coefficients and form factors
from Ref.~\cite{Murgui:2019czp}. Uncertainty bands  as in  Fig.~\ref{fig:dgdwddstar}.}  
\label{fig:poltauasim-ddstar}
\end{figure}

In this appendix we collect in Figs.~\ref{fig:dgdwpoleD}--\ref{fig:poltauasim-ddstar},  results for $\bar B\to D\tau\bar\nu_\tau$ and
$\bar B\to D^*\tau\bar\nu_\tau$ decays where the final $\tau$ has well defined helicity
in the CM or LAB frames. All observables have been evaluated  with the NP  Wilson coefficients of Fits 6 and 7 and form factors
from Ref.~\cite{Murgui:2019czp}.

We obtain predictions that are qualitatively  similar to those discussed in Sec.~\ref{sec:Bc2cc} for  
$\bar B_c\to\eta_c$ and $\bar B_c\to J/\psi$ semileptonic decays. We would like to stress that unlike the unpolarized case, where all the accessible observables could be determined either from the CM  or LAB distributions, in the polarized case, the LAB and CM charged lepton helicity distributions provide complementary  information. Actually both differential distributions $d^2\Gamma/(d\omega d\cos\theta_\ell)$ 
and $d^2\Gamma/(d\omega dE_\ell)$ should be simultaneously used to determine the five new independent functions ${\cal A}_H, {\cal B}_H, {\cal C}_H, {\cal D}_H$ and ${\cal E}_H$, which appear for   the case of a polarized final $\tau$ (see Eq.~(23) of Ref.~\cite{Penalva:2020xup}).


\bibliography{B2Dbib}

\end{document}